\documentclass[prb,twocolumn,floatfix]{revtex4}

\usepackage{amsmath}
\usepackage{epsfig}

\renewcommand{\vec}[1]{{\rm\bf #1}}
\newcommand{\Tr}{\mathop{\mathrm{Tr}}}
\newcommand{\Av}[1]{\left\langle{#1}\right\rangle}

\newcommand{\Div}{\mathop{\mathrm{div}}}

\renewcommand{\Im}{\mathop{\mathrm{Im}}}
\renewcommand{\Re}{\mathop{\mathrm{Re}}}

\newcommand{\vep}{\varepsilon}
\newcommand{\ep}{\epsilon}
\newcommand{\df}{f}
\newcommand{\Stoss}{\mathop{\mathrm{St}}\nolimits}

\newcommand{\Kvec}{\vec{K}}
\newcommand{\Kpnt}{K}
\newcommand{\unitmatrix}{\openone}
\newcommand{\repplus}{+}
\newcommand{\reptimes}{\times}
\newcommand{\parity}{U_{C_2}}
\newcommand{\bloch}{\mathcal{U}}
\newcommand{\blochph}{\underline{\mathcal{X}}}
\newcommand{\phonon}{\mu}
\newcommand{\photon}{\ell}
\newcommand{\deviation}{w}

\begin{document}

\title{Theory of resonant multiphonon Raman scattering in graphene}
\author{D.~M.~Basko}\email{basko@sissa.it}
\affiliation{International School of Advanced Studies (SISSA), via
Beirut~2-4, 34014 Trieste, Italy}

\begin{abstract}
We present a detailed calculation of intensities of two-phonon 
and four-phonon Raman peaks in graphene. Writing the low-energy 
hamiltonian of the interaction of electrons with the crystal vibrations
and the electromagnetic field from pure symmetry considerations,
we describe the system in terms of just a few independent coupling
constants, considered to be parameters of the theory. The electron
scattering rate is introduced phenomenologically as another parameter.

The results of the calculation are used to extract information about
these parameters from the experimentally measured Raman peak
intensities. In particular, the Raman intensities are sensitive to the
electron scattering rate, which is not easy to measure by other techniques.
Also, the Raman intensities depend on electron-phonon coupling constants;
to reproduce the experimental results, one has to take into account
renormalization of these coupling constants by electron-electron interaction.
\end{abstract}

\maketitle

\tableofcontents

\section{Introduction}

In the past decades, Raman spectroscopy\cite{Raman} techniques
were successfully applied to carbon compounds, such as graphite
(see Ref.~\onlinecite{ThomsenReich2004} and references therein)
and carbon nanotubes.\cite{Kastner1994,Rao1997} Upon the
discovery of graphene,\cite{Novoselov2004} Raman spectroscopy
has proven to be a powerful tool to identify the number of layers,
structure, doping, disorder, and to characterize the phonons and
electron-phonon coupling.\cite{Ferrari2006,Gupta2006,Graf2007,
Yan2007,Pisana2007,Casiraghi2007,Das2008} So far, most of the
attention was focused on the position and width of the Raman peaks.

Here we present a detailed calculation of the {\em intensities}
of the multiphonon Raman peaks in graphene. Raman scattering
involves an electron-hole pair as an intermediate state; we
show that the multiphonon Raman peaks are strongly sensitive to
the dynamics of this electron-hole pair. Thus, Raman scattering
can be used as a tool to probe this dynamics.
Writing the low-energy 
hamiltonian of the interaction of electrons with the crystal vibrations
and the electromagnetic field from pure symmetry considerations,
we describe the system in terms of just a few independent coupling
constants, considered to be parameters of the theory. The electron
scattering rate is introduced phenomenologically as another parameter.
The results of the present calculation are used to extract information
about these parameters from the Raman peak intensities, measured
experimentally. 

As shown below, the Raman intensities strongly depend
on the electron scattering rate; moreover, the electron-phonon and
electron-electron contributions to this rate can be separated.
This is especially important as
there are very few techniques giving experimental access to electron
scattering rates, which, in turn, determine the transport properties
of graphene samples.
Besides, the quasiclassical character
of the process imposes a severe restriction on the electron
and hole trajectories which can contribute to the two-phonon
Raman scattering: upon the phonon emission the electron and
the hole must be scattered backwards. This restriction results
in a significant polarization memory: it is almost three times more
probable for the scattered photon to have the same polarization
as the incidend photon than to have the orthogonal polarization.

Also, the Raman intensities depend on electron-phonon coupling
constants; to reproduce the experimental results, one has to take
into account renormalization of these coupling constants by
electron-electron interaction. This renormalization is missed by
local or semi-local approximations to the density-functional theory,
typically used for the {\em ab initio} calculation of the coupling
constants.

\subsection{Fully resonant processes}\label{sec:fullresonance}

Since graphene is a non-polar crystal, Raman scattering involves
electronic excitations as intermediate states: the electromagnetic
field of the incident laser beam interacts primarily with the electronic
subsystem, and emission of phonons occurs due to electron-phonon
interaction. The matrix element of the process can be schematically
represented as
\begin{widetext}\begin{equation}\label{Ramanmatrixelement=}
\mathcal{M}\sim\sum_{s_0,\ldots,s_n}
\frac{\langle{i}|\hat{H}_{e-em}|s_0\rangle\langle{s_0}|\hat{H}_{e-ph}|s_1\rangle
\ldots\langle{s_{n-1}}|\hat{H}_{e-ph}|s_n\rangle\langle{s}_n|\hat{H}_{e-em}|f\rangle}
{(E_i-E_0+2i\gamma)(E_i-E_1+2i\gamma)\ldots(E_i-E_n+2i\gamma)}.
\end{equation}\end{widetext}
Here $|i\rangle$ is the initial state of the process (the incident
photon with a given frequency and polarization, and no excitations
in the crystal), $|f\rangle$ is the final state (the emitted photon
and $n$~phonons left in the crystal), while~$s_k$, $k=0,\ldots,n$,
label the intermediate states where no photons are present, but
an electron-hole pair is created in the crystal and $k$~phonons
have been emitted.
$E_i$~and~$E_k$, $k=0,\ldots,n$ are the energies of these
states, and $2\gamma$ is the inverse lifetime of the electron (hole)
due to collisions.  $\hat{H}_{e-em}$~and~$\hat{H}_{e-ph}$
stand for the terms in the system hamiltonian describing interaction
of electrons with the electromagnetic field and with phonons,
respectively.

In the calculations we do not include the phonon broadening, assuming
the phonon states to have zero width. First, this approximation is
consistent with the available experimental information: the phonon width
is about $10-20\:\mathrm{cm}^{-1}\approx{2}-3\:\mathrm{meV}$
at most,\cite{Yan2007,Pisana2007} while the electronic broadening is
at least an order of magnitude
higher (see the discussion below, Sec.~\ref{sec:2intensities}). Second,
this approximation is irrelevant provided that we calculate the
integrated intensities of the Raman peaks, since they are determined
by the total spectral weight of the phonon state which does not
depend on the phonon broadening.

\begin{figure}
\includegraphics[width=7cm]{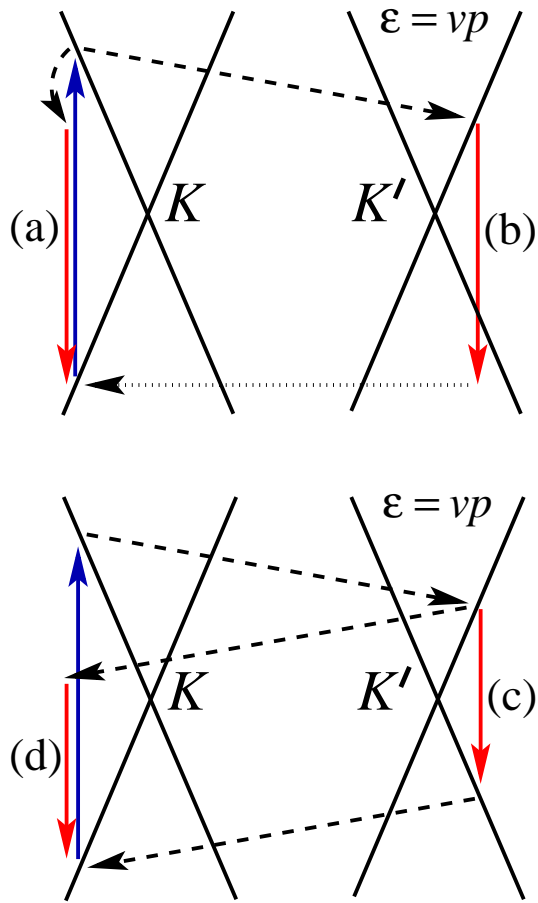}
\caption{\label{fig:res} (color online).
  Schematic representation of the role of
  electron dispersion (Dirac cones, shown by solid lines) in the
  one-phonon (a,b) and two-phonon Raman scattering (c,d). Vertical
  solid
  arrows represent interband electronic transitions accompanied by
  photon absorption or emission (photon wave vector is neglected),
  dashed arrows represent phonon emission, the horisontal dotted
  arrow represents the impurity scattering.}
\end{figure}

The photon wave vector is negligible, so momentum conservation
requires that the sum of the wave vectors of emitted phonons
must vanish (provided that the impurity scattering is neglected).
For the same reason Raman
scattering on one intervalley phonon must be impurity-assisted
[process~(b) in Fig.~\ref{fig:res}, giving rise to the so-called
$D$~Raman peak]. $D$~peak is absent in the experimental Raman
spectrum of graphene,\cite{Ferrari2006} showing that impurity
scattering is indeed negligible in these samples.

Looking at the intermediate electronic states involved in the Raman
scattering (Fig.~\ref{fig:res}), we notice that for one-phonon scattering
[processes (a),~(b)] at least one intermediate state must be virtual,
since energy and momentum conservation cannot be satisfied
simultaneously in all processes. Thus, at least one of the factors in the
denominator of Eq.~(\ref{Ramanmatrixelement=}) must be of the order
of the phonon frequency~$\omega_{ph}$ [for the impurity assisted
scattering one of the electron-phonon matrix elements in the numerator
of Eq.~(\ref{Ramanmatrixelement=}) should be replaced by the
electron-impurity matrix element]. For the two-phonon scattering
[process~(c)] all intermediate states can be real, so that all energy
mismatches in the denominator of Eq.~(\ref{Ramanmatrixelement=})
can be nullified simultaneously and the result is determined by the
electron scattering rate~$2\gamma$. We emphasize the
qualitative difference between the {\em fully resonant}
process~(c) and the double-resonant\cite{ThomsenReich2000}
process~(b), where one intermediate state is still virtual.
We also note the analogous difference between the two-phonon
processes (c)~and~(d) in Fig.~\ref{fig:res}: only the process~(c)
is fully resonant, the other one involves an energy mismatch of
$2\omega_{ph}$. As a result, its amplitude will be smaller by
a factor $\sim\gamma/\omega_{ph}$.

Obviously, these arguments can be extended to all multi-phonon
processes with odd and even number of phonons involved: in order
to annihilate radiatively, the electron and the hole must have
opposite momenta; if the total number of emitted phonons is odd,
the electron and the hole must emit a different number of phonons,
which is incompatible with energy conservation in all processes.
Our main focus will thus be on even-phonon
processes, as their intensities are determined by the electronic
scattering.

The full resonance picture presented above assumes the mirror
symmetry between the electron and the hole spectra. The
electron-hole asymmetry can be included as a correction to the
Dirac spectrum: the electron and the hole energies can be written
as $vp\pm\alpha_0{p}^2$, where $p$~is the momentum counted from the
Dirac point, $v$~is the Dirac velocity, and $\alpha_0$~is the
asymmetry parameter. 
The energy scale~$\Delta_{eh}$, quantifying
the role of the asymmetry in the Raman scattering is defined as
$\Delta_{eh}=\alpha_0(\omega_{in}^2-\omega_{out}^2)/(2v)^2$,
where $\omega_{in}$~and~$\omega_{out}$ are the frequencies of the
incident and the scattered photon
(the details are given in Secs.~\ref{sec:warping},~\ref{sec:Intdev}).
Namely, the arguments of the
previous paragraph hold if $\Delta_{eh}\ll\gamma$.
In the opposite case, it is $\Delta_{eh}$ that determines the smallest
value of the denominators in Eq.~(\ref{Ramanmatrixelement=}).
We will always assume that both $\gamma,\Delta_{eh}\ll\omega_{ph}$.

In the real space, the typical size of the region of space probed by the
electron-hole pair in the fully resonant two-phonon Raman scattering, is
$\sim{v}/\max\{\gamma,\Delta_{eh}\}$. For the doubly resonant
defect-induced one-phonon scattering, the inverse energy
mismatch~$1/\omega_{ph}$ determines the time duration of the
process by virtue of the uncertainty principle, so the length scale of the
process in the real space is $v/\omega_{ph}$.
Although this length scale is much shorter than $v/\gamma$,
it is still much greater than the lattice constant or the electron
wavelength $v/\ep$. 

\subsection{Quasiclassical real-space picture}\label{sec:realspace}

The fully resonant Raman scattering, where the energy is conserved
in each of the elementary scattering processes, admits a simple
quasiclassical description, described qualitatively in this
subsection, and justified rigorously in
Secs.~\ref{sec:2raman},\ref{sec:4raman}.

Let us denote by~$\ep$ the energy of the electron and the
hole in the photoexcited pair. Initially, it is given by the half
of the excitation frequency~$\omega_{in}$,
$\ep=\omega_{in}/2\sim{1}\;\mathrm{eV}$. After the
emission of $n$~phonons it is decreased by~$n\omega_{ph}$;
assuming $\omega_{ph}\leq{0}.2\:\mathrm{eV}\ll\ep$, we
neglect this decrease in the qualitative considerations. Thus,
during all the time taken by the Raman scattering,
electron and hole can be viewed as wave packets of the size
$\sim{v}/\ep$, propagating across the crystal along classical
trajectories.

\begin{figure}
\includegraphics[width=8cm]{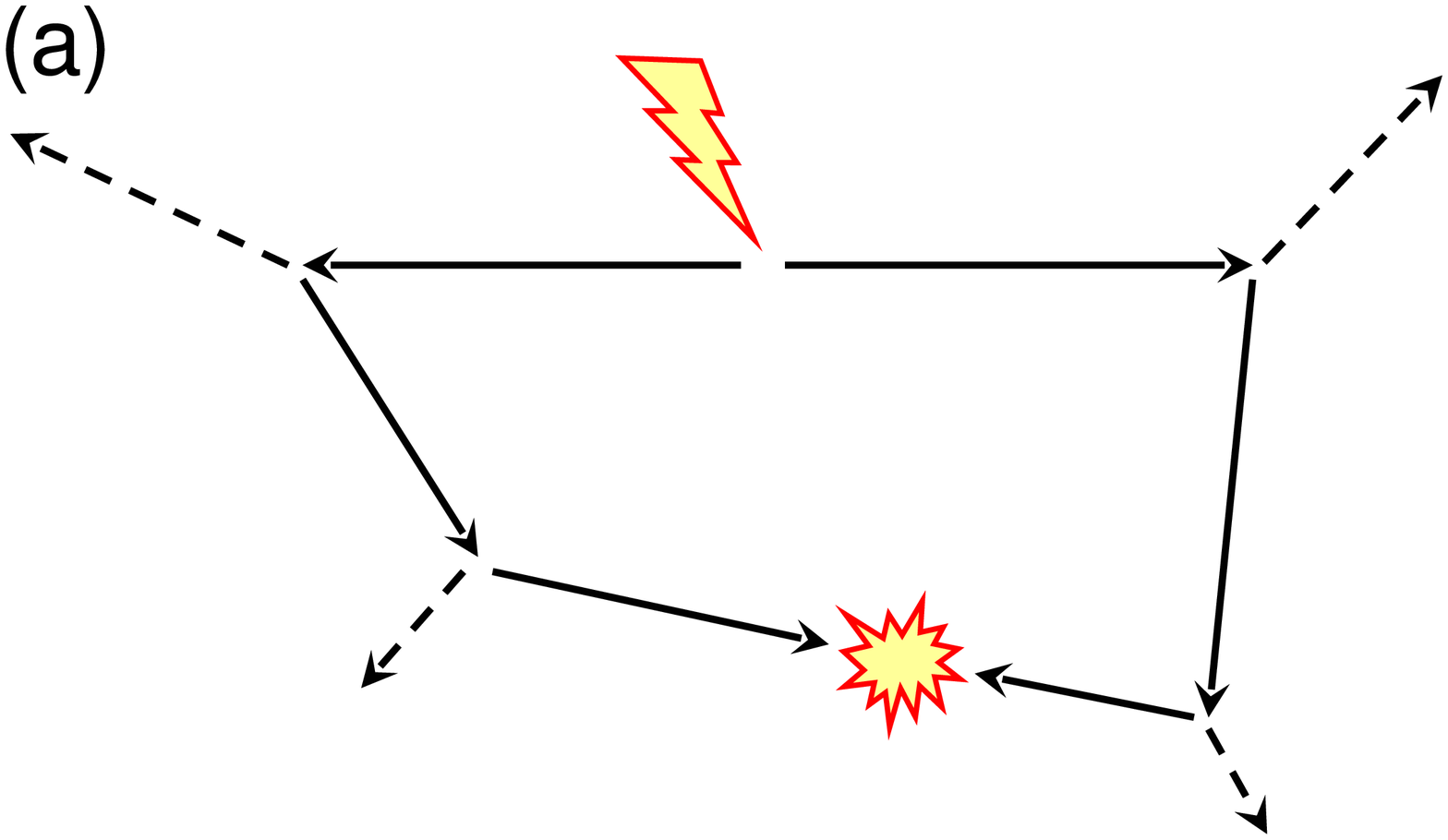}
\includegraphics[width=8cm]{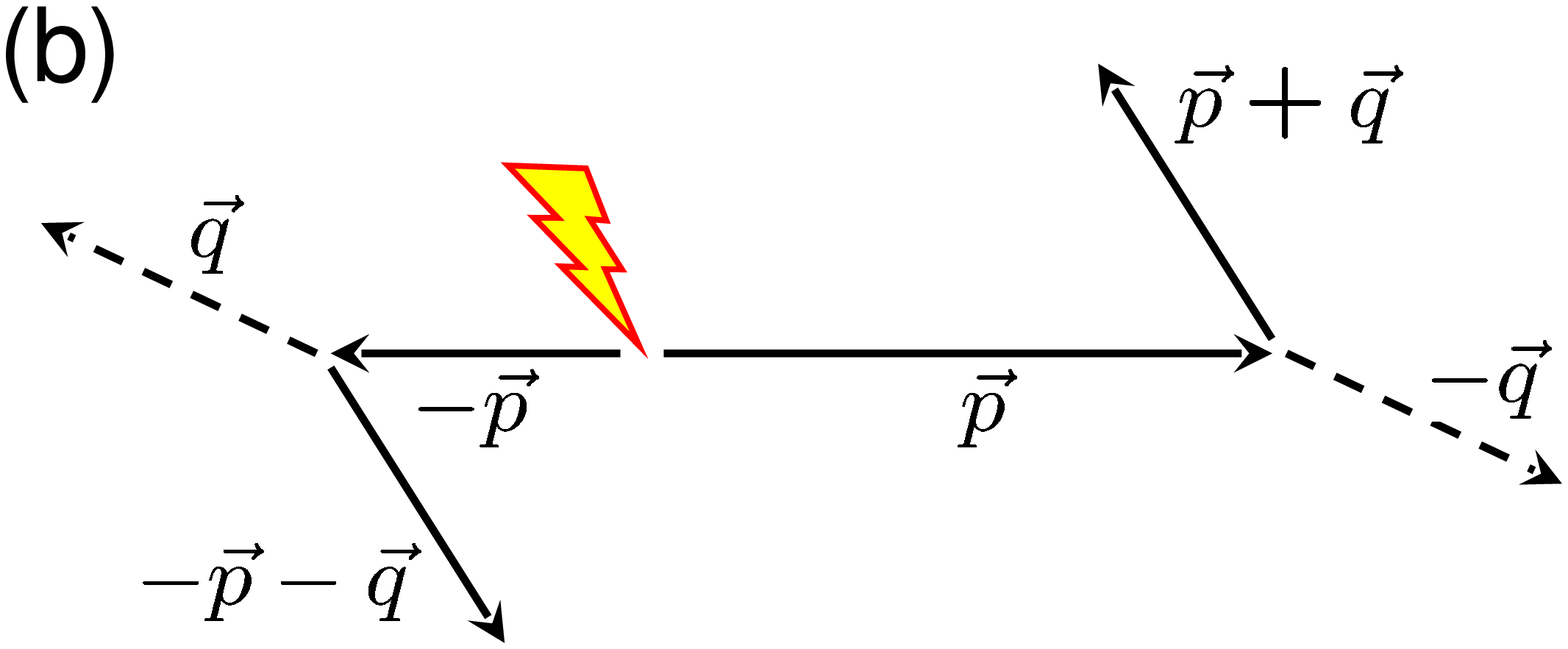}
\includegraphics[width=8cm]{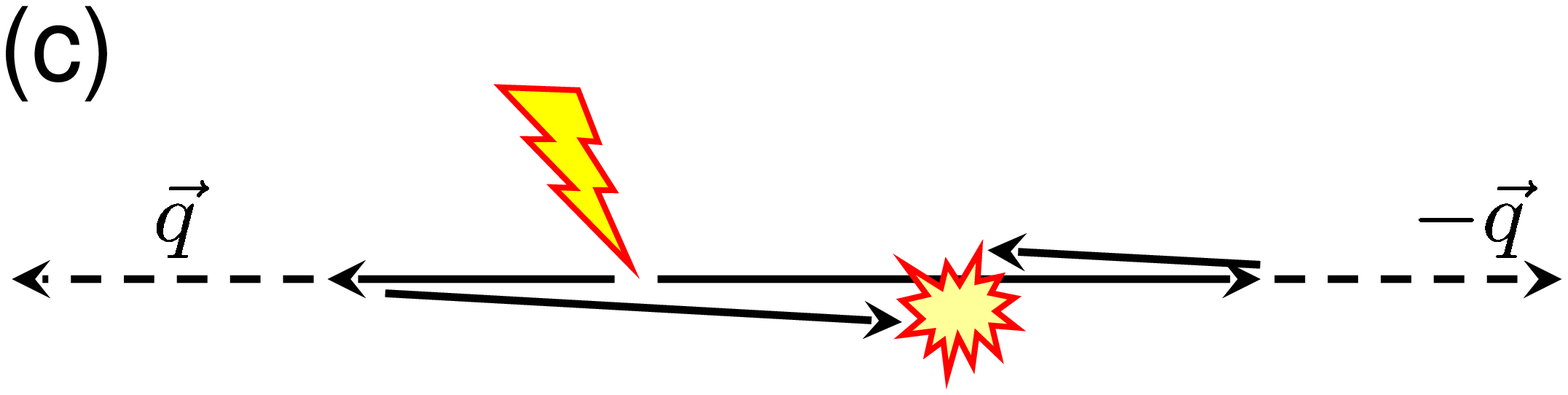}
\caption{\label{fig:trajectories} (color online).
(a) An example of a quasiclassical
electron-hole trajectory contributing to the four-phonon Raman
scattering.
(b,c)~Trajectories with emission of two phonons,
not contributing~(b) and contributing~(c) to the two-phonon
Raman scattering.
In all pictures the lightning represents the incident photon which
creates the pair.
The solid lines denote the free propagation of the electron
and the hole. The flash represents the radiative recombination
of the electron-hole pair.
The dashed lines denote the emitted phonons.}
\end{figure}

The electron and the hole are created in the same region of space
of the size $\sim{v}/\ep$ around some point~$\vec{r}_0$ at the
moment of the arrival of the excitation photon. At this initial
moment they have opposite momenta $\vec{p},-\vec{p}$, and
opposite velocities
$\vec{v},-\vec{v}$ (if the electron-hole asymmetry is taken
into account, the two velocities will have slightly different
magnitude), so they move along the straight lines, their positions
being $\vec{r}_e(t)=\vec{r}_0+\vec{v}t$,
$\vec{r}_h(t)=\vec{r}_0-\vec{v}t$. After a typical time
$t\sim{1}/\gamma$ they undergo some scattering processes
(e.~g., phonon emission), where their momenta and
(generally speaking) energies are changed. Each such elementary
scattering process occurs during a short time
$\sim{1}/\ep\ll{1}/\gamma$.
Thus, the trajectories of the electron and the hole after their creation
are represented
by broken lines, with the typical segment length $\sim{v}/\gamma$
(the electron mean free path).
The crucial point is that in order to recombine radiatively and
contribute to Raman signal, the electron and the hole should meet
again within a spatial region of the size $\sim{v}/\ep$,
and have opposite momenta. The latter condition automatically
implies that the number of the phonons emitted by the electron and
the hole is the same. These considerations are illustrated by
Fig.~\ref{fig:trajectories}.

In the presence of a significant electron-hole asymmetry,
$\Delta_{eh}\gg\gamma$, the described picture is modified. Namely,
one of the segments of either the electron or the hole trajectory
has the length $v/\Delta_{eh}$ instead of $v/\gamma$, the
corresponding time travelling being restricted by the phase mismatch
rather than by collisions.

\section{Summary of the main results}

\subsection{On the labelling of Raman peaks in graphene}

For the single-phonon Raman peaks the commonly accepted notations
are ``$G$'' for the peak at $1580\:\mathrm{cm}^{-1}$~corresponding
to emission of an optical phonon with zero wave vector, and ``$D$''
for the defect-induced peak at $1350\:\mathrm{cm}^{-1}$
corresponding to emission of an optical phonon with the wave vector
near $K$~or~$K'$ points of the Brillouin zone.\cite{ThomsenReich2004}
Sometimes one also distinguishes the so-called $D'$-peak at
$1620\:\mathrm{cm}^{-1}$. This peak is also defect-induced, and
corresponds to emission of an optical phonon with a small wave
vector $q\sim\ep/v$. As mentioned above, in the present work we study
only the clean graphene, hence $D$~and~$D'$ peaks are of no interest
to us.

Unfortunately, there is no single commonly accepted system for
labelling of the multiphonon Raman peaks. The strong peak at
$2700\:\mathrm{cm}^{-1}$, corresponding to emission of two
phonons with the opposite wave vectors near the $K$~and~$K'$
points, was historically called~$G'$ (as it is not defect-induced);
sometimes it is denoted by $2D$ or by~$D^*$ to stress that it
is the second overtone of the $D$~peak. The peak at
$3250\:\mathrm{cm}^{-1}$ corresponding to emission of two phonons
with two opposite wave vectors near the $\Gamma$~point is
sometimes called $G^*$, $2G$ or $2D'$. The latter notation
reflects the fact that the frequency of this peak is not exactly
the double of that of the $G$~peak, but rather the double of the
defect-induced $D'$~peak.

In the following we use the notation $n\Gamma+mK$ to denote
the peak corresponding to emission of $n$~phonons with wave
vectors within $\sim\ep/v$ from the $\Gamma$~point  and of
$m$~phonons with wave vectors within $\sim\ep/v$ from the
$K$~or~$K'$ points. For multiphonon peaks this nomenclature
is unambiguous.
Thus, the peaks at $2700\:\mathrm{cm}^{-1}$
and at $3250\:\mathrm{cm}^{-1}$ will be called $2K$ and
$2\Gamma$, respectively.

\subsection{One-phonon Raman processes}\label{sec:1intensities}

In the clean graphene the only one-phonon Raman process, allowed
by the momentum conservation corresponds to the emission of
the $E_2$~optical phonon with zero wave vector and frequency
$1580\:\mathrm{cm}^{-1}$. For this process
the situation turns out to be drastically different from that described
by the qualitative considerations of Sec.~\ref{sec:fullresonance}.
As shown in Sec.~\ref{sec:1raman}, if one approximates the
electron spectrum by the Dirac cones, the numerator of
Eq.~(\ref{Ramanmatrixelement=}) vanishes due to high symmetry
of the low-energy electronic Dirac hamiltonian (as compared to the
microscopic symmetry of the crystal). Thus, the main
contribution to the Raman amplitude comes from the regions of the
electronic Brillouin zone far from the Dirac points. As a consequence,
the typical energy mismatch
in the denominator of Eq.~(\ref{Ramanmatrixelement=}) is of the
order of the whole electronic bandwidth. Thus, the Raman process,
responsible for the $1580\:\mathrm{cm}^{-1}$~peak, is completely
off-resonant and the picture shown in Fig.~\ref{fig:res} for the
process~(a) is wrong.

As a result, the intensity of the peak is expected to be insensitive
to most external parameters: polarization, electron concentration,
degree of disorder, etc. To characterize this intensity, one has to
introduce an additional parameter into the theory which has no simple
relation to the parameters of the low-energy effective hailtonian.
The resulting intensity of the peak is given by Eq.~(\ref{IG=}); it
is proportional to the fourth power of the excitation frequency
which is the standard result for Raman scattering when the difference
between the frequencies of the incident and scattered photons is small.
This dependence also agrees with the
experimental results of Ref.~\onlinecite{Pimenta}.

Note that the results described above do not hold for the
defect-induced peak at $1350\:\mathrm{cm}^{-1}$. For this
peak the double resonance
picture,\cite{ThomsenReich2000} shown in Fig.~\ref{fig:res},
process~(b), is fully adequate.

\subsection{Two-phonon Raman processes}\label{sec:2intensities}

As the phonons are emitted by electrons with momentum
$\omega_{in}/(2v)$, the largest possible phonon momentum is
$q_{max}=(\omega_{in}+\omega_{out})/2v$,
corresponding to the electron and hole backscattering
(for the $2K$~peak at 2700~cm$^{-1}$ we
count the phonon momenta from the $K$~and~$K'$ points). It would
be natural to expect that any pair of phonons with opposite momenta 
$\vec{q},-\vec{q}$ and $|\vec{q}|\leq{q}_{max}$ can be emitted, the
only exception being $\vec{q}=0$ which is prohibited by
symmetry\cite{ThomsenReich2004} and the nearby ones which are
suppressed due to the smallness of the matrix elements. These arguments
would predict the width of the peak to be of the order of
$(v_{ph}/v)\omega_{in}$, where $v_{ph}$~is the $K$~phonon group
velocity; besides, the shape of the peak would be strongly asymmetric:
a sharp cutoff on the high-energy side at the frequency
$2\omega_{ph}(q_{max})$ due to the resonance restriction,
and a smooth drop-off towards zero at $2\omega_{ph}(q=0)$ due to the
matrix element suppression.
The phonon dispersion can be deduced from the dependence of the
frequency of the impurity-assisted one-phonon $D$~peak in graphite on
the excitation energy~$\omega_{in}$:
$d\omega_{ph}/d\omega_{in}=v_{ph}/v\approx{50}\:\mathrm{cm}^{-1}/\mathrm{eV}$.\cite{Vidano,Wang1990,Pocsik1998,Matthews1999}
Thus, for $\omega_{in}=2$~eV these arguments give the width of the
$2K$~peak to be about $200$~cm$^{-1}$.
However, the experimentally observed width is only about 30~cm$^{-1}$
at $\omega_{in}=2.2$~eV, and its shape is quite symmetric.\cite{Ferrari2006,Yan2007}

The observed small width of the peak is explained by the quasiclassical picture,
presented in Sec.~\ref{sec:realspace}. If upon the emission of phonons the
electron and the hole are scattered by an arbitrary angle, as shown in
Fig.~\ref{fig:trajectories}(b), they will not be able to meet at the same
spatial point in order to recombine radiatively and contribute to the
two-phonon Raman peak. Only if the scattering is backwards, this
event is possible, as illustrated by Fig.~\ref{fig:trajectories}(c). This condition
fixes the wave vectors of the emitted phonons to be $q=q_{max}=2\ep/v$.
The small deviations of the scattering angle from~$\pi$ are restricted by the
quantum diffraction, and the width of the two-phonon Raman peaks, instead
of being $\sim(v_{ph}/v)\omega_{in}$, is determined by a much smaller energy
scale (see the discussion below).

The dominance of the electron and hole backscattering manifests itself
in the polariazion memory of the Raman signal. If the incident light is
linearly polarized, the probability of excitation of the electron-hole
pair with a given direction of momenta is proportional to $\sin^2\varphi$,
where $\varphi$~is the angle between the electric field vector of the
light and the momenta. Thus, upon backscattering and radiative
recombination, the probability to detect a photon of the same polarization
as the original one is $\propto\sin^4\varphi$, and that of the orthogonal
polarization is $\propto\sin^2\varphi\cos^2\varphi$. Averaging
over~$\varphi$, we obtain the ratio of intensities for the detection
of polarization parallel and perpendicular to that of the incident light
to be $I_\|/I_\perp=3$. This ratio may be slightly decreased due to a
finite aperture (see the discussion in Sec.~\ref{sec:ramanprobab}).

The calculation of the intensities of the two-phonon Raman peaks is
performed in Sec.~\ref{sec:2raman}.
The explicit expressions for the intensities of the $2K$~and~$2\Gamma$
peaks, obtained under the assumption of Dirac spectrum for the electrons,
are represented by Eqs.~(\ref{I2Kend=}) and~(\ref{I2Gend=}). Both are
proportional to $1/\gamma^2$, where $2\gamma$~is the electron (hole)
inelastic scattering rate. If the latter is smaller that the electron-hole
asymmetry~$\Delta_{eh}$, then, according to the arguments of
Sec.~\ref{sec:fullresonance}, it is $\Delta_{eh}$ that restricts the
energy denominators from below. Formally, this results in the
replacement~(\ref{I2Ktrig=}) in both
Eqs.~(\ref{I2Kend=}),~(\ref{I2Gend=}).

Numerically,
$\alpha_0(1\;\mbox{eV})^2/v^2\sim{0}.1\;\mbox{eV}$ (see, e.~g.,
Ref.~\onlinecite{DresselhausBook}), so the relative correction to
Eq.~(\ref{I2Kend=}) for small $\Delta_{eh}$ can be estimated as
$-(1/2)\Delta_{eh}^2/(2\gamma)]^2/2%
\sim-{10}^{-4}(\omega_{in}/2\gamma)^2$. The total electronic
broadening $2\gamma$ was measured by time-resolved photoemission
spectroscopy to be 20~meV in Ref.~\onlinecite{Gao} and $25$~meV in
Ref.~\onlinecite{Moos} (all values taken for
$\ep=\omega_{in}/2=1$~eV). A recent angle-resolved photoemission
spectroscopy (ARPES) measurement gives a
significantly larger value for $2\gamma\sim{100}\:\mbox{meV}$
(Ref.~\onlinecite{Rotenberg}).
Thus, the case
$\gamma\gg\Delta_{eh}$ seems to be more relevant for the description
of experiments, than the opposite one.

The Raman matrix element corresponding to emission of two phonons
with given wave vectors $\vec{q}$ and $-\vec{q}$ is given by
Eq.~(\ref{Mq2ph=}) for $\gamma\gg\Delta_{eh}$.
From this dependence one can deduce the lineshape of the two-phonon
peaks $2K,2\Gamma$:
\begin{equation}\label{dI2dw=}
\frac{dI_{2\phonon}}{d\omega}\propto
\frac{1}{[(v/v_{ph,\phonon})^2(\omega/2-\omega_\phonon)^2+4\gamma^2]^{3/2}},
\end{equation}
where $\phonon=K,\Gamma$, $2\omega_\phonon$ is the central
frequency for each peak, and $v_{ph,\phonon}$~is the group velocity
of the corresponding phonon. Thus, the full width at half maximum (FWHM)
of each peak is given by
\begin{equation}
FWHM_{2\phonon}=\sqrt{2^{2/3}-1}\,\frac{v_{ph,\phonon}}v\,8\gamma
\approx{0}.77\,\frac{v_{ph,\phonon}}v\,8\gamma.
\end{equation}
For $\gamma\ll\Delta_{eh}$ the dependence of the Raman matrix element
on~$q$ is described by Eq.~(\ref{warpedbackscattering=}). The lineshape
corresponds to {\em two} peaks separated by
$(v_{ph,\phonon}/v)4\Delta_{eh}$. Experimentally, one sees
just one $2K$ peak with the FWHM about 30~cm$^{-1}$
at the excitation frequency
$\omega_{in}\approx{2}\:\mathrm{eV}$.\cite{Ferrari2006,Yan2007}
This corresponds to an unrealistically large value of
$2\gamma\sim{0.2}\:\mathrm{eV}$. Most likely, this indicates that
two-phonon peaks are broadened by other mechanisms, not taken
into account in the present work. In particular, Eq.~(\ref{dI2dw=})
neglects (i)~the broadening of the phonon states, and (ii)~the anisotropy
of the phonon dispersion (trigonal warping of the phonon spectrum).
A detailed study of these effects would require introduction of additional
parameters into the theory, so we prefer to postpone such study for the
future work. It is worth emphasizing again that the {\em integrated}
intensity of the peaks, which is the main focus of the present study,
does not depend on these details.

In view of the results of the present paper it is worth mentioning the
experimental measurements
of the intensity $I_{2K}$ as a function of doping. While in
Ref.~\onlinecite{Yan2007} no significant dependence was observed,
Ref.~\onlinecite{Das2008}, where higher doping levels were reached,
shows quite a strong dependence of $I_{2K}/I_\Gamma$ on doping. The
intensity $I_\Gamma$ of the off-resonant single-phonon
$1580\:\mathrm{cm}^{-1}$ peak should not depend on doping (although
the phonon width does exhibit such a dependence, the total spectral weight of
the phonon state, determining the integrated intensity of the peak, must be
preserved). At the same time, the intensity~$I_{2K}$, if determined by the
electron inelastic lifetime,
should be sensitive to the concentration of carriers. Indeed, in the
intrinsic graphene at low temperatures the photoexcited carriers do not
participate in electron-electron collisions, as the phase space volume
is restricted.\cite{Guinea96} As the carriers are added to the system,
the electron-electron collisions become possible, thus the total~$\gamma$
increases, and the intensity~$I_{2K}$ is decreased, in qualitative
agreement with the observation of Ref.~\onlinecite{Das2008}.

\subsection{Four-phonon Raman processes}

The motivation to study the four-phonon Raman process comes from the
following picture for the fully resonant processes. The
incident photon creates an electron and a hole -- real
quasiparticles which can participate in various scattering
processes. If the electron emits a phonon with a
momentum~$\vec{q}$, the hole emits a phonon with the momentum
$-\vec{q}$, and after that the electron and the hole recombine
radiatively, the resulting photon will contribute to the
two-phonon Raman peak. If they do not recombine at this stage, but
each of them emits one more phonon, and they recombine afterwards,
the resulting photon will contribute to the four-phonon peak, {\it
etc.} Three-phonon processes, not being fully resonant,
are not interesting in this context.

Besides phonon emission and radiative recombination, electron and hole
are subject to other inelastic scattering processes, which can also
be viewed as emission of some excitations of the system.
In principle, Raman spectrum should also contain the contrubution from
these excitations, which are left in the system after the radiative
recombination of the electron and the hole.
The key point is that for real quasiparticles, the probability to undergo
a scattering process~$\alpha$ is determined by the ratio of corresponding
scattering rate $2\gamma_\alpha$ to the total scattering rate
$2\gamma\equiv\sum_\alpha{2}\gamma_\alpha$, not by
the history. This probability determines the relative
{\em frequency-integrated} intensity of the corresponding feature in the
Raman spectrum. Thus, the ratio of integrated intensity $I_{(2n+2)K}$ of
the Raman peak corresponding to $2n+2$ $K$~phonons to that for
$2n$ $K$~phonons ($I_{2nK}$) must be proportional to
$(\gamma_K/\gamma)^2$, where $2\gamma_K$ is the rate of emission of each
of the two
$K$~phonons, and the square comes from the phonon emission by the
electron and the hole. This conclusion depends weakly on the
relation between $\gamma$~and~$\Delta_{eh}$, only through a logarithmic
factor.

In the doped graphene, the most obvious competitor of the phonon
emission is the electron-electron scattering: the optically excited
electron can kick out another one from the Fermi sea, i.~e., to emit
another electron-hole pair. Thus, Raman spectrum should
contain contribution from electron-hole pairs; however, their
spectrum extends all the way to the energy of the photo-excited
electron (optical energy) in a completely featureless way. Thus,
it cannot be distinguished from the parasitic background which is
always subtracted in the analysis of Raman spectra, and cannot be
seen in the Raman spectrum directly. However, assuming
$\gamma=\gamma_{K}+\gamma_{\Gamma}+\gamma_{ee}$, where
$2\gamma_\Gamma$ is the rate of emission of phonons from the vicinity
of the $\Gamma$~point of the first Brillouin zone, and
$2\gamma_{ee}$ is the electron-electron collision rate, one can
extract the value of $\gamma_{ee}$, relative to phonon emission rates
from the experimental data. More
precisely, in this way one obtains the rate of all inelastic
scattering processes where the electron loses energy far exceeding
the phonon energy.

Note that arguments leading to
$I_{(2n+2)K}/I_{2nK}\propto(\gamma_{ph}/\gamma)^2$ are not specific for
graphene; in fact, this is nothing but Breit-Wigner formula, applied
once for the electron and once for the hole.
Multi-phonon
Raman scattering has been studied in wide-gap semiconductors both
experimentally~\cite{Damen,Porto} (up to ten phonons were seen in
the Raman spectra of CdS), and theoretically~\cite{Varma,Zeyher}.
In a wide-gap semiconductor an optically excited electron does not
have a sufficient energy to excite another electron across the
gap, so the electron-electron channel is absent. In addition,
interaction with only one phonon mode is dominant, so the ratios
of subsequent peaks are represented by a sequence of fixed
numbers. The simple band structure
(one valley for CdS in contrast to two valleys for graphene)
allowed a calculation of the whole
sequence.
A more complicated electronic band structure in graphene
makes it problematic to calculate the whole sequence,
so we restrict ourselves to the calculation of~$I_{4K}$
for the most intense four-phonon peak.

This calculation is performed in Sec.~\ref{sec:4raman}.
Its result depends, besides the relation between $\gamma$~and~$\Delta_{eh}$,
also on their relation to the energy scale $\omega_{in}(v_{ph}/v)$,
characterizing the phonon dispersion. In Sec.~\ref{sec:2intensities} we
have already discussed this energy scale; for $\omega_{in}=2\:\mathrm{eV}$
we have $\omega_{in}(v_{ph}/v)\approx{1}00\:\mathrm{cm}^{-1}
\approx{12}\:\mathrm{meV}$. The meaning
of this energy scale is the difference between the energies of the electron
and the hole after each of them has emitted two phonons with almost arbitrary
momenta (the only restriction is that the sum of all four phonon momenta
must vanish). If $\omega_{in}(v_{ph}/v)\ll\gamma,\Delta_{eh}$, which seems
to be the case (see the discussion in the previous subsection)
then this
difference can be neglected, and the intensity of the $4K$~peak is given
by Eq.~(\ref{I4gammadispless=}) for $\gamma\gg\Delta_{eh}$ [which is
likely to be the case relevant for most experiments, and which was reported
in the short paper by the author (Ref.~\onlinecite{myself})], and by
Eq.~(\ref{I4Dehdispless=}) for $\gamma\ll\Delta_{eh}$; the polarization
memory is lost in both these cases, $I_\|\approx{I}_\perp$.
In the case $\omega_{in}(v_{ph}/v)\gg\gamma,\Delta_{eh}$ the
intensity~$I_{4K}$ is given by Eq.~(\ref{I4Kdisp=}), and a significant
polarization memory is expected, up to $I_\|/I_\perp\approx{3}$.

A thorough experimental study of the intensity $I_{4K}$ (in particular,
its dependence on doping) is still lacking. Our prediction for the case
$\gamma\gg\Delta_{eh},\omega_{in}(v_{ph}/v)$, which we believe to be
the experimentally relevant one, is\cite{myself,mistake}
\begin{equation}\label{answer=}
\frac{I_{4K}}{I_{2K}}\approx{0}.11
\left(\frac{\gamma_K}{\gamma_K+\gamma_\Gamma+\gamma_{ee}}\right)^2.
\end{equation}

\subsection{Renormalization of the coupling constants}

In the calculations of the Raman intensities, described above,
electron-phonon coupling constants entered as parameters of the theory,
without any assumptions about their values, except for the relations
fixed by the symmetry of the crystal. In particular, the two-phonon
peak intensities
$I_{2K}$~and~$I_{2\Gamma}$ are determined by two independent
dimensionless coupling constants which we denote
$\lambda_K$~and~$\lambda_{\Gamma}$ [see Eq.~(\ref{lambdaphonon=})
for the definition]. A simple estimate of the coupling
constants can be obtained from the tight-binding nearest-neighbor model
of the graphene crystal. In this model the only parameter characterizing
the electron spectrum is the nearest-neighbor electronic matrix
element~$t_0$, and the electron-phonon interaction is characterized
by its change with the bond length, $\partial{t}_0/\partial{a}$.
In this model we obtain $\lambda_K/\lambda_{\Gamma}$ to be given by
the inverse ratio of the corresponding phonon frequencies, about~1.2;
the same result up to a few percent is  obtained from the density-functional
theory (DFT) calculations of Ref.~\onlinecite{Piscanec2004}.
At the same time, by comparing the experimentally measured intensities of
the different two-phonon peaks, and using the result of our calculation
performed in Sec.~\ref{sec:2raman}, we can independently extract the
ratio of the coupling constants. According to the data of
Ref.~\onlinecite{Ferrari2006}, $I_{2K}/I_{2\Gamma}\approx{20}$, which
gives $\lambda_K/\lambda_{\Gamma}\approx{3}$.

To explain this discrepancy we first analyzed the effect of the electronic
trigonal band warping, which affects $I_{2K}$ and $I_{2\Gamma}$ 
differently. The corresponding calculation is done in
Sec.~\ref{sec:warping}. For $\omega_{in}=2$~eV,
we estimate the relative contribution of the warping term as
${5}\cdot{10}^{-4}$ for~$I_{2K}$ and $5\cdot{10}^{-2}$
for~$I_{2\Gamma}$, which is far too little to account for the observed 
ratio $I_{2K}/I_{2\Gamma}$.
We are thus led to the conclusion that the observed ratio 
$I_{2K}/I_{2\Gamma}$ must be due to the difference of the coupling
constants, not accounted for by the DFT calculation.
A similar conclusion about the insufficiency of the DFT calculation
of the electron-phonon coupling constants has been drawn in
Ref.~\onlinecite{Calandra2007}, where an attempt was made to explain the
experimental data obtained by ARPES\cite{Rotenberg} using the results
of the DFT calculation.

At the same time, we should note that the  dimensionless
coupling constant $\lambda_{\Gamma}$ for the phonons near the $\Gamma$~point,
as calculated by DFT\cite{Piscanec2004} ($\lambda_{\Gamma}\approx{0}.028$),
agrees reasonably well with the measured one: the measurements of the
linear in the wave vector~$q$ term in the phonon dispersion (Kohn anomaly
due to electron-phonon interaction),
$\omega_{ph}(q)-\omega_{ph}(q=0)\approx(\lambda_{\Gamma}/8)vq$,
give $\lambda_{\Gamma}\approx{0}.024$ (see Ref.~\onlinecite{Maultzschexp});
the measurements of the dependence of the phonon frequency $\omega_{ph}$
on the electron Fermi energy~$\ep_F$:
$\Delta\omega_{ph}\approx(\lambda_{\Gamma}/2\pi)|\ep_F|$ give
$\lambda_{\Gamma}\approx{0}.034$ (Ref.~\onlinecite{Yan2007})
$\lambda_{\Gamma}\approx{0}.027$ (Ref.~\onlinecite{Pisana2007}).

We show that the difference between the ratio
$\lambda_K/\lambda_{\Gamma}\approx{3}$ extracted from the Raman peak
intensities, and $\lambda_K/\lambda_{\Gamma}\approx{1}.2$ obtained
by the DFT calculation,\cite{Piscanec2004} is due to the part of Coulomb
interaction between electrons, not picked up by the DFT when local
approximations are used for the exchange-correlation functional, such
as the local density approximation (LDA) or the generalized gradient
approximation (GGA), namely, logarithmic renormalizations.\cite{us}
Coulomb interaction has been known to be a source of logarithmic
renormalizations for Dirac fermions.\cite{AbrikosovBeneslavskii,Guinea94,Guinea99}
Coulomb renormalizations in graphene subject to a magnetic field have been
considered in Ref.~\onlinecite{AleinerTsvelik}, Coulomb effect on static
disorder has been studied in
Refs.~\onlinecite{Ye,Guinea2005,AleinerFoster}.
Essentially, the idea of the renormalization of the coupling constants
is that the matrix element of the electron-phonon interaction should
be taken not between the non-interacting electronic states, but
between the states dressed by the Coulomb interaction. If the typical
electronic energy in the problem is~$\ep$ ($\sim{1}\:\mathrm{eV}$ in
the case of Raman scattering), the renormalization is determined by
the Coulomb interaction at all length scales from the shortest ones
(lattice constant) to the electron wavelength $v/\ep$. It is this
long-range part of the exchange and correlation that is missed by
the local approximations in the DFT calculation, which take into
account correctly only the short-range correlations (at the distances
of the order of the lattice constant).

In Sec.~\ref{sec:CoulombRG} we calculate the renormalization of the
dimensionless electron-phonon coupling constants
(a preliminary account of this work was given in the short paper\cite{us}),
and show that the coupling constant~$\lambda_{\Gamma}$ for the phonons
near the $\Gamma$~point is not renormalized (hence the agreement between
the value of~$\lambda_{\Gamma}$ calculated by the DFT and measured in
the experiments, as mentioned above), while the coupling
constant~$\lambda_K$ for the phonons near the $\Kpnt$~point, which
is responsible for the $2K$~Raman peak, is enhanced by the Coulomb
interaction. This enhancement depends on the electronic energy, as
shown in Fig.~\ref{fig:RGflow}. For the electronic energy $1\:\mathrm{eV}$
this enhancement is in quantitative agreement with the measured ratio
$I_{2K}/I_{2\Gamma}$, provided that the screening of the Coulomb
interaction by the substrate is weak. The dependence of the enhancement
on the electronic energy translates into the
dependence of $I_{2K}/I_{2\Gamma}$ on the excitation frequency,
which can be checked experimentally. Similarly, as the Coulomb interaction
is screened by the substrate with a dielectric constant~$\vep_\infty$
(its high-frequency value), the dependence of $I_{2K}/I_{2\Gamma}$
on~$\vep_\infty$ can also serve as an experimental check of the theory.

We also show in Sec.~\ref{sec:phononRG} that the electron-phonon
coupling itself is a source of logarithmic renormalizations. However,
due to the smallness of the coupling constants this effect is much
weaker than the effect of the Coulomb interaction.

\subsection{Structure of the paper}\label{sec:structure}

In Sec.~\ref{sec:hamiltonian} the low-energy hamiltonian of the
interaction of electrons with the crystal vibrations and the
electromagnetic field is written from pure symmetry considerations. In
Sec.~\ref{sec:symmetry} the symmetry of the graphene crystal is
reviewed. In Sec.~\ref{sec:electrons} the symmetry considerations 
are used to write the electronic part of the hamiltonian.
Sec.~\ref{sec:addDirac} is dedicated to the symmetry analysis of the
Dirac part of the electron hamiltonian, whose symmetry is significantly
higher than the symmetry of the crystal. Sec.~\ref{sec:parphonons}
is dedicated to the symmetry analysis of the in-plane crystal vibrations.
In Secs.~\ref{Sec:Eopt} and~\ref{Sec:Eac} we write the hamiltonian of
interaction of electrons with the optical and acoustical vibrations,
respectively. Sec.~\ref{sec:perpphonons} is dedicated to the symmetry
analysis of the out-of-plane vibrations of the graphene crystal.
In Sec.~\ref{Sec:emfield} the hamiltonian of the interaction of electrons
with the electromagnetic field is written.

Sec.~\ref{sec:Ramangeneral} describes the general scheme of the
calculation of Raman peak intensities using the standard perturbation
theory.
In Sec.~\ref{sec:propagators} the Green's functions are introduced,
and in Sec.~\ref{sec:ramanprobab} the general expression for the
Raman scattering probability is derived.
In Sec.~\ref{sec:inelastic} we discuss the electron inelastic scattering,
and calculate the electronic self-energy due to the electron-phonon
coupling.

In Sec.~\ref{sec:1raman} one-phonon Raman scattering is discussed,
and it is shown that the calculation of the one-phonon peak intensity cannot be
performed within the low-energy
theory. In Sec.~\ref{sec:2raman} the two-phonon Raman peak intensities
are calculated, first, under the assumption of the Dirac electron
spectrum (Sec.~\ref{sec:2ramanDirac}), and then taking into account
the trigonal band warping and electron-hole asymmetry
(Sec.~\ref{sec:warping}). In Sec.~\ref{sec:4raman} the intensity $I_{4K}$
of the most intense four-phonon peak is calculated.  
Secs.~\ref{sec:CoulombRG},~\ref{sec:phononRG} are dedicated to the
renormalization of the electron-phonon coupling constants due to Coulomb
interaction and due to the electron-phonon interaction, respectively.

\section{Symmetries and hamiltonian\cite{coauthor}}\label{sec:hamiltonian}

\subsection{Symmetry of the crystal}\label{sec:symmetry}

Since the typical energy of the incident photon (about
$2$~eV) is much smaller than the $\pi$-electron bandwidth
($\sim{20}$~eV), one can expect the low-energy excitations to play
the dominant role. In this section we employ standard symmetry
analysis\cite{Lax} to fix the form of the low-energy hamiltonian.
We prefer not to choose any specific basis and use algebraic
properties.

\begin{figure}
\includegraphics[width=8cm]{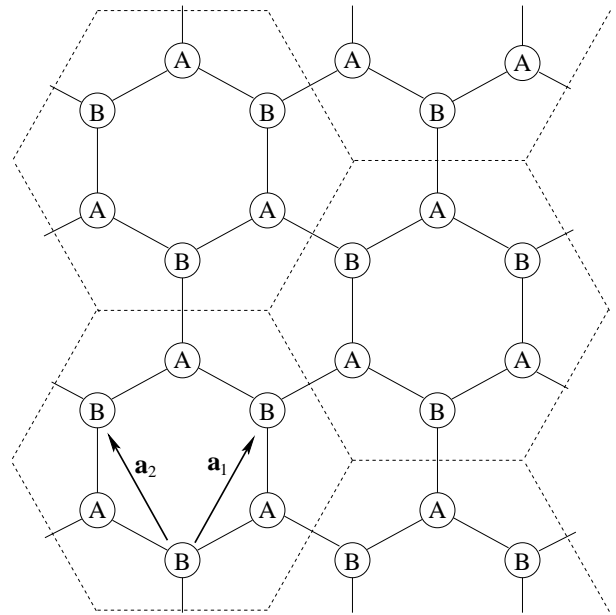}
\caption{\label{fig:lattice} The honeycomb lattice with two atoms
  ($A$~and~$B$) per unit cell. Tripled unit cells are shown by dashed
  hexagons.}
\end{figure}

The carbon atoms of graphene form a honeycomb lattice with two
atoms per unit cell, labeled $A$~and~$B$ (Fig.~\ref{fig:lattice}),
the distance between nearest neighbors being $a=1.42$~{\AA}. Three
out of four electrons of the outer shell of each carbon atom form
strong $\sigma$~bonds with its three nearest neighbors, and
represent no interest to us. The remaining $\pi$~orbitals (one per
each carbon atom) give rise to the half-filled $\pi$~band.

\begin{table}
\begin{tabular}[t]{|c|c|c|c|c|c|c|} \hline
$C_{6v}$ & $E$ & $C_2$ & $2C_3$ & $2C_6$ & $\sigma_{a,b,c}$ &
$\sigma_{a,b,c}'$
\\ \hline\hline $A_1$ & 1 & 1 & 1 & 1 & 1 & 1 \\ \hline $A_2$ & 1
& 1 & 1 & 1 & $-1$ & $-1$ \\ \hline $B_2$ & 1 & $-1$ & 1 & $-1$ &
$1$ & $-1$ \\ \hline $B_1$ & 1 & $-1$ & 1 & $-1$ & $-1$ & $1$ \\
\hline $E_1$ & 2 & $-2$ & $-1$ & $1$ & 0 & 0 \\ \hline $E_2$ & 2 &
2 & $-1$ & $-1$ & 0 & 0 \\ \hline\end{tabular}\hspace{1cm}
\begin{tabular}[t]{|c|c|c|c|}
\hline $C_{3v}$ & $E$ & $2C_3$ & $\sigma_{a,b,c}'$ \\ \hline\hline
$A_1$ & 1 & 1 & 1 \\ \hline $A_2$ & 1 & 1 & $-1$ \\ \hline $E$ & 2
& $-1$ & 0 \\ \hline
\end{tabular}
\caption{Irreducible representations of the groups $C_{6v}$ and
  $C_{3v}$ and their characters.\label{tab:C6vC3v}}
\end{table}

In this paper we will not consider the dimension,
perpendicular to the crystal plane, so the point symmetry group of
the graphene crystal is~$C_{6v}$. It contains 12 elements: the
identity, five rotations $C_6^n$, $n=1,\ldots,5$ ($C_m$~denoting
the rotation by $2\pi/m$) and six reflections in planes
perpendicular to the crystal plane. The three reflections leaving
the $A$~and~$B$ sublattices invariant are denoted by
$\sigma_a,\sigma_b,\sigma_c$, while those swapping the $A$~and~$B$
sublattices points will be denoted
by~$\sigma_a',\sigma_b',\sigma_c'$. Table~\ref{tab:C6vC3v} lists
the irreducible representations of the group~$C_{6v}$ and their
characters.

\begin{figure}
\includegraphics[width=5cm]{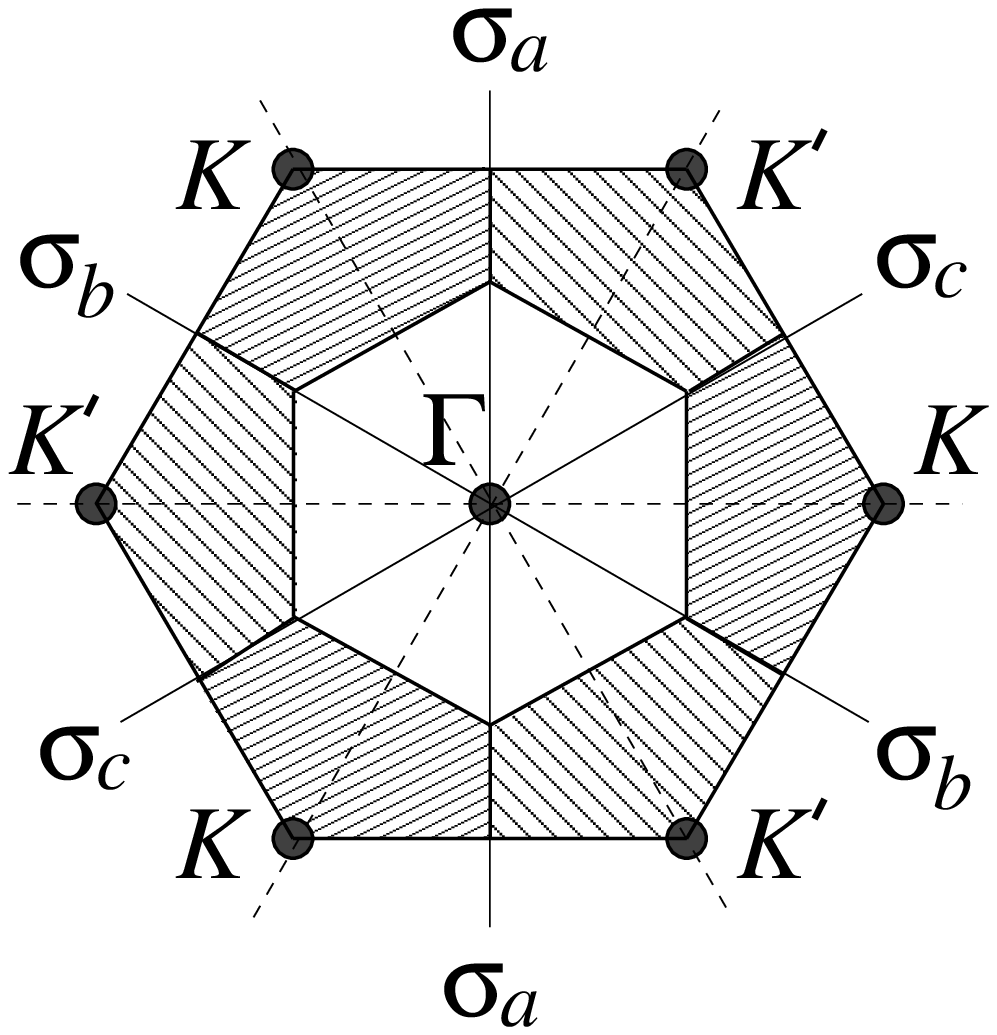}
\caption{\label{fig:B3zone} The first Brillouin zone corresponding to
  the honeycomb lattice, and its wrapping upon tripling of the unit
  cell. Regions with the same shading should be translated so that the
  $\Kpnt,\Kpnt'$ points move to the $\Gamma$~point to form the first
  Brillouin zone of the crystal with the tripled unit cell. Solid
  lines show the $\sigma_a,\sigma_b,\sigma_c$ reflection axes. The
  $\sigma_a',\sigma_b',\sigma_c'$ reflection axes are shown by dashed
  lines (not labeled).}
\end{figure}

The first Brillouin zone of the crystal is a hexagon
(Fig.~\ref{fig:B3zone}). Out of the six corners of the hexagon only
two are inequivalent. They are called $\Kpnt$~and~$\Kpnt'$ points. The
group of the wave vector at these points is~$C_{3v}$.
The states at these points are twice degenerate, transforming
according to the two-dimensional irreducible representation~($E$)
of~$C_{3v}$. Transformations, swapping $\Kpnt$~and~$\Kpnt'$, thus
belonging to~$C_{6v}$ but not to~$C_{3v}$, (reflection~$\sigma_v$ in
the plane, perpendicular to that of~$\sigma_v'$, and rotations
$C_2,C_6$), fix the energies at $\Kpnt,\Kpnt'$ to be equal. One can
form real linear combinations of wave functions from
$\Kpnt$~and~$\Kpnt'$ points which transform according to
$E_1$~and~$E_2$ representations of~$C_{6v}$.
The degeneracy at $\Kpnt,\Kpnt'$ points, in combination with the
absence of any other states in the Brillouin zone with the same
energy,
fixes the Fermi level of a half-filled band to be at
this energy, which is thus natural to choose as $\ep=0$.

Instead of dealing with degenerate states at two different points of
the Brillouin zone ($\Kpnt,\Kpnt'$), one can triple the unit cell of
the crystal. The new unit cell contains 6 atoms which form a hexagon
(Fig.~\ref{fig:lattice}), while the new Brillouin zone is only 1/3 of
the original one (Fig.~\ref{fig:B3zone}). The advantage of this
approach is that now both $\Kpnt$~and~$\Kpnt'$ are mapped onto the
$\Gamma$~point, so one does not have to consider the two of them
separately.

\begin{table*}
\begin{tabular}{|c|c|c|c|c|c|c|c|c|c|} \hline
$C_{6v}''$ & $E$ & $t_{\vec{a}_1},t_{\vec{a}_2}$ &
\parbox{1.5cm}{$C_2,\,t_{\vec{a}_1}C_2$,\\$t_{\vec{a}_2}C_2$} &
$C_3,\,C_3^2$ &
\parbox{2cm}{$t_{\vec{a}_1}C_3,\,t_{\vec{a}_1}C_3^2$,\\
$t_{\vec{a}_2}C_3,\,t_{\vec{a}_2}C_3^2$} &
\parbox{2cm}{$C_6,\,C_6^5$,\\$t_{\vec{a}_1}C_6,\,t_{\vec{a}_1}C_6^5$,
\\$t_{\vec{a}_2}C_6,\,t_{\vec{a}_2}C_6^5$} &
$\sigma_{a,b,c}'$ &
$t_{\vec{a}_1}\sigma_{a,b,c}',\,t_{\vec{a}_2}\sigma_{a,b,c}'$ &
\parbox{2.5cm}{$\sigma_{a,b,c},\,t_{\vec{a}_1}\sigma_{a,b,c}$,\\
$t_{\vec{a}_2}\sigma_{a,b,c}$} \\ \hline\hline $A_1$ & 1 & 1 & 1 &
1 & 1 & 1 & 1 & 1 & 1\\ \hline $A_2$ & 1 & 1 & 1 & 1 & 1 & 1 &
$-1$ & $-1$ & $-1$ \\ \hline $B_2$ & 1 & 1 & $-1$ & 1 & 1 & $-1$ &
$-1$ & $-1$ & $1$ \\ \hline $B_1$ & 1 & 1 & $-1$ & 1 & 1 & $-1$ &
1 & $1$ & $-1$ \\ \hline $E_1$ & 2 & 2 & $-2$ & $-1$ & $-1$ & $1$
& 0 & 0 & 0 \\ \hline $E_2$ & 2 & 2 & 2 & $-1$ & $-1$ & $-1$ & 0 &
0 & 0
\\ \hline $E_1'$ & 2 & $-1$ & 0 & 2 & $-1$ & 0 & 2 & $-1$ & 0 \\
\hline $E_2'$ & 2 & $-1$ & 0 & 2 & $-1$ & 0 & $-2$ & 1 & 0 \\
\hline $G'$ & 4 & $-2$ & 0 & $-2$ & $1$ & 0 & 0 & 0 & 0 \\ \hline
\end{tabular}
\caption{\label{tab:C6vpp} Irreducible representations of the group
$C_{6v}''=C_{6v}\repplus({t}_{\vec{a}_1}C_{6v})\repplus({t}_{\vec{a}_2}C_{6v})$
  and their characters.}
\end{table*}

Tripling of the unit cell means that two translations
$t_{\vec{a}_1}$~and~$t_{\vec{a}_2}$ are factorized out from the
translation group of the crystal, so they should be added to the point
group, which becomes
$C_{6v}''=C_{6v}\repplus(t_{\vec{a}_1}C_{6v})\repplus(t_{\vec{a}_2}C_{6v})$.
Irreducible representations of this group and their characters are
shown on Table~\ref{tab:C6vpp}. The states of the $\pi$-electrons at
the new $\Gamma$~point form a 6-dimensional representation which is
reduced as $A_1\repplus{B}_2\repplus{G'}$, where $A_1$~and $B_2$~states
are the non-degenerate ones corresponding to the old $\Gamma$~point,
while the 4-dimensional irreducible representation~$G'$ contains the
zero-energy states inherited from the old $\Kpnt,\Kpnt'$ points.

\begin{table*}
\begin{tabular}{|c|c|c|c|c|c|c||c|c|c|c|c|c|} \hline
& \multicolumn{6}{|c||}{$\Kpnt\Kpnt'$-diagonal matrices} &
\multicolumn{6}{|c|}{$\Kpnt\Kpnt'$-off-diagonal matrices} \\ \hline
$C_{6v}''$ irreps
& $A_1$ & $B_1$ & $A_2$ & $B_2$ & $E_1$ & $E_2$ &
\multicolumn{2}{|c|}{$E_1'$} & \multicolumn{2}{|c|}{$E_2'$} &
\multicolumn{2}{|c|}{$G'$} \\ \hline
$C_{6v}$ irreps & $A_1$ & $B_1$ & $A_2$ & $B_2$ & $E_1$ & $E_2$ &
$A_1$ & $B_1$ & $A_2$ & $B_2$ & $E_1$ & $E_2$ \\ \hline
%
%
%
notation & $\unitmatrix$ & $\Lambda_z$ &
$\Sigma_z$ & $\Lambda_z\Sigma_z$ & $\Sigma_x,\,\Sigma_y$ &
$-\Lambda_z\Sigma_y,\Lambda_z\Sigma_x$ &
$\Lambda_x\Sigma_z$ & $\Lambda_y\Sigma_z$ & $\Lambda_x$ & $\Lambda_y$
& $\Lambda_x\Sigma_y,-\Lambda_x\Sigma_x$ &
$\Lambda_y\Sigma_x,\Lambda_y\Sigma_y$ \\ \hline
%
%
%
$\mathcal{T}$ & $+$ & $-$ & $-$ & $+$ & $-$ & $+$ & $+$ & $+$ &
$-$ & $-$ & $+$ & $+$ \\ \hline
\end{tabular}
\caption{Classification of $4\times{4}$ hermitian matrices and
their transformation properties under time reversal~$\mathcal{T}$.
\label{tab:matrices}}
\end{table*}

In order to write down the low-energy electronic hamiltonian, we
have to consider $4\times{4}$ hermitian matrices acting in the
4-dimensional space of the zero-energy electronic states. The basis in
the 16-dimensional space of such matrices is provided by the
generators of the $SU(4)$ group forming a 16-dimensional reducible
representation of $C_{6v}$~or~$C_{6v}''$. This representation is
reduced as
\begin{equation}
(E_1\repplus{E}_2)\reptimes(E_1\repplus{E}_2)=
2\,(A_1\repplus{A}_2\repplus{B}_1\repplus{B}_2\repplus{E}_1\repplus{E}_2),
\label{reductionC6v=}
\end{equation}
within the group~$C_{6v}$ (the two sectors corresponding to
matrices either diagonal or off-diagonal in the $\Kpnt\Kpnt'$
subspace), or as
\begin{equation}
G'\reptimes{G}'=A_1\repplus{B}_1\repplus{B}_2\repplus{A}_2
\repplus{E}_1\repplus{E}_2\repplus{E}_1'\repplus{E}_2'\repplus{G}',
\label{reductionC6vpp=}
\end{equation}
within the group~$C_{6v}''$. The correspondence between
Eqs.~(\ref{reductionC6v=}) and~(\ref{reductionC6vpp=}) is given in
Table~\ref{tab:matrices}.

The most convenient way to identify the matrices is by specifying the
irreducible representation according to which they transform, rather
by specifying their explicit form in some particular basis. So, the
matrix which transforms according to the $A_2$~representation
of~$C_{6v}''$ will be denoted by~$\Sigma_z$ and called the
$z$-component of the isospin (the only arbitrariness in this
definition is the overall sign). The two matrices which transform
according to the vector $E_1$~representation of $C_{6v}''$ will be
denoted by $\{\Sigma_x,\Sigma_y\}\equiv\vec\Sigma$ [defined up to an
arbitrary rotation, see Eq.~(\ref{CinfS=}) below], and so on. The full
list of definitions and notations is given in
Table~\ref{tab:matrices}.

Explicit expressions for the electronic matrices are not needed,
as long as their algebraic rules are specified. The simplest way
to specify these rules is to express all the 16 matrices in terms
of two sets, $\{\Sigma_x,\Sigma_y,\Sigma_z\}$ and
$\{\Lambda_x,\Lambda_y,\Lambda_z\}$, and their products, where the
matrices from the same set satisfy the Pauli matrix algebra, while
matrices from different sets just commute. In
Appendix~\ref{app:algebra} we show how these rules can be
established, and give the explicit expressions for the matrices
for some specific choices of the basis. We also note that
$e^{(2\pi{i}/3)\Sigma_z}$ is the matrix of the $C_3$~rotation,
$\Lambda_x\Sigma_z$ -- of the $C_2$~rotation,
$\Lambda_z\Sigma_x$ -- of the $\sigma_a'$~reflection,
$\Lambda_y\Sigma_y$ -- of the $\sigma_a$~reflection,
and $e^{\pm({2}\pi{i}/3)\Lambda_z}$ are the matrices of the two
elementary translations.

\subsection{Electronic wave functions and hamiltonian}\label{sec:electrons}

Since there are two $\pi$-orbitals per unit cell, the
eigenfunctions for each wave vector~$\vec{k}$ in the Brillouin
zone are given by
$e^{i\vec{k}\vec{r}}\bloch_{\vec{k},\pm}(\vec{r},z)$, where the two
Bloch functions $\bloch_{\vec{k},\pm}(\vec{r},z)$, periodic in
$\vec{r}=(x,y)$, are even
and odd with respect to the reflection~$\sigma_a'$, and are
normalized as
\begin{equation}\label{Blochnorm=}
\int\bloch_{\vec{k},i}^*(\vec{r},z)\,\bloch_{\vec{k},j}(\vec{r},z)\,
d^2\vec{r}\,dz=L_xL_y\delta_{ij}\,,\quad i,j=+,-,
\end{equation}
where $L_xL_y$ is the crystal area. It is more convenient to
choose their linear combinations, localized near each carbon atom:
$\bloch_{\vec{k},A}=(\bloch_{\vec{k},+}+\bloch_{\vec{k},-})/\sqrt{2}$
and
$\bloch_{\vec{k},B}=(\bloch_{\vec{k},+}-\bloch_{\vec{k},-})/\sqrt{2}$.
In this basis an arbitrary wave function $\Psi(\vec{r},z)$ involving
only low-energy states can be written in terms of a four-component
smooth envelope function $\psi(\vec{r})$, or its Fourier transform
$\psi(\vec{p})$, $pa\ll{1}$, as
\begin{eqnarray}
\Psi(\vec{r},z)=\int\frac{d^2\vec{p}}{(2\pi)^2}\left[
\psi_{\Kpnt,A}(\vec{p})\,e^{i(\Kvec+\vec{p})\vec{r}}\,
\bloch_{\Kvec,A}(\vec{r},z)\right.+\nonumber\\
+\left.\psi_{\Kpnt,B}(\vec{p})\,e^{i(\Kvec+\vec{p})\vec{r}}\,
\bloch_{\Kvec,B}(\vec{r},z)\right.+\nonumber\\
+\left.\psi_{\Kpnt',A}(\vec{p})\,e^{i(\Kvec'+\vec{p})\vec{r}}\,
\bloch_{\Kvec',A}(\vec{r},z)\right.+\nonumber\\
+\left.\psi_{\Kpnt',B}(\vec{p})\,e^{i(\Kvec'+\vec{p})\vec{r}}\,
\bloch_{\Kvec',B}(\vec{r},z)\right]. \label{Psi=}
\end{eqnarray}
The low-energy effective electronic hamiltonian
$H_{\mathrm{el}}(\vec{p})$ is defined as a $4\times{4}$ matrix
whose matrix element between any two smooth envelope functions
$\psi(\vec{r})$ and $\tilde\psi(\vec{r})$ coincides with the matrix
element of the microscopic hamiltonian
$\mathcal{H}_{\mathrm{el}}(-i\vec\nabla,-i\partial_z;\vec{r},z)$
including the periodic crystal potential, between the corresponding full wave
functions $\Psi(\vec{r},z)$ and $\tilde\Psi(\vec{r},z)$
(see Appendix~\ref{app:extfield} for details):
\begin{eqnarray}
&&\int\psi^\dagger(\vec{r})\,H_{\mathrm{el}}(-i\vec\nabla)\,
\tilde\psi(\vec{r})\,d^2\vec{r}=\nonumber\\
&&=\int\Psi^*(\vec{r},z)\,
\mathcal{H}_{\mathrm{el}}(-i\vec\nabla,-i\partial_z;\vec{r},z)\,
\tilde\Psi(\vec{r},z)\,d^2\vec{r}\,dz.
\label{Heff=}
\end{eqnarray}

Strictly speaking, the spin index should also be attached to the
enevlope function~$\psi(\vec{r})$. However, it would make the
formulas more cumbersome, and we prefer to omit it, as none of
the calculations of the present paper will concern a non-trivial
spin structure. The only role of the spin will be to provide an
additional degeneracy, which will be accounted for, and mentioned
separately every time it will enter the calculations.

We expand the effective hamiltonian in powers of~$\vec{p}$:
$H_{\mathrm{el}}(\vec{p})=H_1(\vec{p})+H_2(\vec{p})+\ldots$, where
$H_n(\vec{p})=O(p^n)$. One can write down different terms from
symmetry considerations, taking into account that momentum
components~$p_x,p_y$ transform according to $E_1$~(vector)
representation of~$C_{6v}$. The leading term in the hamiltonian,
$H_1(\vec{p})$, must have the Dirac form:\cite{Wallace}
\begin{equation}\label{Hel=}
H_1(\vec{p})=vp_x\Sigma_x+vp_y\Sigma_y\equiv v\vec{p}\vec\Sigma\,.
\end{equation}
The coefficient~$v$ turns out to be equal to
$v\approx{10}^8\:\mathrm{cm/s}\:\approx{7}\:\mathrm{eV\cdot\mbox{\AA}}$;
it can be related to the nearest-neighbor coupling matrix element~$t$
of the tight-binding model as $v=3ta/2$.
The four eigenstates of the hamiltonian~(\ref{Hel=}) for
each~$\vec{p}$ can be classified by the value ($\pm{1}$) of
projection of the isospin~$\vec\Sigma$ on~$\vec{p}$, corresponding
to the energies $\pm{v}|\vec{p}|$. In other words, the
hamiltonian~(\ref{Hel=}) is diagonalized by a unitary
transformation:
\begin{equation}\label{diagDirac=}
H_1(\vec{p})=e^{-i\Sigma_z\varphi_{\vec{p}}/2}
e^{-i\Sigma_y\pi/4}\,vp\Sigma_z\,e^{i\Sigma_y\pi/4}
e^{i\Sigma_z\varphi_{\vec{p}}/2},
\end{equation}
$\varphi_{\vec{p}}=\arctan(p_y/p_x)$ being the polar angle of the
vector~$\vec{p}$.

Various perturbations of the Dirac hamiltonian~(\ref{Hel=}) should
also be classified according to Table~\ref{tab:matrices}. Only those
containing the matrix~$\Sigma_z$ will open a gap in the electron
spectrum. Perturbations, not containing~$\Sigma_i$, correspond to an
energy shift of the whole spectrum (which may be accompanied by valley
mixing, if $\Lambda$~matrices are involved). Perturbations,
proportional to~$\vec\Sigma$, correspond to a momentum shift of the
Dirac points, which can be viewed as a gauge vector potential.

The next term in the hamiltonian can be written by taking into
account that the symmetric tensor $p_ip_j$ can be decomposed into
$p^2$, transforming according to~$A_1$, and $2p_xp_y,p_x^2-p_y^2$,
transforming according to~$E_2$:
\begin{equation}\label{H2=}
H_2(\vec{p})=\alpha_0{p}^2\unitmatrix
+\alpha_3\left[-2p_xp_y\Sigma_y+(p_x^2-p_y^2)\Sigma_x\right]\Lambda_z.
\end{equation}
The first term describes electron-hole asymmetry; it vanishes in the
nearest-neighbor tight-binding model, and appears only if coupling to
the second nearest neighbors is included.
The second term, corresponding to $E_2$~representation, is
responsible for the so-called trigonal band warping. In the
nearest-neighbor tight-binding model its value is given by $-3ta^2/8$.

At this point it is convenient to introduce the time-reversal
operation whose action on the microscopic spinless wave function
is defined by $\Psi(\vec{r})\mapsto\Psi^*(\vec{r})$. For the
four-component envelope function $\psi(\vec{r})$ this definition
translates into
\begin{equation}\label{UTpsi=}
\psi(\vec{r})\mathop{\mapsto}_{\mathcal{T}}
{U}_{\mathcal{T}}\psi^*(\vec{r}),
\end{equation}
where $U_{\mathcal{T}}$~is a unitary $4\times{4}$ matrix, with an
additional requirement
$U_{\mathcal{T}}U_{\mathcal{T}}^*=\unitmatrix$, whose explicit
form depends on the choice of the basis. The action of time
reversal on the effective electronic hamiltonian is defined by
\begin{equation}
H_{\mathrm{el}}(\vec{p},\vec{r})\mathop{\mapsto}_{\mathcal{T}}
U_{\mathcal{T}}
{H}_{\mathrm{el}}^*(-\vec{p},\vec{r})\,U_{\mathcal{T}}^\dagger.
\end{equation}
Behavior of electronic matrices under the time reversal is listed
in Table~\ref{tab:matrices}.

Time reversal symmetry of the hamiltonian does not add any new
symmetries to the spectrum, as compared to those imposed by the
spatial symmetry~$C_{6v}$ (namely, $C_{3v}$ symmetry of the
spectrum around each of $\Kpnt,\Kpnt'$ points, and the mirror
symmetry between the spectra at $\Kpnt$ and $\Kpnt'$ points),
because the action of the time reversal on the wave vector is
identical to that of the $C_2$~rotation,
$\Kvec+\vec{p}\mapsto\Kvec'-\vec{p}$. However, some perturbations
may lift the $C_2$~symmetry, while still preserving the time
reversal one (see Secs. \ref{Sec:Eopt} and~\ref{Sec:emfield})


\subsection{Additional symmetries of the Dirac hamiltonian}
\label{sec:addDirac}

The Dirac hamiltonian~(\ref{Hel=}) has a higher symmetry than the
microscopic symmetry~$C_{6v}$. The additional symmetries are
(i)~the full intravalley rotational symmetry $C_{\infty{v}}$:
\begin{subequations}
\begin{eqnarray}
&&\vec\Sigma\mapsto
{e}^{-i\Sigma_z\varphi/2}\vec\Sigma\,{e}^{i\Sigma_z\varphi/2},
\label{CinfS=}\\
&&\left(\begin{array}{c} p_x \\ p_y \end{array}\right)\mapsto
\left(\begin{array}{cc} \cos\varphi & -\sin\varphi \\
\sin\varphi & \cos\varphi \end{array}\right)
\left(\begin{array}{c} p_x \\ p_y \end{array}\right),
\label{Cinfk=}
\end{eqnarray}
\label{Cinf=}
\end{subequations}
which leaves $H_1(\vec{p})$ invariant; (ii)~the chiral property:
\begin{equation}\label{chiral=}
U_{\mathcal{C}}H_1U_{\mathcal{C}}=-H_1,\quad
U_{\mathcal{C}}\equiv\Sigma_z,
\end{equation}
which ensures the symmetry of the spectrum with respect to
$\ep\to-\ep$ (i.~e., particle-hole symmetry).

The intravalley ``time reversal'' symmetry, mentioned in
Ref.~\onlinecite{Ando2002}, can be represented as a combination of
the time reversal~(\ref{UTpsi=}), the intravalley
rotation~(\ref{CinfS=}) by~$\pi$, and the $C_2$~rotation:
\begin{equation}
\psi\mapsto\Lambda_x\Sigma_z{e}^{-i\Sigma_z\pi/2}U_{\mathcal{T}}\psi^*
=-i\Lambda_xU_{\mathcal{T}}\psi^*.
\end{equation}
Applying this operation twice results in a minus sign, since the
matrix~$\Lambda_x$ is odd under time reversal.

If one wishes to include the second-order hamiltonian~(\ref{H2=}),
the $A_1$~term preserves only the rotational
symmetry~$C_{\infty{v}}$, while the $E_2$~term preserves only the
chiral property~(\ref{chiral=}).


\subsection{In-plane phonon modes}\label{sec:parphonons}

There is quite extensive literature dedicated to the phonon modes of
graphene and graphite, and their symmetry analysis (see, e.~g.,
Ref.~\onlinecite{ThomsenReich2004} and references therein). To make
the presentation self-contained, we briefly repeat the facts which are
necessary for the subsequent considerations.

\begin{figure*}
\includegraphics[width=13cm]{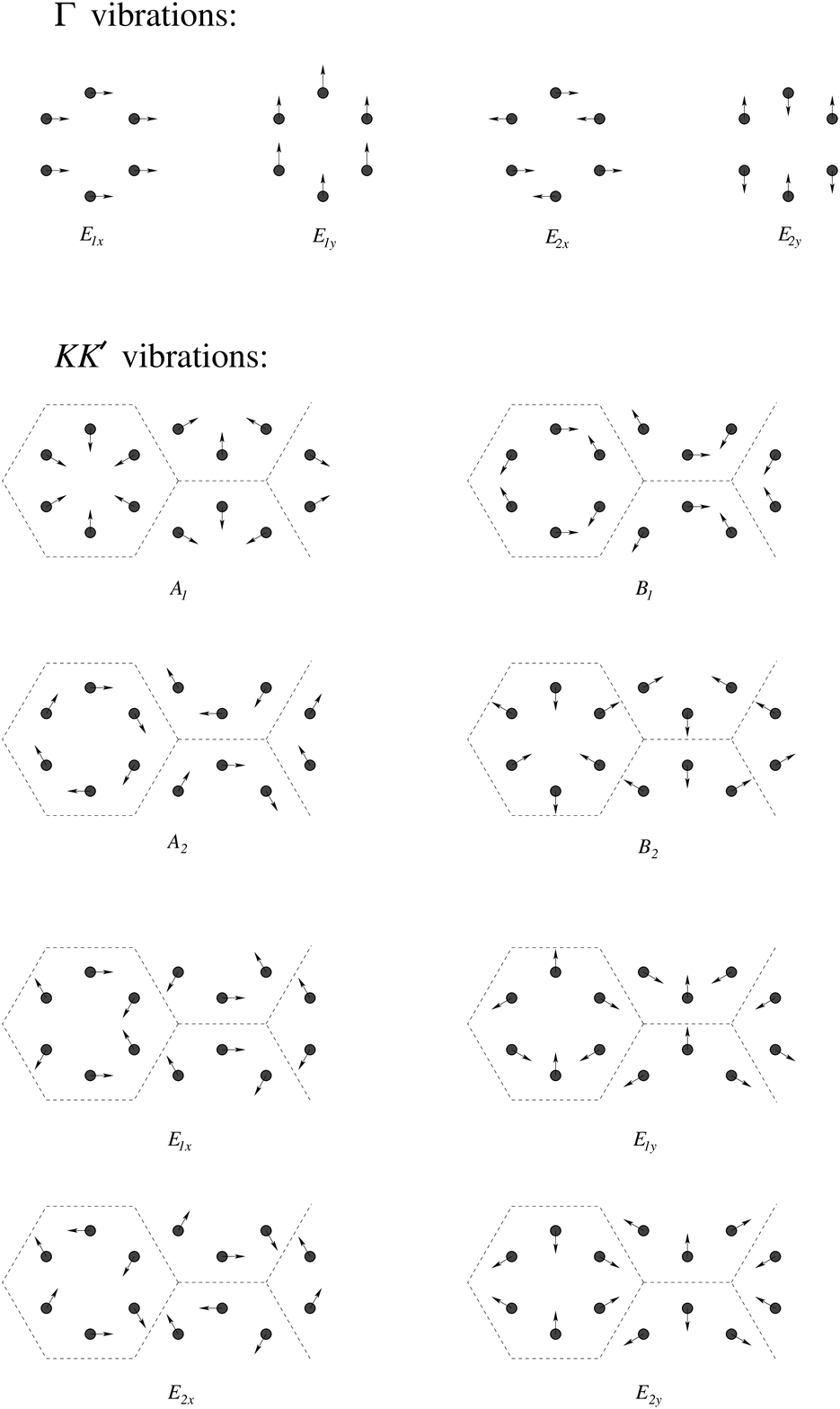}
\caption{\label{fig:phonons} In-plane phonon modes at $\Gamma$~point
  and real linear combinations of the phonon modes at $\Kpnt,\Kpnt'$
  points corresponding to different irreducible representations
  of~$C_{6v}$. The dashed lines show tripled unit cells.
}
\end{figure*}

The only phonons that can efficiently couple to the low-energy
electronic states are those near the $\Gamma$~point (coupling
electronic states in the same valley), and near the $\Kpnt,\Kpnt'$
points (coupling electronic states in different valleys; note that
$\Kvec-\Kvec'$ is equivalent to~$\Kvec'$). As a result, each unit
cell has four in-plane degrees of freedom (two per each carbon
atom). At $\Gamma$~point one has two acoustic and two optical
modes. Coupling of the acoustical modes with a wave
vector~$\vec{q}$ to the electron motion must vanish as $q\to{0}$,
and thus be small in the parameter~$qa$, but we consider them for
the sake of completeness. For eight modes at $\Kpnt,\Kpnt'$ points
it is more convenient to consider their real linear combinations
which transform according to
$A_1\repplus{B}_1\repplus{A}_2\repplus{B}_2\repplus{E}_1\repplus{E}_2$
representations of the group~$C_{6v}$.
All twelve modes are shown in Fig.~\ref{fig:phonons}. They are
linearly polarized, in contrast to the basis with a definite wave
vector, which would have a circular polarization.

After having mixed the $\Kpnt$ and $\Kpnt'$ modes, it is natural
to switch to the tripled unit cell representation. Analogously to
Eq.~(\ref{Psi=}) for electrons, an arbitrary lattice displacement
pattern, involving phonon states with wave vectors close either to
$\Gamma$, or~$\Kpnt$, or~$\Kpnt'$ points, can be expressed in
terms of a smooth envelope function $u(\vec{r})$ or its Fourier
transform $u(\vec{q})$. Since we have to specify two cartesian
components of displacements for each of the six atoms in the
tripled unit cell,
$\delta\underline{x}
=(\delta{x}_1,\delta{y}_1,\ldots,\delta{x}_6,\delta{y}_6)^T$,
the envelope of normal coordinates $u(\vec{q})$~is a
twelve-component vector:
\begin{equation}
\delta\underline{x}(\vec{R})=\int\frac{d^2\vec{q}}{(2\pi)^2}
\sum_{\phonon=1}^{12}u_\phonon(\vec{q})\,e^{i\vec{q}\vec{R}}
\blochph_{\vec{q},\phonon}\,,
\end{equation}
where $\vec{R}$ spans a discrete set of points, labeling different
unit cells, and $\blochph_{\vec{q},\phonon}$~is the pattern of
displacements for the $\phonon$th normal mode, normalized as
\begin{equation}
\blochph_{-\vec{q},\phonon}^T\blochph_{\vec{q},\phonon'}=
6\delta_{\phonon\phonon'}.
\end{equation}
For $\vec{q}=0$ the twelve normal modes
$\blochph_{\vec{q}=0,\phonon}$ are shown in
Fig.~\ref{fig:phonons}.
Introducing $\Pi_{\phonon}(\vec{r})$, the canonically conjugate
momentum to $u_{\phonon}(\vec{r})$, we write the bare phonon
hamiltonian as
\begin{equation}\label{Hph=}
H_{\mathrm{ph}}=
\sum_{\phonon=1}^{12}\int\frac{N\,d^2\vec{r}}{L_xL_y}
\left[\frac{\Pi_{\phonon}^2(\vec{r})}{2M}
+\frac{M}{2}\,u_{\phonon}(\vec{r})\,
\omega_{\phonon}^2(-i\vec\nabla)\,u_{\phonon}(\vec{r})\right]
\end{equation}
with $M=1.993\cdot{10}^{-23}$~g being the carbon atom mass,
$N$~the total number of the carbon atoms in the crystal, and
$\omega_{\phonon}(\vec{q})$ the frequency of the $\phonon$th
normal mode.

At $\vec{q}=0$ the modes belonging to the same irreducible
representation of $C_{6v}''$ are degenerate. Expansion of the
phonon potential energy at small~$\vec{q}$, describing the
splitting, can be done analogously to that of the electronic
hamiltonian. It is convenient to preserve the form~(\ref{Hph=}) of
the hamiltonian, but instead of summing over normal
modes~$\phonon$ one should sum over irreducible representations
of~$C_{6v}''$. For each irreducible representation
$\omega^2(-i\vec\nabla)$ becomes a matrix acting in the space of
the degenerate modes belonging to this representation.

For two-dimensional representations ($E_1$~and~$E_2$ at the
$\Gamma$~point, and $E_1',E_2'$ at $\Kpnt,\Kpnt'$ points) we have
to classify $2\times{2}$ hermitian matrices according to
irreducible representations in the corresponding
decomposition:
\begin{subequations}\begin{eqnarray}
&&E_1\reptimes{E}_1=A_1+A_2+E_2,\\
&&E_2\reptimes{E}_2=A_1+A_2+E_2,\\
&&E_1'\reptimes{E}_1'=A_1+A_2+E_1',\\
&&E_2'\reptimes{E}_2'=A_1+A_2+E_2'.
\end{eqnarray}\end{subequations}
The basis in the space of such matrices is provided by the unit
matrix~$\sigma_0$, transforming according to~$A_1$, the Pauli
matrix~$\sigma_z$ transforming according to~$A_2$, and the Pauli
matrices $\sigma_x,\sigma_y$ transforming according to the third term
in each decomposition. Since the components $q_x,q_y$ transform
according to~$E_1$, the linear in $\vec{q}$ term is absent for all
two-dimensional representations. For the four-dimensional
representation~$G'$ we have to deal with $4\times{4}$ matrices, so
everything is fully analogous to the electronic case, and these
phonons also have Dirac spectrum.
In the second order we have $q^2\sim{A}_1$ and
$q_x^2-q_y^2,2q_xq_y\sim{E}_2$, so we can write
\begin{subequations}
\begin{widetext}
\begin{eqnarray}
\omega_{E_1}^2(\vec{q})&=&
\frac{v_L^2+v_T^2}{2}\,q^2+\frac{v_L^2-v_T^2}{2}
\left[(q_x^2-q_y^2)\sigma_x-2q_xq_y\sigma_y\right],\label{omE1=}\\
\omega_{E_2}^2(\vec{q})&=&\omega_{E_2}^2 +\omega_{E_2}(J_L+J_T)q^2
+\omega_{E_2}(J_L-J_T)
\left[(q_x^2-q_y^2)\sigma_x-2q_xq_y\sigma_y\right],\label{omE2=}\\
\omega_{E_1'}^2(\vec{q})&=&\omega_{E_1'}^2+2\omega_{E_1'}J_{E_1'}q^2,\\
\omega_{E_2'}^2(\vec{q})&=&\omega_{E_2'}^2+2\omega_{E_2'}J_{E_2'}q^2,\\
\omega_{G'}^2(\vec{q})&=&\omega_{G'}^2
+2\omega_{G'}v_{G'}\vec{q}\vec\Sigma+O(q^2).\label{omGp=}
\end{eqnarray}
\end{widetext}
\end{subequations}
The $2\times{2}$ matrices appearing in Eqs.~(\ref{omE1=}),~(\ref{omE2=}),
can be diagonalized to yield the dispersion of the longitudinal and transverse
phonons: $\omega_{E_1,L(T)}(\vec{q})=v_{L(T)}q$,
$\omega_{E_2,L(T)}(\vec{q})=\omega_{E_2}+J_{L(T)}q^2$.

\subsection{Electronic coupling to in-plane optical phonons}\label{Sec:Eopt}

In the electron-phonon interaction hamiltonian the normal mode
displacements should  be paired with the electronic matrices,
corresponding to the same irreducible representation. For optical
phonons it is sufficient to take the leading $\vec{q}=0$ term, so
we have one independent coupling constant for each irreducible
representation of the group~$C_{6v}''$:
\begin{eqnarray}
H_{\mathrm{e-opt}}&=&
F_{\Gamma}[\vec{u}_{E_2}\times\Lambda_z\vec\Sigma]_z+\nonumber\\
&+&F_K\left(u_{A_1}\Lambda_x\Sigma_z
+u_{B_1}\Lambda_y\Sigma_z\right)+\nonumber\\
&+&F_K'\left([\vec{u}_{E_1}\times\Lambda_x\vec\Sigma]_z
+\vec{u}_{E_2}\Lambda_y\vec\Sigma\right). \label{Heph=}
\end{eqnarray}
Note that $E_2'$~phonons cannot couple to the electron motion in this
approximation since the corresponding matrices $\Lambda_x,\Lambda_y$
change sign under the time reversal, while the electronic hamiltonian
must preserve time reversal symmetry even in the crystal with
displaced atoms. Electrons will couple to (i)~momentum $\Pi_{E_2'}$,
whose effect is small in the ratio of the phonon
frequency~$\omega_{E_2'}$ to the electron bandwidth, (ii)~gradient
$\vec\nabla{u}_{E_2'}$, whose effect is small in the parameter~$qa$,
(iii)~the squares of displacements~$u_{E_2'}^2$, whose effect is small
in the parameter $1/(Ma^2\omega_{E_2'})$.


The three constants in Eq.~(\ref{Heph=}) can be evaluated in the
tight-binding approximation, where they are expressed in terms of
$F\equiv\partial{t}/\partial{a\approx{6}\:\mathrm{eV/\mbox{\AA}}}$
 -- the change of the nearest-neighbor
coupling matrix element with the distance between the atoms. We obtain
$F_{\Gamma}=F_K=3F$, while $F_K'$~vanishes. This vanishing is an
artifact of the nearest-neighbor bond-stretching approximation. If
one takes into account the dependence of the electronic hamiltonian on
the angles between the bonds, $F_K'$~becomes different from
zero. Nevertheless, it is an order of magnitude smaller
than~$F_K$ (e.~g., density-functional theory
calculations\cite{Piscanec2004} give $F_K'/F_K\approx{0}.13$), and
we neglect the $G'$~phonons for the rest of the paper.
Thus, our attention will be focused on the two degenerate modes
$E_2$~and~$E_1'$, which will be referred to simply as
$\Gamma$~and~$\Kpnt$ phonons, and their frequencies will be
denoted by $\omega_{E_2,L}(\vec{q})=\omega_{\Gamma,L}(\vec{q})$,
$\omega_{E_2,T}(\vec{q})=\omega_{\Gamma,T}(\vec{q})$,
$\omega_{E_1'}(\vec{q})=\omega_K(\vec{q})$.

The strength of electron-phonon interaction can be conveniently
characterized by the dimensionless coupling constants
\begin{equation}\label{lambdaphonon=}
\lambda_{\Gamma}=
\frac{F_{\Gamma}^2}{M\omega_{\Gamma}(0)v^2}\frac{\sqrt{27}a^2}4,\quad
\lambda_K=
\frac{F_K^2}{M\omega_K(0)v^2}\frac{\sqrt{27}a^2}4,
\end{equation}
where $\sqrt{27}a^2/4$ is the area per carbon atom.
For $F_{\Gamma}=F_K=3F$, $F=6\:\mathrm{eV}/\mbox{\AA}$,
$M=2.00\cdot{10}^{-23}\:\mathrm{g}=%
2.88\cdot{10}^3\:(\mathrm{eV}\cdot\mbox{\AA}^2)^{-1}$,
$v=10^6\,\mbox{m}/\mbox{s}=6.58\,\mbox{eV}\cdot\mbox{\AA}$,
$\omega_{\Gamma}=1580\:\mathrm{cm}^{-1}=0.196\:\mathrm{eV}$,
$\omega_K=1370\:\mathrm{cm}^{-1}=0.170\:\mathrm{eV}$,
$a=1.42\:\mbox{\AA}$ we have
\begin{equation}\label{phononnumbers=}
\frac{F_{\Gamma}^2}{M\omega_{\Gamma}v^2}\frac{\sqrt{27}a^2}4\approx{0}.035,
\quad
\frac{F_K^2}{M\omega_Kv^2}\frac{\sqrt{27}a^2}4\approx{0}.040.
\end{equation}
The easiest way to match the notations for the electron-phonon
coupling constants, used in the present paper, to those used in
other works, is to compare observable quantities. For example,
adding electrons to the system leads to a shift of the phonon
frequencies. For the $\Gamma$~phonons at $\vec{q}=0$ in
the present notations this shift is expressed in terms of the
Fermi energy~$\ep_F$ as
\begin{equation}
\delta\omega_\Gamma=\frac{\lambda_\Gamma}{2\pi}
\left(|\ep_F|+\frac{\omega_\Gamma}4
\ln\frac{\omega_\Gamma-2|\ep_F|}{\omega_\Gamma-2|\ep_F|}\right).
\end{equation}
Alternatively, one can look at the correction to the phonon
dispersion as a function of the wave vector~$\vec{q}$ (see
Appendix~\ref{app:polarization}).

\subsection{Electronic coupling to in-plane acoustical phonons}\label{Sec:Eac}

Since a uniform translation of the crystal cannot affect the electron
motion, it can couple to acoustic phonons only through spatial and
time derivatives of the corresponding displacements:
\begin{eqnarray}
H_{\mathrm{e-ac}}&=&\frac{\vec\Pi_{E_1}}M\,(\vec{p}+m_ev\vec\Sigma)+\nonumber\\
&+&\Xi_0(\partial_x{u}_{E_1x}+\partial_yu_{E_1y})\unitmatrix+\nonumber\\
&+&\Xi_1(\partial_x{u}_{E_1y}-\partial_yu_{E_1x})\Sigma_z+\nonumber\\
&+&\Xi_2\left[-(\partial_x{u}_{E_1y}+\partial_yu_{E_1x})\Sigma_y
\right.+\nonumber\\ &&\quad\left.+
(\partial_x{u}_{E_1x}-\partial_yu_{E_1y})\Sigma_x\right]\Lambda_z.
\end{eqnarray}
The general form of the coupling to $\vec\Pi_{E_1}$ in first line
follows from the symmetry considerations (the coefficient at
$\vec\Pi_{E_1}$ should transform according to $E_1$ and change
sign under time reversal). However, its exact form follows from
the $\vec{k}\cdot\vec{p}$ perturbation theory (which identifies
$m_e$ with the free electron mass) in combination with the
requirement of Galilean invariance: if $\Psi(\vec{r},t)$ satisfied
the Schr\"odinger equation for the stationary lattice, then for
the lattice moving at a constant velocity
$\dot{\vec{u}}_{E_1}=\vec\Pi_{E_1}/M$, the solution should be
$\Psi(\vec{r}-\dot{\vec{u}}_{E_1}t,t)$ (we neglect the
contribution of electrons to the total kinetic energy of the
crystal). Coupling to the strain is written using the
decomposition $E_1\reptimes{E}_1=A_1+A_2+E_2$.

\subsection{Out-of-plane phonons}\label{sec:perpphonons}

\begin{figure}
\includegraphics[width=8cm]{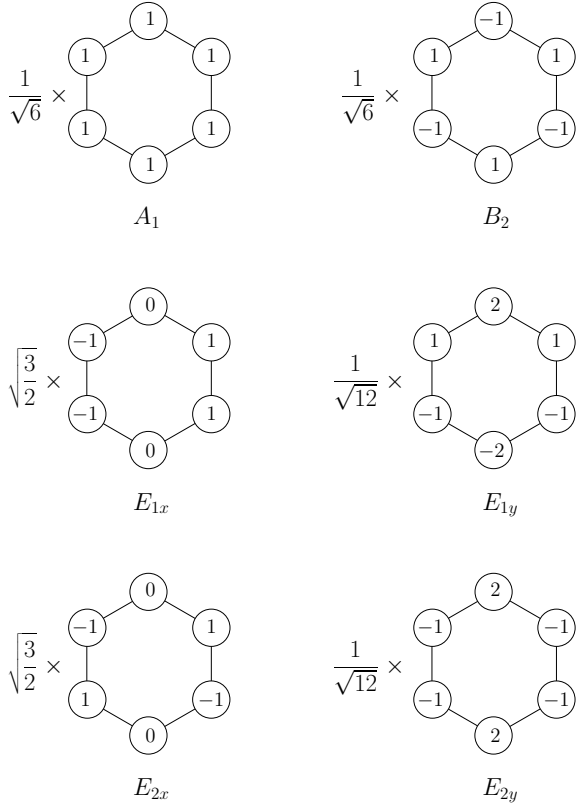}
\caption{\label{fig:perp} Out-of-plane phonon modes at $\Gamma$~point
  and real linear combinations of the phonon modes at $\Kpnt,\Kpnt'$
  points corresponding to different irreducible representations
  of~$C_{6v}$.}
\end{figure}

The basis vectors for out-of plane displacements coincide with those
for the electronic wave function amplitudes in the tight-binding
picture (since in both cases a single number is associated with each
lattice site). At the $\Gamma$~point the basis vectors represent
$A$~and~$B$ atoms shifting in the same direction or opposite
directions, corresponding to $A_1$~and~$B_2$ representations. For the
$\Kpnt,\Kpnt'$ phonons we have $E_1+E_2$ representations of~$C_{6v}$,
or the four-dimensional $G'$~representation of $C_{6v}''$ in the
tripled unit cell picture. All six modes are shown in
Fig.~\ref{fig:perp}.

For a suspended graphene sheet the $A_1$~mode has a frequency
$\omega_{A_1}(\vec{q})\propto{q}^2$ (see, e.~g.,
Ref.~\onlinecite{LL7}). The frequencies of other modes are finite.
All frequencies can be strongly affected by the interaction with the
substrate (in particular, there will be no reason for the frequency of
the $A_1$~mode to vanish at $\vec{q}=0$).

Out-of-plane displacements can couple to the
electron motion only quadratically, if the crystal is symmetric
with respect to reflection in the crystal plane. While it is the
case for a suspended graphene sheet, presence of a substrate may
break this symmetry.\footnote{Note that in this case the symmetry
$C_{6v}$ becomes distinguishable from~$D_6$.}

\subsection{Coupling to electromagnetic field}\label{Sec:emfield}

The hamiltonian of interaction of electrons with the
electromagnetic field, described by the long-wavelength
scalar and vector potentials $\varphi(\vec{r})$~and~$\vec{A}(\vec{r})$,
can be obtained from the requirement of gauge invariance:
$H_{\mathrm{el}}(\vec{p})\to%
{H}_{\mathrm{el}}(\vec{p}-(e/c)\vec{A})+e\varphi$.
To the linear order in~$\vec{A}$ we have
\begin{equation}\label{Heem=}
H_{\mathrm{e-em}}=e\varphi-\frac{ev}{c}\,\vec{A}\vec\Sigma.
\end{equation}

Besides coupling via gauge potentials, electrons can couple directly to
electric and magnetic
field. Such terms are gauge invariant and cannot be deduced from
the bare electronic hamiltonian in the envelope function
approximation, as they correspond to the effect of electromagnetic
field on the microscopic Bloch functions. They should be
introduced into the effective low-energy theory directly, and can
be either calculated microscopically (see Appendix~\ref{app:extfield}
for the calculation by $\vec{k}\cdot\vec{p}$ perturbation theory), or
written from symmetry considerations.

The electric field vector is invariant under time reversal, and
its cartesian components
$\mathcal{E}_x,\mathcal{E}_y,\mathcal{E}_z$ transform according to
$\mathcal{E}_z\sim{A}_1$, $(\mathcal{E}_x,\mathcal{E}_y)\sim{E}_1$
under $C_{6v}$. The magnetic field vector changes sign under time
reversal, and its components
$\mathcal{B}_x,\mathcal{B}_y,\mathcal{B}_z$ transform as
$\mathcal{B}_z\sim{A}_2$, $(-\mathcal{B}_y,\mathcal{B}_x)\sim{E}_1$
under $C_{6v}$.
They can couple only to valley-diagonal matrices. Indeed, for
uniform electric and magnetic fields the effective hamiltonian must
be invariant under translations, represented by the
matrices~$e^{\pm(2\pi{i}/3)\Lambda_z}$. This requirement is not
satisfied by any matrix of the form
$\Sigma_i\Lambda_x,\Sigma_i\Lambda_y$, $i=0,x,y,z$.
In particular, the term proportional to
$\mathcal{B}_x\Lambda_x\Sigma_x+\mathcal{B}_y\Lambda_x\Sigma_y$,
considered in Ref.~\onlinecite{Lukyanchuk}, must be absent (see
Appendix~\ref{app:extfield} for a microscopic calculation).

These considerations enables us to write the hamiltonian to
the first order in the fields as
\begin{equation}\label{Heemp=}
H_{\mathrm{e-em}}^\prime=
-d_z\mathcal{E}_z\unitmatrix
+\mu_{xy}(\mathcal{B}_x\Sigma_y-\mathcal{B}_y\Sigma_x)
-\mu_z\mathcal{B}_z\Sigma_z.
\end{equation}
The first term in this hamiltonian represents the coupling to the
$z$-component of electric dipole moment of $\pi$~electrons in each
unit cell. The second term represents the coupling of the magnetic
field to the in-plane component of the magnetic moment.
These terms are
forbidden for a suspended graphene sheet due to the symmetry with
respect to reflection in the crystal plane, but may be allowed if
this symmetry is broken due to the presence of a substrate. On the
contrary, the third term, corresponding to the $z$-component of the
magnetic moment of the
unit cell, is allowed for a suspended graphene sheet as well (the
$z$-component of the magnetic moment does not change sign under
reflection in the crystal plane).

Terms quadratic in $\mathcal{E}$~and~$\mathcal{B}$ correspond to
polarizabilities of the unit cell. The corresponding hamiltonian
can be written following the same lines as above. This is,
however, beyond the scope of our interest.

\section{Raman scattering: general expressions}
\label{sec:Ramangeneral}

In this section we derive the general expressions for the Raman scattering
probability using the standard perturbation theory.

\subsection{Green's functions}\label{sec:propagators}

The second-quantized version of the Dirac hamiltonian~(\ref{Hel=})
reads as
\begin{equation}\label{hatH1=}
\hat{H}_1=\int{d}^2\vec{r}\,\hat\psi^\dagger(\vec{r})\,
(-iv\vec\Sigma\cdot\vec\nabla)\,\hat\psi(\vec{r})\,.
\end{equation}
Since all energies we are interested in ($\sim{1}\:\mbox{eV}$) are
much higher than temperature, we set the latter equal to zero.
The zero-temperature electronic Green's function, corresponding to
hamiltonian~(\ref{hatH1=}), is given by
\begin{eqnarray}\label{Gpe=}
&&G(\vec{p},\ep)=-i\int\left\langle\mathrm{T}
\hat\psi(\vec{r},t)\,\hat\psi^\dagger(0,0)\right\rangle
e^{-i\vec{p}\vec{r}+i\ep{t}}d^2\vec{r}\,dt=\nonumber\\
&&=\frac{\ep+v\vec{p}\cdot\vec\Sigma}{\ep^2-(vp-io)^2}\,,
\end{eqnarray}
where $io$ is the infinitesimal imaginary shift of the pole.
$\mathrm{T}$~is the sign of the chronological ordering, and
the average is taken over the ground state of the system.

Upon quantization of the phonon field based on
hamiltonian~(\ref{Hph=}), the normal mode displacement operator
and the bare phonon hamiltonian become
\begin{subequations}\begin{eqnarray}
&&\hat{u}_\phonon(\vec{r})=\sum_{\vec{q}}
\frac{\hat{b}_{\vec{q},\phonon}e^{i\vec{q}\vec{r}}
+\hat{b}_{\vec{q},\phonon}^\dagger e^{-i\vec{q}\vec{r}}}
{\sqrt{2NM\omega_{\phonon}(\vec{q})}},\\ &&\hat{H}_{\mathrm{ph}}=
\sum_{\vec{q},\phonon}\omega_{\vec{q},\phonon}
\left(\hat{b}_{\vec{q},\phonon}^\dagger\hat{b}_{\vec{q},\phonon}
+\frac{1}{2}\right),\\ &&\sum_{\vec{q}}\equiv
L_xL_y\int\frac{d^2\vec{q}}{(2\pi)^2}.\label{sumintegral=}
\end{eqnarray}\end{subequations}
The phonon Green's function is defined as
\begin{eqnarray}
&&D_{\phonon}(\vec{q},\omega)=\nonumber\\
&&=-i\frac{2NM\omega_\phonon(\vec{q})}{L_xL_y}\int\langle\mathrm{T}
\hat{u}_\phonon(\vec{r},t)\,\hat{u}_{\phonon}(0,0)\rangle
e^{-i\vec{q}\vec{r}+i\omega{t}}d^2\vec{r}\,dt\nonumber\\
&&=\frac{2\omega_{\phonon}(\vec{q})}
{\omega^2-[\omega_{\phonon}(\vec{q})-io]^2}.
\end{eqnarray}
This definition implies that each electron-phonon vertex
corresponding to the second-quantized version of the interaction
hamiltonian~(\ref{Heph=}) contains, besides the coupling constant
$F_\phonon$ and the corresponding electronic matrix
$\Lambda_i\Sigma_j=(\Lambda\Sigma)_\phonon$, a factor
$\sqrt{L_xL_y/[2NM\omega_\phonon(\vec{q})]}$. Thus, the
overall factor appearing in each vertex at small~$q$ is just
$v\sqrt{\lambda_\phonon/2}$, where $\lambda_\phonon$~is
the dimensionless coupling constant defined in
Eq.~(\ref{lambdaphonon=}). The factor $L_xL_y/N=\sqrt{27}a^2/4$
is the area per one carbon atom.

Upon quantization of the electromagnetic field the operator of the
vector potential is expressed in terms of creation and
annihilation operators
$\hat{a}^\dagger_{\vec{Q},\photon},\hat{a}_{\vec{Q},\photon}$ of
three-dimensional photons in the quantization volume $V=L_xL_yL_z$
with the wave vector~$\vec{Q}$ and two transverse polarizations
$\photon=1,2$ with unit vectors $\vec{e}_{\vec{Q},\photon}$:
\begin{equation}
\hat{\vec{A}}(\vec{r})=
\sum_{\vec{Q},\photon}\sqrt{\frac{2\pi{c}}{VQ}}
\left(\vec{e}_{\vec{Q},\photon}\hat{a}_{\vec{Q},\photon}e^{i\vec{Q}\vec{r}}
+\vec{e}^*_{\vec{Q},\photon}\hat{a}^\dagger_{\vec{Q},\photon}
e^{-i\vec{Q}\vec{r}}\right).
\end{equation}
The photon propagator is defined analogously to the phonon one:
\begin{eqnarray}
&&\Upsilon_\photon(\vec{Q},\Omega)=\nonumber\\
&&=-i\frac{Q}{2\pi{c}}\int\left\langle\mathrm{T}
\hat{A}_\photon(\vec{r},t)\,\hat{A}_{\photon}(0,0)\right\rangle
e^{-i\vec{Q}\vec{r}+i\Omega{t}}d^2\vec{r}\,dt=\nonumber\\
&&=\frac{2cQ}{\Omega^2-[cQ-io]^2},
\end{eqnarray}
so that each electron-photon vertex corresponding to the
second-quantized version of the interaction hamiltonian~(\ref{Heem=})
contains the factor $e(v/c)$, playing the role of the coupling
constant, the electronic matrix
$(\vec{e}_{\vec{Q},\photon}\cdot\vec\Sigma)$,
and a factor $\sqrt{2\pi{c}/Q}$.

Let us diagonalize each of the two electronic Green's functions,
entering and leaving the electron-photon vertex, by
transformation~(\ref{diagDirac=}). Then, neglecting the change
of the electronic momentum~$\vec{p}$ upon emission of the photon,
we transform the electronic matrix in the vertex as
\begin{eqnarray}
&&e^{i\Sigma_z\varphi_{\vec{p}}i/2}e^{i\Sigma_y\pi/4}\,
(\vec{e}_{\vec{Q},\photon}\cdot\vec{\Sigma})\,
e^{-i\Sigma_z\varphi_{\vec{p}}/2}e^{-i\Sigma_y\pi/4}=\nonumber\\
&&=\frac{(\vec{p}\cdot\vec{e}_{\vec{Q},\photon})}{|\vec{p}|}\,\Sigma_z
+\frac{[\vec{p}\times\vec{e}_{\vec{Q},\photon}]_z}{|\vec{p}|}\,\Sigma_y,
\end{eqnarray}
where $\varphi_{\vec{p}}$ is the polar angle of the wave
vector~$\vec{p}$.
Since we are interested in the interband transition, we need the
$\Sigma_y$~part of this expression, which tells us that the
transition dipole moment is perpendicular to the electron
momentum.

\begin{figure}
\includegraphics[width=7.5cm]{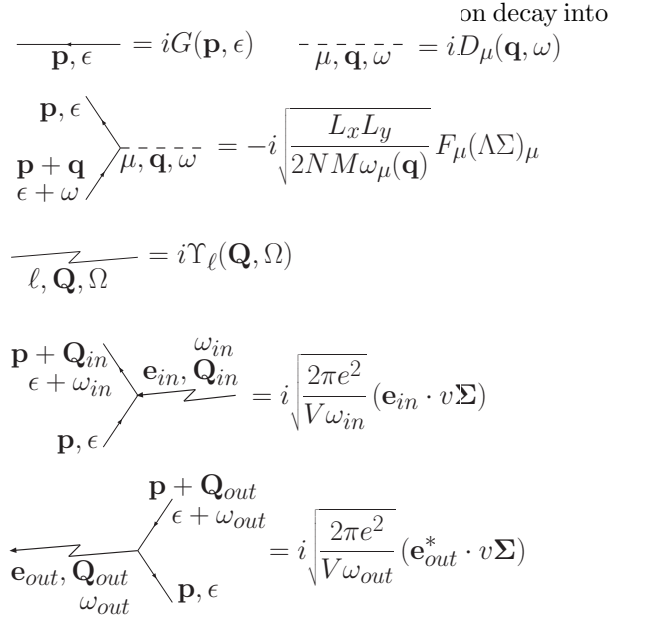}
\caption{\label{fig:diagrams} Graphical representation of
electron, phonon, and photon Green's functions and vertices.}
\end{figure}

The graphical representation of the Green's functions and vertices,
introduced in this subsection, is shown on Fig.~\ref{fig:diagrams}.

\subsection{Raman scattering probability}\label{sec:ramanprobab}

\begin{figure}
\includegraphics[width=7.5cm]{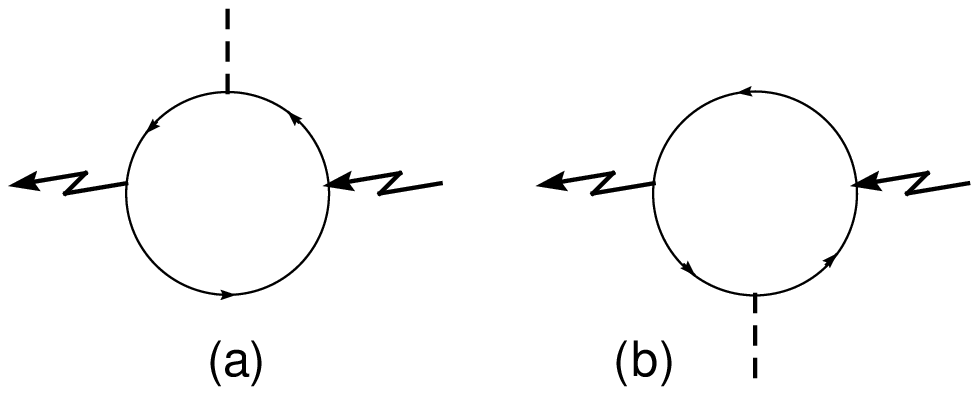}
\caption{\label{fig:Raman1ph} Diagrams for the one-phonon Raman
amplitude in the leading order.}
\end{figure}

Formally, $n$-phonon Raman scattering is a quantum-mechanical
transition from the initial state with the crystal
in the ground state and one incoming photon with the wave
vector~$\vec{Q}_{in}$, frequency~$\omega_{in}=c|\vec{Q}_{in}|$,
and polarization~$\vec{e}_{in}$, into the final state with one
outgoing photon with the wave vector~$\vec{Q}_{out}$,
frequency~$\omega_{out}=c|\vec{Q}_{out}|$,
polarization~$\vec{e}_{out}$, and $n$~phonons corresponding to
normal modes $\phonon_1,\ldots,\phonon_n$ with wave vectors
$\vec{q}_1,\ldots,\vec{q}_n$. Denoting the ground state of the
system by $|\mbox{vac}\rangle$, we represent these two states
as $\hat{a}_{in}^\dagger|\mbox{vac}\rangle$ and
$\hat{a}_{out}^\dagger\hat{b}^\dagger_{\phonon_1\vec{q}_1}
\ldots\hat{b}^\dagger_{\phonon_n\vec{q}_n}|\mbox{vac}\rangle$. Let
us introduce the $S$-matrix $\hat{S}(\infty)$ and define the
amplitude of the $n$-phonon Raman scattering as
\begin{widetext}
\begin{equation}
\mathcal{A}_{\vec{q}_1\ldots\vec{q}_n}^{\phonon_1\ldots\phonon_n}\!%
\left(\Omega;\{\omega_i\}_{i=1}^{i=n}\right)
=\int\frac
{\Av{\mathrm{T}\,\hat{a}_{out}(t)\,
\hat{b}_{\phonon_1\vec{q}_1}(t_1)\ldots\hat{b}_{\phonon_n\vec{q}_n}(t_n)\,
\hat{S}(\infty)\,\hat{a}^\dagger_{in}(0)}}
{\Av{\hat{S}(\infty)}}\,
e^{i\Omega{t}}dt\prod_{i=1}^ne^{i\omega_it_i}dt_i\,.
\label{RamanA=}
\end{equation}
Here the operators are taken in the interaction representation,
and $\mathrm{T}$~represents the chronological ordering.

The diagrammatic representation of the amplitude~(\ref{RamanA=})
in the leading order in electron-photon and electron-phonon
coupling is shown in Figs.~\ref{fig:Raman1ph} and~\ref{fig:Raman2ph}
for $n=1$ and $n=2$,
respectively. The incoming photon line corresponds to pairing
of the operator $\hat{a}_{in}^\dagger$, the outgoing lines -- to
pairings of the operators $\hat{a}_{out}$,~$\hat{b}_{\phonon_i,\vec{q}_i}$.
The loop represents the intermediate states of the electron-hole
pair, summation over the momentum circulating in the loop and
integration over energy should be performed.

Let us separate the Green's functions of the scattering particles
[only the positive-frequency parts of Green's functions enter, as
shown by the $(+)$ superscripts],
the momentum-conserving $\delta$-function,
and explicitly introduce the permutations~$\mathcal{P}$ of the
phonon indices:
\begin{eqnarray}
\mathcal{A}_{\vec{q}_1\ldots\vec{q}_n}^{\phonon_1\ldots\phonon_n}\!%
\left(\Omega;\{\omega_i\}_{i=1}^{i=n}\right)
&=&\frac{i}{\sqrt{V}}\,\Upsilon_{in}^{(+)}\!%
\left(\vec{Q}_{in},\Omega+\sum_{i=1}^n\omega_i\right)
\frac{i}{\sqrt{V}}\,\Upsilon_{out}^{(+)}(\vec{Q}_{out},\Omega)
\prod_{j=1}^n\frac{i}{\sqrt{L_xL_y}}\,
D_{\phonon_j}^{(+)}(\vec{q}_j,\omega_j)\times\nonumber\\
&&{}\times
(2\pi)^2\delta\!\left(\sum_{i=1}^n\vec{q}_i+\vec{Q}_{out}-\vec{Q}_{in}
\right)
\sum_{\mathcal{P}}(-i)\mathcal{M}_{\mathcal{P}\{\vec{q}_i\}}%
^{\mathcal{P}\{\phonon_i\}}
(\Omega;\mathcal{P}\{\omega_i\})\,.
\end{eqnarray}
Then $-i\mathcal{M}$~is given by the sum of all topologically
inequivalent diagrams with amputated external lines. Equivalently,
in the leading order
$\sum_{\mathcal{P}}(-i)\mathcal{M}_{\mathcal{P}}$ is given by the sum
of all $(n+1)!$ connected pairings of electronic $\psi$-operators
while electron-photon and electron-phonon vertices are held fixed.

The transition probability per unit time is given by
\begin{eqnarray}
&&\lim_{t\to+\infty}\frac{1}{t}\left|\int
\mathcal{A}_{\vec{q}_1\ldots\vec{q}_n}^{\phonon_1\ldots\phonon_n}%
(\Omega;\{\omega_i\}_{i=1}^{i=n})\,
e^{-i\Omega{t}}\,\frac{d\Omega}{2\pi}
\prod_{i=1}^ne^{-i\omega_it}\,\frac{d\omega_i}{2\pi}\right|^2=
\nonumber\\ {}&&=\frac{1}{V^2(L_xL_y)^{n}}\,{2}\pi\delta\!
\left(\sum_{i=1}^n\omega_{\vec{q}_i}+\omega_{out}-\omega_{in}\right)
\left[(2\pi)^2\delta\!
\left(\sum_{i=1}^n{\vec{q}_i}+\vec{Q}_{out}-\vec{Q}_{in}\right)\right]^2
\left|\sum_{\mathcal{P}}
\mathcal{M}_{\mathcal{P}\{\vec{q}_i\}}^{\mathcal{P}\{\phonon_i\}}
(\omega_{out};\mathcal{P}\{\omega_{\vec{q}_i}\})\right|^2,
\end{eqnarray}
where we used the relation
\begin{equation}
\lim_{t\to+\infty}\frac{1}{t}\left|\frac{e^{izt}-1}z\right|^2=
2\pi\delta(z)\,.
\end{equation}
The absolute dimensionless probability for the incoming photon to
scatter with emission of $n$~phonons of any kind is obtained by
summing over all final states (here one should remember that a
permutation of phonon arguments represents the same state) and
multiplying by the photon
attempt period $L_z/c$ (at this point we also recall about the
electron spin and multiply the matrix element by a factor of~2
which appears after tracing the fermion loop with respect to
spin indices):
\begin{eqnarray}
I_n=\frac{1}{V^2}\frac{L_z}{c}\sum_{\vec{Q}_{out},\photon_{out}}
\frac{1}{(L_xL_y)^{n}}\frac{1}{n!}\sum_{\{\phonon_i,\vec{q}_i\}}
2\pi\delta\!\left(\sum_{i=1}^n\omega_{\phonon_i}({\vec{q}_i})
+c|\vec{Q}_{out}|-c|\vec{Q}_{in}|\right) \times\nonumber\\\times\left[(2\pi)^2\delta\!
\left(\sum_{i=1}^n{\vec{q}_i}+\vec{Q}_{out}-\vec{Q}_{in}\right)\right]^2
\left|2\mathcal{M}^{\{\phonon_i\}}_{\{\vec{q}_i\}}\right|^2.\label{In=}
\end{eqnarray}
The square of the momentum $\delta$-function is taken care of by
the relation $(2\pi)^2\delta(\vec{q}=0)=L_xL_y$, consistent with
Eq.~(\ref{sumintegral=}). We can also pass
to the spectrally resolved probability by inserting
$1=\int{d}\omega_{out}\,\delta(\omega_{out}-c|\vec{Q}_{out}|)$:
\begin{equation}\label{dIndw=}
\frac{dI_n}{d\omega_{out}}=
\frac{1}c\sum_{\photon_{out}}\int\frac{d^3\vec{Q}_{out}}{(2\pi)^3}\,
\delta(c|\vec{Q}_{out}|-\omega_{out})\,
\frac{1}{(L_xL_y)^{n-1}}\frac{1}{n!}\sum_{\{\phonon_i\}}\sum_{\vec{q}_1+\ldots+\vec{q}_n=0}
2\pi\delta\!\left(\sum_{i=1}^n\omega_{\phonon_i}({\vec{q}_i})
+\omega_{out}-\omega_{in}\right)
\left|2\mathcal{M}^{\{\phonon_i\}}_{\{\vec{q}_i\}}\right|^2.
\end{equation}
At first glance, the matrix
element~$\mathcal{M}^{\{\phonon_i\}}_{\{\vec{q}_i\}}$ does not seem to
depend on the outgoing photon wave vector~$\vec{Q}_{out}$, since
the latter is negligible in comparison with electron and phonon
momenta contributing to~$\mathcal{M}^{\{\phonon_i\}}_{\{\vec{q}_i\}}$,
so the integration over the photon wave vectors gives just
$\omega_{out}^2/(2\pi^2c^3)$.
However, $\mathcal{M}^{\{\phonon_i\}}_{\{\vec{q}_i\}}$ depends
on the orientation of the polarization vector $\vec{e}_{out}$, which, in
turn, depends on the direction of~$\vec{Q}_{out}$. Thus, we should
consider the differential probability of emission into the elementary solid
angle $do_{out}=\sin\Theta\,d\Theta\,d\Phi$, where the spherical angles
$\Theta\in[0,\pi]$ and $\Phi\in[0,2\pi]$ parametrize the direction
of~$\vec{Q}_{out}$. For this differential probability we can write
\begin{eqnarray}
4\pi\,\frac{dI_n}{do_{out}}&=&
\frac{2\pi}c\,\frac{\omega_{out}^2}{2\pi^2c^3}\,
\frac{1}{(L_xL_y)^{n-1}}\frac{1}{n!}\sum_{\{\phonon_i\}}\sum_{\vec{q}_1+\ldots+\vec{q}_n=0}
\left|2\mathcal{M}^{\{\phonon_i\}}_{\{\vec{q}_i\}}\right|^2=\nonumber\\
&=&I_n^\perp
(\vec{e}_{in}\cdot\vec{e}_{in}^*)(\vec{e}_{out}\cdot\vec{e}_{out}^*)
+(I_n^\|-I_n^\perp)\,
\frac{(\vec{e}_{in}\cdot\vec{e}_{out})(\vec{e}_{in}^*\cdot\vec{e}_{out}^*)
+(\vec{e}_{in}\cdot\vec{e}_{out}^*)(\vec{e}_{in}^*\cdot\vec{e}_{out})}2.
\label{dIdout=}
\end{eqnarray}
We stress that only the {\em in-plane} components of the polarization
vectors participate in these scalar products. In writing Eq.~(\ref{dIdout=})
we have assumed the crystal itself to be isotropic, which is true as long as
the calculation is done for the Dirac spectrum. As soon as the trigonal
warping term in Eq.~(\ref{H2=}) is taken into account, the orienation of
the polarization vectors with respect to the crystal directions will enter.
The corrections due to the trigonal warping will be analyzed in
Sec.~\ref{sec:warping}, and will be shown to be small.

For each direction of $\vec{Q}_{out}$ we can choose the basis of
$s$- and $p$-polarizations:
\begin{equation}
\vec{e}_s=(-\sin\Phi,\cos\Phi,0),\quad
\vec{e}_p=(-\cos\Theta\cos\Phi,-\cos\Theta\sin\Phi,\sin\Theta).
\end{equation}
Suppose the light coming out from the sample is collected by a lens
within a cone with the aperture $2\Theta_{det}$. Upon passing through
the lens the polarization vectors change into
\begin{equation}
\vec{e}_s\to(-\sin\Phi,\cos\Phi,0),\quad
\vec{e}_p\to(-\cos\Phi,-\sin\Phi,0),
\end{equation}
so a linearly polarized detector oriented
at an angle~$\Phi_{det}$ will detect only those photons whose
polarization before the lens was
\begin{equation}
\vec{e}_\|=\vec{e}_s\sin(\Phi-\Phi_{det})+\vec{e}_p\cos(\Phi-\Phi_{det}).
\end{equation}
Averaging over the directions gives
\begin{eqnarray}
I_n&=&\int\limits_0^{\Theta_{det}}\frac{\sin\Theta\,d\Theta}2
\int\limits_0^{2\pi}\frac{d\Phi}{2\pi}\,e_\|^ie_\|^j
\left[I_n^\perp(\vec{e}_{in}\cdot\vec{e}_{in}^*)\,\delta_{ij}+
(I_n^\|-I_n^\perp)\,\frac{e_{in}^i(e_{in}^j)^*
+(e_{in}^i)^*e_{in}^j}2\right]
=\nonumber\\
&=&
\int\limits_0^{\Theta_{det}}\frac{\sin\Theta\,d\Theta}{16}
\left[\delta_{ij}(1-\cos\Theta)^2
+2e_{det}^ie_{det}^j(1+\cos\Theta)^2\right]
\left[I_n^\perp(\vec{e}_{in}\cdot\vec{e}_{in}^*)\,\delta_{ij}+
(I_n^\|-I_n^\perp)\,\frac{e_{in}^i(e_{in}^j)^*
+(e_{in}^i)^*e_{in}^j}2\right]=\nonumber\\
&=&\frac{1}{48}
\left[(I_n^\|+I_n^\perp)(1-\cos\Theta_{det})^3
+2I_n^\perp(8-(1+\cos\Theta_{det})^3)\right]
(\vec{e}_{in}\cdot\vec{e}_{in}^*)+\nonumber\\
&&{}+\frac{1}{24}(I_n^\|-I_n^\perp)(8-(1+\cos\Theta_{det})^3)
\left|(\vec{e}_{in}\cdot\vec{e}_{det})\right|^2.\label{Inpolar=}
\end{eqnarray}
\end{widetext}
where $\vec{e}_{det}=(\cos\Phi_{det},\sin\Phi_{det},0)$. Unpolarized
detection corresponds to adding the contributions from two mutually
perpendicular polarizations~$\vec{e}_{det}$ and results in
\begin{equation}\label{Inisotr=}
I_n=\frac{4-3\cos\Theta_{det}-\cos^3\Theta_{det}}{12}\,
(I_n^\|+I_n^\perp)(\vec{e}_{in}\cdot\vec{e}_{in}^*).
\end{equation}

\subsection{Inelastic broadening}\label{sec:inelastic}

As discussed in Sec.~\ref{sec:fullresonance}, when the number of
emitted phonons is even, energy and momentum conservation can be
satisfied in all elementary processes, represented by vertices on
the diagrams for the Raman amplitude. As a consequence, the
energy denominators of all electronic Green's functions forming
the electron-hole loop can be nullified simultaneously, and the
integral over the internal momentum and energy diverges. To cure
this divergence it is essential to include broadening of the
electronic states. In other words, the infinitesimal imaginary
part $-io$ in the denominator of the Green's function~(\ref{Gpe=})
should be replaced by the actual broadening
$-i\gamma_{\vec{p}}=i\Im\Sigma(\vec{p},vp)$, where
$\Sigma(\vec{p},\ep)$ is the electronic self-energy (the effect of
$\Re\Sigma$ will be studied in Sec.~\ref{sec:RG}).

\begin{figure}
\includegraphics[width=7.5cm]{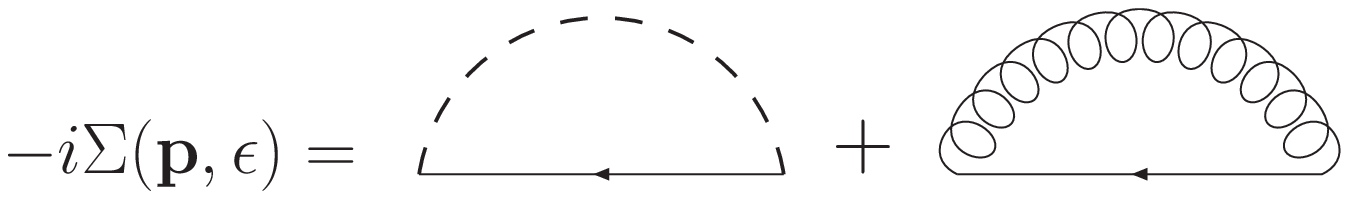}
\caption{\label{fig:selfenrg} Electron self-energy due to
the interactions with phonons and with electron-hole pairs.
The propagator of electron-hole pairs is shown by the
wiggly line, its explicit form is not important for our
calculation.}
\end{figure}

$\Im\Sigma(\vec{p},vp)$ corresponds to emission of some excitations
by the electron (hole). One obvious candidate is the phonon itself,
the corresponding contribution to $\Sigma(\vec{p},\ep)$ is
represented by the first term in Fig.~\ref{fig:selfenrg}.
Besides phonons, an electron can emit other kinds of excitations,
the most important contribution can be expected to come from
emission of electron-hole pairs. Emission of electron-hole pairs
in the undoped graphene must be impurity-assisted,\cite{Guinea96}
while in the doped case electrons can collide without impurities
involved, so the electron-electron collision rate strongly depends
on doping. The propagator of electron-hole pairs is
represented in Fig.~\ref{fig:selfenrg} by the wiggly line. We do
not need the explicit form of the propagator, being interested
only in the contribution to $\Im\Sigma$, introduced
phenomenologically in the denominator of $G(\vec{p},\ep)$.
The only property which is crucial for our consideration is that
the spectrum of electron-hole pairs at each fixed wave vector is
very broad as compared to Raman peak widths, as well as
to~$\gamma_{\vec{p}}$.

The phonon contribution to $\Sigma(\vec{p},\ep)$ (the first term
in Fig.~\ref{fig:selfenrg}) can be calculated explicitly.
Let us do it for the case of $E_2$~phonons:
\begin{eqnarray}\nonumber
&&-i\Sigma^{\Gamma}(\vec{p},\ep)=\nonumber\\
&&=F_{\Gamma}^2\int\frac{d\ep'}{2\pi}\frac{d^2\vec{p}'}{(2\pi)^2}\,
\frac{\sqrt{27a^2}}{4}
\frac{D_{\Gamma}(\vec{p}-\vec{p}',\ep-\ep')}
{2M\omega_{\Gamma}(\vec{p}-\vec{p}')}\times\nonumber\\
&&\quad\times\left[\Sigma_x\Lambda_z\,G(\vec{p}',\ep')\,\Sigma_x\Lambda_z+
\Sigma_y\Lambda_z\,G(\vec{p}',\ep')\,\Sigma_y\Lambda_z\right].\quad
\end{eqnarray}
The two terms in the square brackets correspond to two phonon
polarizations.
Integrating over~$\ep'$ and neglecting the phonon dispersion, we
obtain:
\begin{eqnarray}
&&\Sigma^{\Gamma}(\vec{p},\ep)=\nonumber\\
&&=\frac{F_{\Gamma}^2}{M\omega_{\Gamma}v^2}\frac{\sqrt{27}a^2}4\,\frac{\ep}{2\pi}
\int\limits_0^{\xi_{max}}\frac{\xi'\,d\xi'}{\ep^2-(\xi'+\omega_{\Gamma}-io)^2}
=\nonumber\\
&&=\lambda_{\Gamma}
\left[\frac{\omega_{\Gamma}-\ep}{4\pi}\ln\frac{\xi_{max}}{\omega_{\Gamma}-\ep}
-(\ep\to-\ep) \right], \label{SigmaE2=}
\end{eqnarray}
where $\xi_{max}$ is an ultraviolet cutoff of the order of the
electronic bandwidth.
If we consider scattering on $E_1'$~phonons ($A_1$~representation
of $C_{3v}$~in the $\Kpnt$~point), we obtain the same expression
as Eq.~(\ref{SigmaE2=}), but with $F_{\Gamma}$ and $\omega_{\Gamma}$
replaced by $F_K$ and $\omega_K$. The dimensionless coupling constants
$\lambda_\Gamma$ and $\lambda_K$ are defined in Eq.~(\ref{lambdaphonon=}).

\begin{figure}
\includegraphics[width=7.5cm]{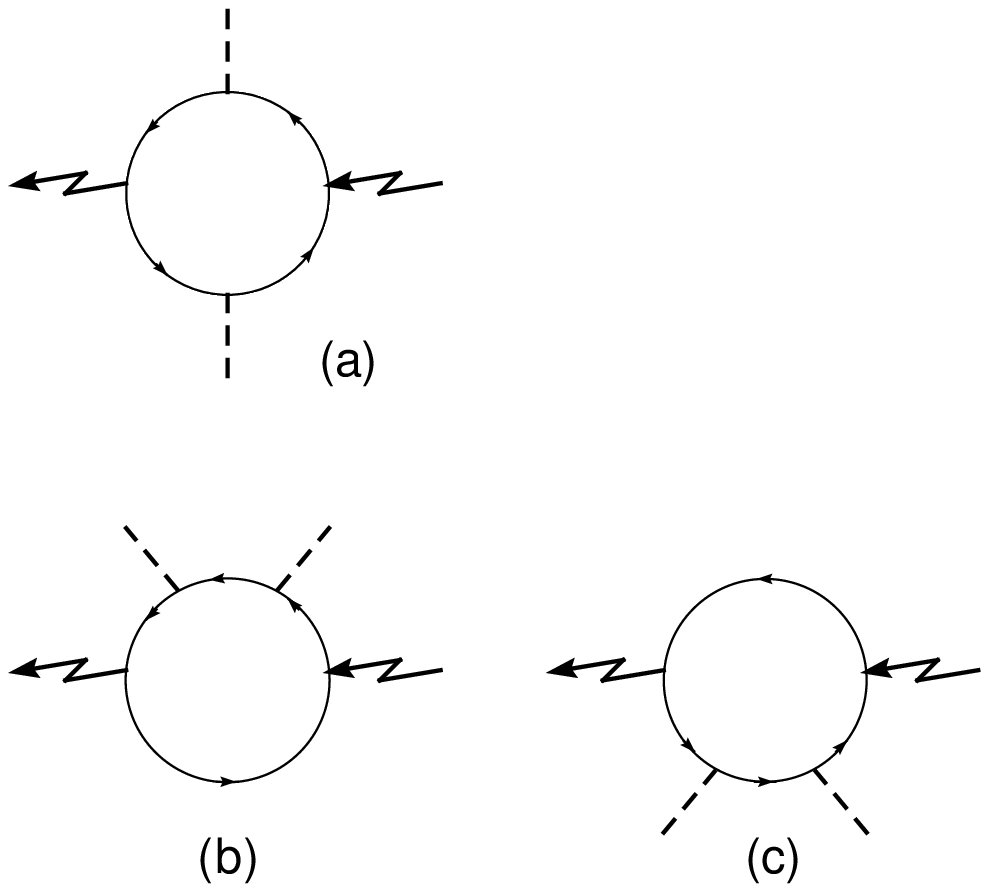}
\caption{\label{fig:Raman2ph} Diagrams for the two-phonon Raman
amplitude in the leading order.}
\end{figure}

\begin{figure}
\includegraphics[width=7cm]{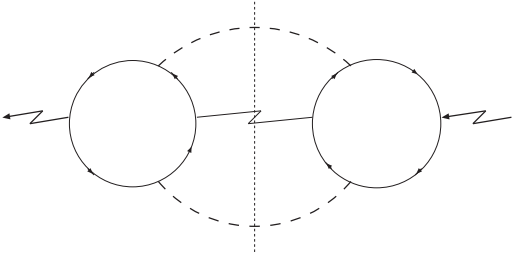}
\caption{\label{fig:cutphoton}
Raman scattering rate as the imaginary part of the photon
self-energy: diagram~(a) on Fig.~\ref{fig:Raman2ph} for the
Raman matrix element is obtained by cutting the photon self-energy
diagram in two, as shown by the vertical short-dashed line.}
\end{figure}

\begin{figure}
\includegraphics[width=7cm]{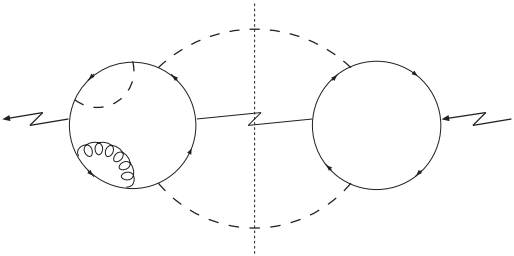}
\caption{\label{fig:cutphotonSE}
A diagram corresponding to self-energy insertions in
the electronic Green's functions.}
\end{figure}

\begin{figure}
\includegraphics[width=7cm]{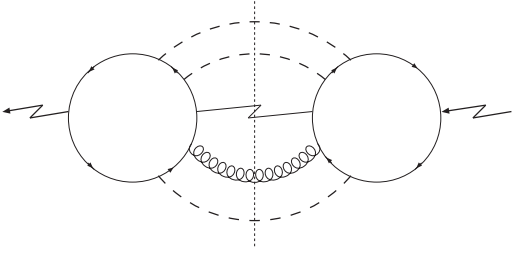}
\caption{\label{fig:cutphotonEXC}
A diagram describing emission of extra excitations.}
\end{figure}

\begin{figure}
\includegraphics[width=7cm]{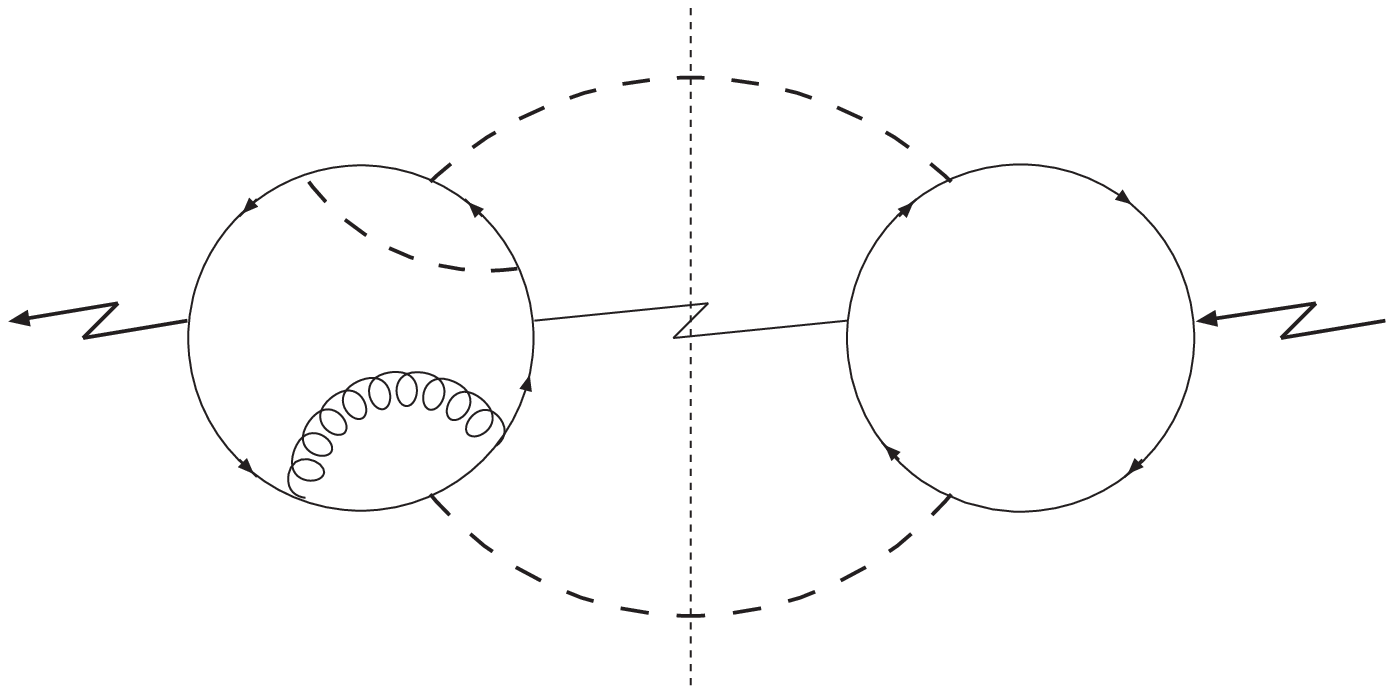}
\caption{\label{fig:cutphotonV}
A diagram representing vertex corrections.}
\end{figure}

To conclude this section, we note that Raman scattering rate
can be viewed as the imaginary part of the incoming photon
self-energy; indeed, any diagram for the Raman matrix element,
e.~g., the first one in Fig.~\ref{fig:Raman2ph}, can be obtained
by cutting a photon self-energy diagram in two, as shown in
Fig.~\ref{fig:cutphoton}. Dressing the
electronic Green's function with self-energy corrections corresponds
to going to higher orders of perturbation theory and taking into
account diagrams of the type shown in Fig.~\ref{fig:cutphotonSE}.
One may ask, what other diagrams may appear in the higher orders
of perturbation theory, and how important they are.
\begin{itemize}
\item
Diagrams where the two electronic loops are connected with more
phonon lines (Fig.~\ref{fig:cutphotonEXC}) describe Raman
scattering on the corresponding number of phonons. If the loops
are connected with an electron-hole propagator, this corresponds to the
contribution of electron-hole pairs to the Raman spectrum. As
discussed in Sec.~\ref{sec:fullresonance}, the resulting spectrum
is broad and featureless, so we are not interested
in these processes.
\item
One can also insert phonon and electron-hole propagators as shown
in Fig.~\ref{fig:cutphotonV}. Such diagrams represent vertex
corrections, their effect is analyzed in Sec.~\ref{sec:RG}.
\item
All diagrams involving more electron-photon vertices have an extra
smallness in the parameter $e^2/c\approx{1}/137$, and are neglected.
\item
Inserting electronic loops into phonon propagators
describes the phonon frequency shift due to electron-electron
interaction and phonon decay into the continuum of electron-hole
pairs. They have been studied before, both
theoretically\cite{Piscanec2004,Lazzeri2006,LazzeriMauri2006,Guinea2007}
and experimentally.\cite{Gruneis,Maultzschexp,Yan2007,Pisana2007}
However, shift and broadening of the phonon states are not important
for our calculation, as we are interested in the
{\em frequency-integrated} intensities of the Raman peaks. They
are determined by the total phonon spectral weight, which is not
changed by shift and broadening.
\end{itemize}

\section{One-phonon Raman scattering\cite{coauthor}}\label{sec:1raman}

Since the photon wave vector is negligibly small compared to all
other scales in the problem, the in-plane momentum conservation
requires the emitted phonon to belong to the $\Gamma$~point. This fact
explains the small width of the peak at 1580~cm$^{-1}$. The matrix
element for the one-phonon process is given by two diagrams in
Fig.~\ref{fig:Raman1ph}:
\begin{widetext}
\begin{eqnarray}
\mathcal{M}^{x,y}=
\frac{2\pi{e}^2v^2}{i\sqrt{\omega_{in}\omega_{out}}}
\sqrt{\frac{L_xL_yF_{\Gamma}^2}{2NM\omega_\Gamma(0)}}
\int\frac{d\ep}{2\pi}\frac{d^2\vec{p}}{(2\pi)^2}
\Tr_{4\times{4}}\left\{(\vec{e}_{in}\vec\Sigma)\,
G(\vec{p},\ep_--\omega_{\Gamma})\,T_{x,y}\,
G(\vec{p},\ep_-)\,(\vec{e}_{out}^*\vec\Sigma)\,
G(\vec{p},\ep_+)\right.+\nonumber\\ +\left. G(\vec{p},\ep_-)\,
(\vec{e}_{out}^*\vec\Sigma)\,G(\vec{p},\ep_+)\,T_{x,y}\,
G(\vec{p},\ep_++\omega_{\Gamma})\,(\vec{e}_{in}\vec\Sigma)
\right\},\label{ampl1ph=}
\end{eqnarray}
\end{widetext}
where we denoted $T_x=\Lambda_z\Sigma_y$, $T_y=-\Lambda_z\Sigma_x$,
$\ep_\pm\equiv\ep\pm\omega_{out}/2$ and
$\omega_{\Gamma}\equiv\omega_{\Gamma}(\vec{q}=0)$.

At first glance, elementary power counting (we set $\ep\sim{v}p$
and note that each Green's function $G\sim{1}/\ep$) tells us that
the integral is logarithmically divergent at high energies.
However, let us recall the continuous
symmetry~(\ref{Cinf=}) 
of the Dirac hamiltonian~(\ref{Hel=}). A rotation~(\ref{Cinfk=})
of the dummy integration variable~$\vec{p}$ by an arbitrary fixed
angle~$\varphi$ cannot change the value of the integral. Thus, the
integral must not change if each Green's function is replaced by
$G\to{e}^{-i\Sigma_z\varphi/2}\,G\,{e}^{i\Sigma_z\varphi/2}$.
Application of these $e^{\pm{i}\Sigma_z\varphi/2}$ to the vertices
is equivalent to the inverse rotation~(\ref{Cinfk=}) of the
polarization vectors $\vec{e}_{in},\vec{e}^*_{out}$ and of the
matrices $\Sigma_x,\Sigma_y$. This gives us a {\em cubic}
combination of $\sin\varphi,\cos\varphi$ which has no
$\varphi$-independent terms. This may be viewed as conservation of
the $z$-component of the angular momentum: indeed, transformation
properties under rotations~$C_{\infty{v}}$ for both the photon
polarization vectors $\vec{e}_{in},\vec{e}_{out}$ and the
$E_2$~phonon displacements $u_{E_2x},u_{E_2y}$ correspond to the
angular momentum $m=\pm{1}$. As a result, for the Dirac spectrum
the one-phonon matrix element~(\ref{ampl1ph=}) {\em must vanish}.

In fact, Eq.~(\ref{ampl1ph=}) gives zero even prior to the
$\vec{p}$-integration. We just notice that
\begin{subequations}\begin{eqnarray}
&&-i\Lambda_x\,G(\vec{p},\ep)\,i\Lambda_x=G(\vec{p},\ep),\label{CPconjG=}\\
&&-i\Lambda_x\,\vec\Sigma\,i\Lambda_x=\vec\Sigma,\label{CPconjS=}\\
&&-i\Lambda_x(\Lambda_z\vec\Sigma)\,i\Lambda_x=-\Lambda_z\vec\Sigma.
\label{CPconjT=}
\end{eqnarray}\label{CPconj=}\end{subequations}
The matrix $-i\Lambda_x$ is the combination of the matrix
$\Lambda_x\Sigma_z$ (the full $C_2$~rotation) and
$e^{-i\Sigma_z\pi/2}$ (rotation by~$\pi$ within each valley). This
symmetry relates the spectra at the points $\Kvec+\vec{p}$ and
$\Kvec'+\vec{p}$.

As a consequence, in order to describe the one-phonon Raman peak
at 1580~cm$^{-1}$, one has to go beyond the leading order in the
small-$p$ expansion of the hamiltonian [i.~e., to take into
account~$H_2$ from Eq.~(\ref{H2=})] and interaction vertices.
Since $\Lambda_x$-symmetry (\ref{CPconjG=}) is broken by the
second term of the hamiltonian $H_2(\vec{p})$ [Eq.~(\ref{H2=})],
the result will be different from zero already at the next order
in~$p$. Then an additional power of~$p$ appears in the integrand
of Eq.~(\ref{ampl1ph=}), so the divergence at the upper limit
becomes linear rather than logarithmic. This means that the
small-$p$ expansion is inapplicable (i.~e., all of its terms give
the contribution of the same order), and the whole Brillouin zone
is responsible for the 1580~cm$^{-1}$ peak. We stress that this
statement is valid in the clean limit, when impurity scattering
can be neglected. Impurity scattering can allow the one-phonon
process in the leading order and make the integral convergent in
the upper limit.\cite{ThomsenReich2004} In the clean limit the proper
tool for the calculation is thus not the effective low-energy
electronic theory, but {\em ab initio} band structure methods,
which are, of course, beyond the scope of the present work. For
low photon energies we are interested in, we can simply add a term
to the system
hamiltonian, corresponding to a direct photon-phonon interaction.
The form of this hamiltonian is fixed by the $C_{6v}$ symmetry
($E_1\reptimes{E}_1\reptimes{E}_2$ contains only one
$A_1$~representation):
\begin{equation}
H_{\mathrm{em-ph}}=\frac{e^2}{v_{\mathrm{R}}}
\int{d}^2\vec{r}
\left[(\mathcal{E}_x^2-\mathcal{E}_y^2)u_{E_2y}
-2\mathcal{E}_x\mathcal{E}_yu_{E_2x}\right],
\end{equation}
where $\mathcal{E}_x,\mathcal{E}_y$ are the in-plane cartesian
components of the electric
field vector $\vec{\mathcal{E}}=-(1/c)\partial\vec{A}/\partial{t}$,
$e$~is the electron charge, and $v_{\mathrm{R}}$~is
the unknown constant of the dimensionality of velocity. Since it
originates from the electronic $\pi$-band, its magnitude is roughly
given by the product of the electronic bandwidth and the lattice
constant, i.~e., it should be of the same order as the electronic
velocity~$v$.

This hamiltonian leads to the following expression for the one-phonon
Raman matrix element:
\begin{eqnarray}
\left.\begin{array}{c} \mathcal{M}^x \\ \mathcal{M}^y \end{array}\right\}
&=&\frac{2e^2}{v_{\mathrm{R}}}
\frac{2\pi\sqrt{\omega_{in}\omega_{out}}}{\sqrt{2NM\omega_\Gamma(0)/(L_xL_y)}}
\times\nonumber\\
&&\times\left\{\begin{array}{c}
-e_{in,x}e_{out,y}^*-e_{in,y}e_{out,x}^*,\\
e_{in,x}e_{out,x}^*-e_{in,y}e_{out,y}^*.
\end{array}\right.
\end{eqnarray}
The resulting intensity does not depend on the polarization of the detector.
The sum over the two polarizations, calculated according to the prescription 
of Sec.~\ref{sec:ramanprobab}, is given by
\begin{eqnarray}
I_\Gamma&=&8\pi\left(\frac{e^2}c\right)^2
\frac{\omega_{in}\omega_{out}^3}{v_\mathrm{R}^2c^2}
\frac{\sqrt{27a^2}}{4M\omega_\Gamma(0)}\times\nonumber\\
&&\times\frac{4-3\cos\Theta_{det}-\cos^3\Theta_{det}}{6}\,
(\vec{e}_{in}\cdot\vec{e}_{in}^*).\label{IG=}
\end{eqnarray}

Furthermore, Eqs.~(\ref{CPconj=}) 
lead to vanishing of the Raman amplitude for any odd number of
$\Gamma$~phonons. This property resembles Furry's theorem in the
spinor quantum electrodynamics.\cite{LL4} There is, however, a
difference: Furry's theorem holds for the sum of two diagrams
containing an odd number of photon lines and differing by the
direction of the electronic loop which cancel each other, while in
Eq.~(\ref{ampl1ph=}) each of the two terms vanishes separately. We
also note that $\Kpnt$~phonons can be emitted only in pairs: the
excited electron-hole pair should switch valleys an even number of
times in order to annihilate.\footnote{In principle, momentum
conservation allows
  emission of just three $\Kpnt$~phonons, as $\Kvec+\Kvec+\Kvec$ is
  equivalent to~$\vec{0}$. However, this process cannot be resonant.}
The difference from the one-phonon process is that the 
integral in the matrix element converges at large~$k$ even after the
next term in the small-$k$ expansion have been picked up. Thus, the
three-phonon process can be described within the low energy theory,
but the corresponding amplitude will contain an additional smallness
$\sim\omega_{in}a/c$.

\section{Two-phonon Raman scattering}\label{sec:2raman}

\subsection{Calculation for the Dirac
spectrum}\label{sec:2ramanDirac}

First, let us focus on the $2K$~peak at 2700~cm$^{-1}$, which
corresponds to emission
of two scalar phonons ($A_1$~in terms of the $C_{3v}$~symmetry,
or~$E_1'$ in terms of the $C_{6v}''$ symmetry) from the vicinity of
the $\Kpnt$ and $\Kpnt'$ points.
The integrated intensity of the $2K$~peak is given by
\begin{equation}\label{I2Kstart=}
4\pi\,\frac{dI_{2K}}{do_{out}}
=\frac{2\pi}{c^2}\,\frac{\omega_{out}^2}{2\pi^2c^2}\,
\frac{1}{2}\cdot 2\int\frac{d^2\vec{q}}{(2\pi)^2}
\left|2\mathcal{M}_{\vec{q}}^K\right|^2.
\end{equation}
The factor $1/2$ in front of the sum eliminates double counting in
the summation over the final states (phonon permutations), the
factor of~$2$ comes from the sum over $A_1$ and $B_1$~modes (no
cross-terms arise, as they would yield a traceless combination
$\Lambda_x\Lambda_y$), the factor of~2 inside the square takes
care of the spin degeneracy.

The diagrams for the two-phonon matrix element are shown in
Fig.~\ref{fig:Raman2ph}. Only the first diagram corresponds to
the fully resonant process, described in Sec.~\ref{sec:fullresonance},
the other two give a contribution, smaller by a factor
$\gamma/\omega_K$. Thus, the matrix element is given by
\begin{eqnarray}
\mathcal{M}_{\vec{q}}^K&=&
\frac{2\pi{e}^2v^2}{\sqrt{\omega_{in}\omega_{out}}}\,
\frac{L_xL_yF^2_K}{2NM\omega_K(\vec{q})}
\int\frac{d\ep}{2\pi}\frac{d^2\vec{p}}{(2\pi)^2}\times\nonumber\\
&&\times\Tr_{4\times{4}}\left\{(\vec{e}_{in}\cdot\Sigma)\,
G(\vec{p},\ep_-)\,\Lambda_x\Sigma_zG(\vec{p}+\vec{q},\ep_-')\right.
\times\nonumber\\ &&\times\left. (\vec{e}_{out}^*\cdot\Sigma)\,
G(\vec{p}+\vec{q},\ep_+')\,\Lambda_x\Sigma_z
G(\vec{p},\ep_+)\right\}+\nonumber\\ &&{}+(\vec{q}\to-\vec{q}),
\end{eqnarray}
where $\ep_\pm=\ep\pm\omega_{in}/2$, $\ep_\pm'=\ep\pm\omega_{out}/2$.
We diagonalize the Green's functions by the unitary
transformation~(\ref{diagDirac=}) and neglect the off-resonant
contribution:
\begin{equation}\label{UGU=}
G(\vec{p},\ep_\pm)=U_{\vec{p}}^\dagger\,
\frac{\ep_\pm+\Sigma_z\xi_{\vec{p}}}{\ep_\pm^2-\xi_{\vec{p}}^2}\,
\,U_{\vec{p}}\approx{U}_{\vec{p}}^\dagger\,
\frac{(\unitmatrix\pm\Sigma_z)/2}{\ep_\pm\mp\xi_{\vec{p}}}\,
U_{\vec{p}}\,,
\end{equation}
where $\xi_{\vec{p}}\equiv{v}|\vec{p}|-i\gamma_{\vec{p}}$,
$U_{\vec{p}}=e^{i\Sigma_y\pi/4}e^{i\Sigma_z\varphi_{\vec{p}}/2}$, and
$\varphi_{\vec{p}}=\arctan(p_y/p_x)$ is the polar angle of the
vector~$\vec{p}$. The unitary matrices rotate the vertices as
\begin{subequations}\begin{eqnarray}
&&U_{\vec{p}}\,\Sigma_x\,U_{\vec{p}'}^\dagger=
\Sigma_z\cos\frac{\varphi_{\vec{p}}+\varphi_{\vec{p}'}}2
-\Sigma_y\sin\frac{\varphi_{\vec{p}}+\varphi_{\vec{p}'}}2,
\label{USigmaxU=}\\
&&U_{\vec{p}}\,\Sigma_y\,U_{\vec{p}'}^\dagger=
\Sigma_z\sin\frac{\varphi_{\vec{p}}+\varphi_{\vec{p}'}}2
+\Sigma_y\cos\frac{\varphi_{\vec{p}}+\varphi_{\vec{p}'}}2,
\label{USigmayU=}\\
&&U_{\vec{p}}\,\Sigma_z\,U_{\vec{p}'}^\dagger=
i\unitmatrix\sin\frac{\varphi_{\vec{p}}-\varphi_{\vec{p}'}}2
-\Sigma_x\cos\frac{\varphi_{\vec{p}}-\varphi_{\vec{p}'}}2,
\label{USigmazU=}\\
&&U_{\vec{p}}\,(\vec{e}_{in}\cdot\vec{\Sigma})\,U_{\vec{p}}^\dagger
=(\vec{e}_{\vec{p}}\cdot\vec{e}_{in})\,\Sigma_z
+[\vec{e}_{\vec{p}}\times\vec{e}_{in}]_z\,\Sigma_y,
\qquad\label{UvecSigmaU=}
\end{eqnarray}\label{USigmaU=}\end{subequations}
where $\vec{e}_{\vec{p}}=\vec{p}/|\vec{p}|$. We are interested in the
$\Sigma_y$~term in Eq.~(\ref{UvecSigmaU=}), which
corresponds to interband transitions, and also tells us that the
transition dipole moment is perpendicular to the electron momentum.
In Eq.~(\ref{USigmazU=}), the rotated phonon vertex, we
need the intraband term $\propto\unitmatrix$. Evaluation of the trace
gives
\begin{equation}\label{2phonontrace=}
2\,\frac{[\vec{e}_{\vec{p}}\times\vec{e}_{in}]_z
[\vec{e}_{\vec{p}+\vec{q}}\times\vec{e}_{out}^*]_z
\sin^2(\varphi_{\vec{p}+\vec{q}}/2-\varphi_{\vec{p}}/2)}
{(\ep_-+\xi_{\vec{p}})(\ep_-'+\xi_{\vec{p}+\vec{q}})
(\ep_+'-\xi_{\vec{p}+\vec{q}})(\ep_+-\xi_{\vec{p}})}\,.
\end{equation}
%
%
We can rewrite each of the factors in the denominator of this
expression as
\begin{eqnarray}
\frac{1}{\ep-\omega_{in}/2+\xi_{\vec{p}}}&=&
i\int\limits_0^\infty{d}\bar{t}_0\,
e^{-i\ep\bar{t}_0+i(\omega_{in}/2-\xi_{\vec{p}})\bar{t}_0},\nonumber\\
\frac{1}{\ep-\omega_{out}/2+\xi_{\vec{p}+\vec{q}}}&=&
i\int\limits_0^\infty{d}\bar{t}_1\,
e^{-i\ep\bar{t}_1+i(\omega_{out}/2-\xi_{\vec{p}+\vec{q}})\bar{t}_1},\nonumber\\
\frac{1}{\ep+\omega_{in}/2-\xi_{\vec{p}}}&=&
\frac{1}i\int\limits_0^\infty{d}t_0\,
e^{i\ep{t}_0+i(\omega_{in}/2-\xi_{\vec{p}}){t}_0},\nonumber\\
\frac{1}{\ep+\omega_{out}/2-\xi_{\vec{p}+\vec{q}}}&=&
\frac{1}i\int\limits_0^\infty{d}{t}_1\,
e^{i\ep{t}_1+i(\omega_{out}/2-\xi_{\vec{p}+\vec{q}}){t}_1}.\nonumber\\
\label{timerep=}
\end{eqnarray}
Integration over $d\ep/(2\pi)$ produces
$\delta(t_0+t_1-\bar{t}_0-\bar{t}_1)$. Thus the times $t_0,t_1$ and
$\bar{t}_0,\bar{t}_1$ can be interpreted as times spend in the
corresponding intermediate states by the electron and the hole,
respectively.

Let us denote $p_0=\omega_{in}/(2v)$, $p_1=\omega_{out}/(2v)$,
$\tilde{q}=q-p_0-p_1$, and let $\tilde{p}_\|,\tilde{p}_\perp$ be
the components of the deviation
$\tilde{\vec{p}}=\vec{p}+p_0\vec{e}_{\vec{q}}$ along~$\vec{q}$ and
perpendicular to~$\vec{q}$, respectively. We expect the deviations
to be small, so we approximate
\begin{subequations}\begin{eqnarray}
\Re\xi_{\vec{p}}&=&
v\sqrt{(-p_0+\tilde{p}_\|)^2+\tilde{p}_\perp^2}\approx\nonumber\\
&\approx& vp_0-v\tilde{p}_\|+\frac{v\tilde{p}_\perp^2}{2p_0},\\
\Re\xi_{\vec{p}+\vec{q}}&=&
v\sqrt{(p_1+\tilde{p}_\|+\tilde{q})^2+\tilde{p}_\perp^2}
\approx\nonumber\\
&\approx&vp_1+v\tilde{p}_\|+v\tilde{q}+\frac{v\tilde{p}_\perp^2}{2p_1},
\end{eqnarray}\label{diffraction=}\end{subequations}
and $\vec{e}_{\vec{p}}\approx-\vec{e}_{\vec{q}}$,
$\vec{e}_{\vec{p}+\vec{q}}\approx\vec{e}_{\vec{q}}$,
$\varphi_{\vec{p}+\vec{q}}/2-\varphi_{\vec{p}}/2\approx\pi/2$.
Integration over momentum deviation~$\tilde{\vec{p}}$ is performed
as (we denote
$\gamma_{\vec{p}_0}+\gamma_{\vec{p}_0+\vec{q}}=2\gamma$ for
brevity):
\begin{widetext}
\begin{eqnarray}
&&\int\limits_0^\infty{d}t_0\,dt_1\,d\bar{t}_0\,d\bar{t}_1\,
\delta(t_0+t_1-\bar{t}_0-\bar{t}_1)\,e^{-iv\tilde{q}(t_1+\bar{t}_1)}
\int\limits_{-\infty}^\infty\frac{d\tilde{p}_\|}{2\pi}\,
e^{iv\tilde{p}_\|(t_0+\bar{t}_0-t_1-\bar{t}_1)} \times\nonumber\\
&&\quad{}\times\int\limits_{-\infty}^\infty\frac{d\tilde{p}_\perp}{2\pi}\,
e^{-[\gamma_{\vec{p}_0}+iv\tilde{p}_\perp^2/(2p_0)](t_0+\bar{t}_0)
-[\gamma_{\vec{p}_0+\vec{q}}+iv\tilde{p}_\perp^2/(2p_1)](t_1+\bar{t}_1)}
=-\frac{1}{8v^2}\sqrt{\frac{\omega_{in}\omega_{out}}
{\omega_{in}+\omega_{out}}}
\frac{1}{(v\tilde{q}-2i\gamma)^{3/2}},\label{backscattering=}
\end{eqnarray}
\end{widetext}
which gives the matrix element:
\begin{eqnarray}
\mathcal{M}_{\vec{q}}^K&=&
\frac{2\pi{e}^2v^2}{\sqrt{\omega_{in}+\omega_{out}}}\,
\frac{L_xL_yF_K^2}{2NM\omega_K(\vec{q})}\, \frac{1}{8v^2}
\times\nonumber\\ &&\times
\frac{2[\vec{e}_{\vec{q}}\times\vec{e}_{in}]_z
[\vec{e}_{\vec{q}}\times\vec{e}_{out}^*]_z}
{(v\tilde{q}-2i\gamma)^{3/2}} +(\vec{q}\to-\vec{q}).\qquad
\label{Mq2ph=}
\end{eqnarray}
This expression describes a strongly peaked $q$~dependence of the
matrix element $\mathcal{M}(\vec{q})$ around the value
$q_{bs}=(\omega_{in}+\omega_{out})/(2v)$, corresponding to
backscattering of the electron and the hole by the phonons. The
width of the peak is determined by the {\em electron lifetime}:
$\delta{q}\sim\gamma/{v}$. Such sharply peaked dependence cannot
be derived from pure symmetry considerations, which just prescribe
vanishing of the phonon matrix element for the forward
scattering.\cite{ThomsenReich2004} It is a consequence of the
fully resonant character of the two-phonon Raman scattering, which
should be contrasted to the double-resonant impurity-assisted
single-phonon scattering. For the latter case, the scale
determining the width of the peak is $\delta{q}\sim\omega_{in}/v$
[see Eq.~(2) of Ref.~\onlinecite{Maultzsch}].

The origin of such peaked dependence lies in the quasiclassical
nature of the electron and hole motion. Namely, the incoming
photon creates and electron with momentum $\vec{p}_0$, moving with
the velocity
$\vec{v}_0=\partial\Re\xi_{\vec{p}}/\partial\vec{p}|_{\vec{p}=\vec{p}_0}$,
and a hole with momentum $-\vec{p}_0$, moving with the velocity
$-\vec{v}_0$. After a time $t_0$~the electron emits a phonon of
momentum~$-\vec{q}$, and after a time~$\bar{t}_0$ the hole emits a
phonon of momentum~$\vec{q}$. Afterwards, at time
$t_0+t_1=\bar{t}_0+\bar{t}_1$ they recombine and emit a photon. In
order to recombine, they must meet at the same spatial point (as
prescribed by the spatial $\delta$-function resulting from the
integration over the deviations~$\tilde{\vec{p}}$) with opposite
momenta $\vec{p}_1=\vec{p}_0+\vec{q}$ and $-\vec{p}_1$ (as
prescribed by the momentum conservation). It is possible only if
the velocity
$\vec{v}_1=\partial\Re\xi_{\vec{p}}/\partial\vec{p}|_{\vec{p}=\vec{p}_1}$
is directed oppositely to the initial velocity~$\vec{v}_0$. The
deviation of the typical scattering angle~$\varphi$ from~$\pi$ is
$|\varphi-\pi|\sim\sqrt{\gamma/\omega_{in}}\ll{1}$, determined by
the quantum diffraction [note that we had to expand the energy to
the second order in~$\tilde{p}_\perp$ in
Eqs.~(\ref{diffraction=})].

Proceeding with the calculation, we substitute Eq.~(\ref{Mq2ph=}) into
Eq.~(\ref{I2Kstart=}). Angular averaging gives
\begin{widetext}
\begin{equation}
\int\limits_0^{2\pi}\frac{d\varphi_{\vec{q}}}{2\pi}\,
\left|[\vec{e}_{\vec{q}}\times\vec{e}_{in}]_z
[\vec{e}_{\vec{q}}\times\vec{e}_{out}^*]_z\right|^2
=\frac{1}8
\left[(\vec{e}_{in}\cdot\vec{e}_{out})(\vec{e}_{in}^*\cdot\vec{e}_{out}^*)
+(\vec{e}_{in}\cdot\vec{e}_{out}^*)(\vec{e}_{in}^*\cdot\vec{e}_{out})
+(\vec{e}_{in}\cdot\vec{e}_{in}^*)(\vec{e}_{out}\cdot\vec{e}_{out}^*)\right],
\end{equation}
so that $I_{2K}^\|=3I_{2K}^\perp$, as seen from Eq.~(\ref{dIdout=}).
We stress that only the {\em in-plane} components of the polarization
vectors participate in these scalar products. Evaluating the $q$~integral
and using Eq.~(\ref{Inpolar=}), we obtain the final result:
\begin{eqnarray}\nonumber
I_{2K}&=&\left(\frac{e^2}{c}\right)^2
\frac{v^2}{c^2}\,\frac{\omega_{out}^2}
{8\gamma^2} \left(\frac{F^2_K}{Mv^2\omega_K(q_{bs})}
\frac{\sqrt{27}a^2}4\right)^2 \times\\ &&\times
\frac{1}8\left[\frac{|\vec{e}_{in}|^2}8
(1-\cos\Theta_{det})(3+\cos^2\Theta_{det})
+\frac{8-(1+\cos\Theta_{det})^3}{12}
|(\vec{e}_{in}\cdot\vec{e}_{det})|^2\right], \label{I2Kend=}
\end{eqnarray}
\end{widetext}
where $q_{bs}=(\omega_{in}+\omega_{out})/(2v)$ is the phonon
wave vector corresponding to the backscattering, and $\sqrt{27}a^2/4$
is the area per carbon atom. Eq.~(8) of Ref.~\onlinecite{myself}
corresponds\cite{mistake} to normal incidence ($|\vec{e}_{in}|=1$),
collection in the full solid angle~$4\pi$, and summation over the two
orthogonal directions of~$\vec{e}_{det}$, which makes the second line
of Eq.~(\ref{I2Kend=}) equal to 1/3.

Eq.~(\ref{I2Kend=}) shows how the dominant role of backscattering
manifests itself in the polarization memory. Indeed, linearly
polarized light preferentially excites electrons and holes with
momenta perpendicular to the electric field vector. After the phonon
emission these momenta change to the opposite, and the photon emitted
after the annihilation has a preferred direction for the polarization,
perpendicular to the electron and hole momenta. Quantitatively,
Eq.~(\ref{I2Kend=}) gives the ratio of intensities
for $\vec{e}_{det}\|\vec{e}_{in}$ and $\vec{e}_{det}\perp\vec{e}_{in}$
to be 3 at $\Theta_{det}\to{0}$ and $23/9$ at $\Theta_{det}=\pi/2$
(collection into the solid angle~$2\pi$).

Let us now turn to the $2\Gamma$ phonon peak at 3250~cm$^{-1}$,
corresponding to emission of two pseudovector $E_2$~phonons from the
vicinity of the $\Gamma$~point. Its integrated intensity can be
calculated analogously:
\begin{equation}
4\pi\,\frac{dI_{2\Gamma}}{do_{out}}=
\frac{2\pi}{c^2}\,\frac{\omega_{out}^2}{2\pi^2c^2}\,
\frac{1}{2}\sum_{i,j=x,y}\int\frac{d^2\vec{q}}{(2\pi)^2}
|2\mathcal{M}^{ij}_{\vec{q}}|^2.
\end{equation}
As in Eq.~(\ref{I2Kstart=}), the factor $1/2$ in front of the sum
eliminates double counting in the summation over the final states
(phonon permutations), the factor of~2 inside the square takes
care of the spin degeneracy. Here, however, the summation over the
two phonon modes is less simple. First, let us neglect the phonon
dispersion, so that the longitudinal and transverse phonons are
degenerate. The matrix element is given by
\begin{eqnarray}
\mathcal{M}^{ij}_{\vec{q}}&=&
\frac{2\pi{e}^2v^2}{\sqrt{\omega_{in}\omega_{out}}}\,
\frac{L_xL_yF_{\Gamma}^2}{2NM\omega_{\Gamma}}
\int\frac{d\ep}{2\pi}\frac{d^2\vec{p}}{(2\pi)^2}\times\nonumber\\
&&\times\Tr_{4\times{4}}\left\{(\vec{e}_{in}\cdot\Sigma)\,
G(\vec{p},\ep_-)\,T_i\,G(\vec{p}+\vec{q},\ep_-')\right.
\times\nonumber\\ &&\times\left. (\vec{e}_{out}^*\cdot\Sigma)\,
G(\vec{p}+\vec{q},\ep_+')\,T_j\,
G(\vec{p},\ep_+)\right\}+\nonumber\\ &&{}+(\vec{q}\to-\vec{q}),
\end{eqnarray}
where $T_x=\Lambda_z\Sigma_y$, $T_y=-\Lambda_z\Sigma_x$. The unitary
transformation~(\ref{diagDirac=}) rotates the isospin as
given by Eqs.~(\ref{USigmaxU=}), (\ref{USigmayU=}).
Thus, the trace will be given by an expression, analogous to
Eq.~(\ref{2phonontrace=}), but with the replacement
\begin{eqnarray*}
&&\sin^2(\varphi_{\vec{p}+\vec{q}}/2-\varphi_{\vec{p}}/2)\to
-\bar{e}^i\bar{e}^j,\\
&&\bar{e}^x=-\sin\frac{\varphi_{\vec{p}}+\varphi_{\vec{p}+\vec{q}}}2,
\quad
\bar{e}^y=\cos\frac{\varphi_{\vec{p}}+\varphi_{\vec{p}+\vec{q}}}2.
\end{eqnarray*}
All arguments leading to the dominance of backscattering remain
valid, so we obtain $\bar{\vec{e}}=-\vec{e}_{\vec{q}}$ which
means that only the longitudinal phonons are emitted. Summation
of the probability over the two phonon polarizations gives
$(\bar{e}^x)^4+(\bar{e}^y)^4+2(\bar{e}^x)^2(\bar{e}^y)^2=1$, so
the intensity of the $2\Gamma$ peak is given by one half of the
expression~(\ref{I2Kend=}) with replacements $F_K\to{F}_{\Gamma}$,
$\omega_K\to\omega_{\Gamma,L}$:
\begin{widetext}
\begin{eqnarray}\nonumber
I_{2\Gamma}&=&\frac{1}{2}
\left(\frac{e^2}{c}\right)^2\frac{v^2}{c^2}\,\frac{\omega_{out}^2}
{8\gamma^2} \left(\frac{F^2_{\Gamma}}{Mv^2\omega_{\Gamma,L}(q_{bs})}
\frac{\sqrt{27}a^2}4\right)^2 \times\\ &&\times
\frac{1}8\left[\frac{|\vec{e}_{in}|^2}8
(1-\cos\Theta_{det})(3+\cos^2\Theta_{det})
+\frac{8-(1+\cos\Theta_{det})^3}{12}
|(\vec{e}_{in}\cdot\vec{e}_{det})|^2\right]. \label{I2Gend=}
\end{eqnarray}
\end{widetext}

\subsection{Effect of trigonal warping and electron-hole asymmetry}
\label{sec:warping}

In the previous subsection we have calculated the integrated intensities
of $2K$ and $2\Gamma$ peaks at 2700~cm$^{-1}$ and 3250~cm$^{-1}$,
respectively. According to the data of Ref.~\onlinecite{Ferrari2006},
$I_{2K}/I_{2\Gamma}\approx{20}$. At the same time, for the corresponding
electron-phonon coupling constants the nearest-neighbor
bond-stretching approximation gives $F_K/F_{\Gamma}=1$, which agrees
with DFT calculations of Ref.~\onlinecite{Piscanec2004} with the precision
of~1\%. Then, what is the origin of such huge difference in the
intensities of the two peaks? In this section we consider the effect of
electron-hole asymmetry and trigonal band warping on the intensities of
the two-phonon Raman peaks with the purpose to check whether the
trigonal warping can explain the observed large difference of intensities
$I_{2K}$~and~$I_{2\Gamma}$.

Typically, one neglects corrections to the Dirac spectrum
$v|\vec{p}|$, arising from the quadratic term $H_2(\vec{p})$ in the
electronic hamiltonian, given by Eq.~(\ref{H2=}), as they are smaller
than $v|\vec{p}|$ by
a factor $pa\ll{1}$. However, according to the results of the previous
subsection, two-phonon scattering is sensitive to the directions of
electronic velocities and momenta on the angular scale
$\delta\varphi\sim\sqrt{\gamma/\omega_{in}}$. This means that effects
of (i)~phase mismatch between the electron and the hole, introduced by
the first term in Eq.~(\ref{H2=}), and (ii)~non-collinearity of
velocity and momentum, introduced by the second term in
Eq.~(\ref{H2=}), may become important already when
$H_2(\vec{p})\sim\gamma$, which happens at a smaller energy scale
than $H_2(\vec{p})\sim{H}_1(\vec{p})$. This means that, while
including $H_2(\vec{p})$, we still can neglect higher-order
terms of the expansion ($H_3,H_4,\ldots$).

Thus, we repeat the calculations of the previous subsection, taking
into account the $H_2(\vec{p})$ term only in the denominator of
expression~(\ref{2phonontrace=}) (the numerator is a smooth
function of~$\vec{p}$, so corrections to it will be small as $pa$
indeed).
The quadratic term~(\ref{H2=}) in the hamiltonian modifies the
electron dispersion as
\begin{subequations}\begin{eqnarray}
\Re\xi_{\vec{p}}&=&vp+\alpha_0p^2
\pm\alpha_3\nu_\|(\vec{p})\,{p^2}\label{Rexip=}\\
\Re\bar\xi_{\vec{p}}&=&vp-\alpha_0p^2
\pm\alpha_3\nu_\|(\vec{p})\,{p^2}\label{Rebarxip=}\\
\frac{\partial\Re\xi_{\vec{p}}}{\partial\vec{p}}&=&
\left[\frac{v}p+{2}\alpha_0
\pm{2}\alpha_3\nu_\|(\vec{p})\right]\vec{p}\pm\nonumber\\
&&{}\pm{2}\alpha_3\nu_\perp(\vec{p})[\vec{e}_z\times\vec{p}],\\
\nu_\|(\vec{p})&=&
\frac{p_x^3-3p_xp_y^2}{p^3}=\cos{3}\varphi_{\vec{p}},\\
\nu_\perp(\vec{p})&=&
\frac{3p_y^3-9p_x^2p_y}{p^3}=-3\sin{3}\varphi_{\vec{p}},
\end{eqnarray}\label{warpdisp=}\end{subequations}
where the sign of the $\alpha_0$ terms is ``$+$'' in~$\xi_{\vec{p}}$
(electron dispersion) and ``$-$'' in~$\bar\xi_{\vec{p}}$ (hole
dispersion), and the sign of the $\alpha_3$ terms is ``$+$''
and ``$-$'' for $\Kpnt$~and~$\Kpnt'$ valleys, respectively.


For a given direction $\vec{e}_{\vec{q}}$ we choose
$\vec{p}_0=-p_0\vec{e}_{\vec{q}}$,
$\vec{p}_1=p_1\vec{e}_{\vec{q}}$, with $p_0$ and $p_1$ such that
$\Re\xi_{\vec{p}_0}+\Re\bar\xi_{\vec{p}_0}=\omega_{in}$,
$\Re\xi_{\vec{p}_1}+\Re\bar\xi_{\vec{p}_1}=\omega_{out}$, and
denote $\tilde{q}=|\vec{q}|-p_0-p_1$,
$\tilde{\vec{p}}=\vec{p}-\vec{p}_0$. The energies of the
intermediate states to be substituted in Eqs.~(\ref{timerep=}) at
the corresponding times, can be approximated as
\begin{widetext}
\begin{subequations}\begin{eqnarray}
\Re\bar\xi_0 &\approx& \frac{\omega_{in}}2-\alpha_0p_0^2
+\left[\frac{v}p_0-2\alpha_0+2\alpha_3\nu_\|(\vec{p}_0)\right]
(\vec{p}_0\cdot\tilde{\vec{p}})
+2\alpha_3\nu_\perp(\vec{p}_0)[\vec{p}_0\times\tilde{\vec{p}}]_z
+\frac{v}{2p_0^3}[\vec{p}_0\times\tilde{\vec{p}}]_z^2,\\
\Re\bar\xi_1&\approx& \frac{\omega_{out}}2-\alpha_0p_1^2
+\left[\frac{v}p_1-2\alpha_0\mp{2}\alpha_3\nu_\|(\vec{p}_1)\right]
(\vec{p}_1\cdot\tilde{\vec{p}})\mp{2}\alpha_3\nu_\perp(\vec{p}_1)
[\vec{p}_1\times\tilde{\vec{p}}]_z
+\frac{v}{2p_1^3}[\vec{p}_1\times\tilde{\vec{p}}]_z^2+v\tilde{q},\\
\Re\xi_0&\approx& \frac{\omega_{in}}2+\alpha_0p_0^2
+\left[\frac{v}p_0+2\alpha_0+2\alpha_3\nu_\|(\vec{p}_0)\right]
(\vec{p}_0\cdot\tilde{\vec{p}})
+2\alpha_3\nu_\perp(\vec{p}_0)[\vec{p}_0\times\tilde{\vec{p}}]_z
+\frac{v}{2p_0^3}[\vec{p}_0\times\tilde{\vec{p}}]_z^2,\\
\Re\xi_1&\approx& \frac{\omega_{out}}2+\alpha_0p_1^2
+\left[\frac{v}p_1+2\alpha_0\mp{2}\alpha_3\nu_\|(\vec{p}_1)\right]
(\vec{p}_1\cdot\tilde{\vec{p}})
\mp{2}\alpha_3\nu_\perp(\vec{p}_1)[\vec{p}_1\times\tilde{\vec{p}}]_z
+\frac{v}{2p_1^3}[\vec{p}_1\times\tilde{\vec{p}}]_z^2+v\tilde{q}.
\end{eqnarray}\label{warpedexpand=}\end{subequations}
As before, the upper (lower) sign corresponds to emission of
scalar~$E_1'$ (pseudovector~$E_2$) phonons, accompanied by
intervalley (intravalley) electron scattering. Then instead of
Eq.~(\ref{backscattering=}) we have
\begin{eqnarray}
&&\int\limits_0^\infty{d}t_0\,dt_1\,d\bar{t}_0\,d\bar{t}_1\,
\delta(t_0+t_1-\bar{t}_0-\bar{t}_1)\,
e^{i\alpha_0(p_0^2-p_1^2)(t_1-\bar{t}_1)-\gamma_{\vec{p}_0}(t_0+\bar{t}_0)
-\gamma_{\vec{p}_1}(t_1+\bar{t}_1)-iv\tilde{q}(t_1+\bar{t}_1)}\times\nonumber\\
&&{}\times\int\frac{dp_\|}{2\pi}\,
{e}^{i\tilde{p}_\|{v}(t_0+\bar{t}_0-t_1-\bar{t}_1)
+2i\tilde{p}_\|\alpha_0[p_0(t_0-\bar{t}_0)-p_1(t_1-\bar{t}_1)]
-2i\tilde{p}_\|\alpha_3\nu_\|(\vec{q})[p_0(t_0+\bar{t}_0)\mp{p}_1(t_1+\bar{t}_1)]}
\times\nonumber\\ &&{}\times \int\frac{dp_\perp}{2\pi}\,
e^{-2i\tilde{p}_\perp\alpha_3\nu_\perp(\vec{q})
[p_0(t_0+\bar{t}_0)\mp{p}_1(t_1+\bar{t}_1)]
-i(v/2)\tilde{p}_\perp^2[(t_0+\bar{t}_0)/p_0+(t_1+\bar{t}_1)/p_1]}.
\end{eqnarray}
Let us denote 
$\gamma_{\vec{p}_0}+\gamma_{\vec{p}_1}=2\gamma$,
and introduce
$T=t_0+\bar{t}_0=t_1+\bar{t}_1$ (the latter equality follows from
the longitudinal spatial $\delta$-function in the leading order),
$\tau=t_0-\bar{t}_0=-t_1+\bar{t}_1$, and rewrite the integral as
\begin{eqnarray}
&&\int\limits_{-\infty}^{\infty}d\tau
\int\limits_{|\tau|}^\infty\frac{dT}{4v}\,
e^{i\alpha_0(p_0^2-p_1^2)\tau-(2\gamma+iv\tilde{q})T}
\int\limits_{-\infty}^\infty\frac{d\tilde{p}_\perp}{2\pi}\,
e^{-2i\tilde{p}_\perp\alpha_3\nu_\perp(\vec{q})(p_0\mp{p}_1){T}
-i(v/p_0+v/p_1)\tilde{p}_\perp^2T/2}
=\nonumber\\
&&=\frac{1}{8iv^2}
\sqrt{\frac{\omega_{in}\omega_{out}}{\omega_{in}+\omega_{out}}}\,
\frac{1}{\Delta_{eh}}
\left(\frac{1}{\sqrt{\Delta_\perp+\Delta_{eh}+2i\gamma-{v}\tilde{q}}}
-\frac{1}{\sqrt{\Delta_\perp-\Delta_{eh}+2i\gamma-{v}\tilde{q}}}\right),
\label{warpedbackscattering=}
\end{eqnarray}
\end{widetext}
where we also denoted
\begin{subequations}
\begin{eqnarray}
&&\Delta_\perp=
2[\alpha_3\nu_\perp(\vec{q})(p_0\mp{p}_1)]^2\frac{p_0p_1}{v(p_0+p_1)},\\
&&\Delta_{eh}=\alpha_0(p_0^2-p_1^2).\label{Deltaeh=}
\end{eqnarray}
\end{subequations}
Squaring the matrix element, and performing the final integration, we
obtain Eq.~(\ref{I2Kend=}) with the replacement
\begin{equation}\label{I2Ktrig=}
\frac{1}{\gamma^2}\to\frac{4}{\Delta_{eh}^2}
\ln\frac{\gamma^2+\Delta_{eh}^2/4}{\gamma^2}.
\end{equation}
The meaning of this replacement is that when the electron-hole
asymmetry becomes greater than the level broadening, it starts to
play the main role in restricting the energy denominators from
below. Numerically,
$\alpha_0(1\;\mbox{eV})^2/v^2\sim{0}.1\;\mbox{eV}$ (see, e.~g.,
Ref.~\onlinecite{DresselhausBook}), so the relative correction to
Eq.~(\ref{I2Kend=}) for small $\Delta_{eh}$ can be estimated as
$-(1/2)\Delta_{eh}^2/(2\gamma)]^2/2%
\sim-{10}^{-4}(\omega_{in}/2\gamma)^2$. The total electronic
broadening $2\gamma$ was measured by time-resolved photoemission
spectroscopy to be 20~meV in Ref.~\onlinecite{Gao} and $25$~meV in
Ref.~\onlinecite{Moos} (all values taken for
$\ep=\omega_{in}/2=1$~eV). A recent ARPES measurement gives a
significantly larger value for $2\gamma\sim{100}\:\mbox{meV}$
(Ref.~\onlinecite{Rotenberg}), which agrees better with Eqs.
(\ref{SigmaE2=}),~(\ref{phononnumbers=}).

The effect of the trigonal warping term (the only one sensitive to the
difference between intervalley and intravalley scattering) turns out
to be small in the parameter $\alpha_3\omega_{in}/v^2$ [in
Eq.~(\ref{warpedbackscattering=}) warping enters
through~$\Delta_\perp$, which results only in a small shift of the
integration variable~$\tilde{q}$]. For $\omega_{in}=2$~eV,
$\alpha_3=-va/4$ (tight-binding model), we estimate the relative
contribution of the warping term as
$\Delta_\perp/\omega_{in}\approx{5}\cdot{10}^{-4},\;5\cdot{10}^{-2}$
for intervalley and intravalley scattering, respectively.
Thus, trigonal warping
cannot account for the observed ratio $I_{2K}/I_{2\Gamma}$.

\section{Four-phonon Raman scattering}\label{sec:4raman}

\subsection{Resonant manifold}\label{sec:resonant}

We will calculate only the intensity of the peak which we will call
$4K$, the double of the $2K$ peak at 2700~cm$^{-1}$.
The $4K$ peak corresponds to emission
of four scalar phonons ($A_1$~in terms of the $C_{3v}$~symmetry,
or~$E_1'$ in terms of the $C_{6v}''$ symmetry) from the vicinity of
the $\Kpnt$ and $\Kpnt'$ points. We will not consider other
four-phonon peaks, as their intensity is smaller.

\begin{figure}
\includegraphics[width=5cm]{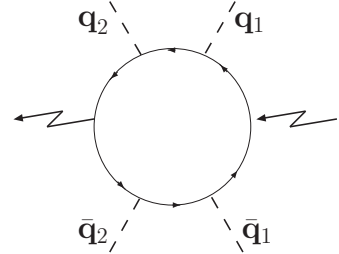}
\caption{\label{fig:Raman4ph} The only fully resonant diagram for
the four-phonon Raman amplitude in the leading order.}
\end{figure}

The integrated intensity of the $4K$~peak is given by
\begin{eqnarray}
4\pi\frac{dI_{4K}}{do_{out}}&=&
\frac{2\pi}{c^2}\,\frac{\omega_{out}^2}{2\pi^2c^2}\,
\frac{1}{(L_xL_y)^3}\,\frac{8}{4!}
\times\nonumber\\ &&{}\times
\sum_{\vec{q}_1+\vec{q}_2=\bar{\vec{q}}_1+\bar{\vec{q}}_2}
|2\mathcal{M}_{-\vec{q}_1,-\vec{q}_2,\bar{\vec{q}}_2,\bar{\vec{q}}_1}|^2.
\label{I4start=}
\end{eqnarray}
The factor $1/4!$ in front of the sum ensures the proper summation
over the final states (phonon permutations), the factor of~$8$ comes
from the sum over $A_1$ and $B_1$~modes (one can have all four
phonons~$A_1$, all four~$B_1$, or $4!/(2!\cdot{2}!)=6$ combinations
of two~$A_1$ and two~$B_1$ phonons, all other combinations yielding
traceless products of $\Lambda_x$~and~$\Lambda_y$), the factor
of~2 inside the square takes care of the spin degeneracy. The
only fully resonant diagram for the four-phonon matrix element is shown
in Fig.~\ref{fig:Raman4ph}. The matrix element is
given by the sum of $4!=24$ permutations
of the emitted phonon wave vectors
$-\vec{q}_1,-\vec{q}_2,\bar{\vec{q}}_2,\bar{\vec{q}}_1$
[we denote $\omega_K(\vec{q})=\omega_{\vec{q}}$ for the sake of
 compactness]:
\begin{widetext}
\begin{eqnarray}
&&\mathcal{M}_{-\vec{q}_1,-\vec{q}_2,\bar{\vec{q}}_2,\bar{\vec{q}}_1}=
\frac{2\pi{e}^2v^2}{\sqrt{\omega_{in}\omega_{out}}}\,
\frac{[L_xL_yF_K^2/(2NM)]^2}{\sqrt{\omega_{\vec{q}_1}\omega_{\vec{q}_2}
\omega_{\bar{\vec{q}}_1}\omega_{\bar{\vec{q}}_2}}}
\int\frac{d\ep}{2\pi}\frac{d^2\vec{p}}{(2\pi)^2} \times\nonumber\\
&&\qquad\times \Tr_{4\times{4}}\left\{(\vec{e}_{in}\cdot\Sigma)\,
G(\vec{p},\ep-\omega_{in}/2)\,\Sigma_z
G(\vec{p}+\bar{\vec{q}}_1,\ep-\omega_{in}/2+\omega_{\bar{\vec{q}}_1})\,
\Sigma_z G(\vec{p}+\bar{\vec{q}}_1+\bar{\vec{q}}_2,
\ep-\omega_{in}/2+\omega_{\bar{\vec{q}}_1}+\omega_{\bar{\vec{q}}_2})
\times\nonumber\right.\\
&&\qquad\qquad\left.\times(\vec{e}_{out}^*\cdot\Sigma)\,
G(\vec{p}+\vec{q}_1+\vec{q}_2,
\ep+\omega_{in}/2-\omega_{\vec{q}_1}-\omega_{\vec{q}_2})\,\Sigma_z
G(\vec{p}+\vec{q}_1,\ep+\omega_{in}/2-\omega_{\vec{q}_1})\,\Sigma_z
G(\vec{p},\ep+\omega_{in}/2)\right\}+\nonumber\\
&&{}+(\mbox{23 other permutations of $\vec{q}$'s}).
\end{eqnarray}
Again, we switch to the basis of Dirac eigenstates according to
Eqs.~(\ref{UGU=}),~(\ref{USigmaU=}), and evaluate the trace:
\begin{eqnarray}
\Tr_{4\times{4}}\left\{\ldots\right\}&=&2\,\frac{[\vec{e}_{\vec{p}}\times\vec{e}_{in}]_z
\sin(\varphi_{\vec{p}}/2-\varphi_{\vec{p}+\bar{\vec{q}}_1}/2)
\sin(\varphi_{\vec{p}+\bar{\vec{q}}_1}/2
-\varphi_{\vec{p}+\bar{\vec{q}}_1+\bar{\vec{q}}_2}/2)}
{(\ep-\omega_{in}/2+\bar\xi_{\vec{p}})
(\ep-\omega_{in}/2+\omega_{\bar{\vec{q}}_1}+\bar\xi_{\vec{p}+\bar{\vec{q}}_1})
(\ep-\omega_{in}/2+\omega_{\bar{\vec{q}}_1}+\omega_{\bar{\vec{q}}_2}
+\bar\xi_{\vec{p}+\bar{\vec{q}}_1+\bar{\vec{q}}_2})}\times\nonumber\\
&&{}\times\frac{[\vec{e}_{\vec{p}+\vec{q}_1+\vec{q}_2}\times\vec{e}_{out}^*]_z
\sin(\varphi_{\vec{p}+\vec{q}_1+\vec{q}_2}/2-\varphi_{\vec{p}+\vec{q}_1}/2)
\sin(\varphi_{\vec{p}+\vec{q}_1}/2-\varphi_{\vec{p}}/2)}
{(\ep+\omega_{in}/2-\omega_{\vec{q}_1}-\omega_{\vec{q}_2}
-\xi_{\vec{p}+\vec{q}_1+\vec{q}_2})
(\ep+\omega_{in}/2-\omega_{\vec{q}_1}-\xi_{\vec{p}+\vec{q}_1})
(\ep+\omega_{in}/2-\xi_{\vec{p}})}\,.
\label{4phonontrace=}
\end{eqnarray}
\end{widetext}
Just like for the two-phonon scattering, the dominant contribution
to the integrals will come from those momenta which make small all
the factors in the denominators of the above expression. The real
parts of the electron and hole dispersion,
$\Re\xi_{\vec{p}}$, $\Re\bar\xi_{\vec{p}}$, are given by
Eqs.~(\ref{Rexip=}),~(\ref{Rebarxip=}), and
$\Im\xi_{\vec{p}}=\Im\bar\xi_{\vec{p}}=-\gamma_{\vec{p}}$.

Four phonon wave vectors
$-\vec{q}_1,-\vec{q}_2,\bar{\vec{q}}_2,\bar{\vec{q}}_1$, such that
$\vec{q}_1+\vec{q}_2=\bar{\vec{q}}_2+\bar{\vec{q}}_1$, will be said to
satisfy resonance conditions if exists a vector $\vec{p}_0$ such that the
following equalities hold:
\begin{subequations}\begin{eqnarray}
&& \Re\xi_{\vec{p}_0}=\frac{\omega_{in}}2\,,\\
&&\Re\xi_{\vec{p}_0+\vec{q}_1}=\frac{\omega_{in}}2-\omega_{\vec{q}_1},\\
&&\Re\xi_{\vec{p}_0+\bar{\vec{q}}_1}=\frac{\omega_{in}}2-\omega_{\bar{\vec{q}}_1},\\
&&\Re\xi_{\vec{p}_0+{\vec{q}}_1+{\vec{q}}_2}
=\frac{\omega_{in}}2-\omega_{{\vec{q}}_1}-\omega_{{\vec{q}}_2},
\label{resonance3=}\\
&&\Re\xi_{\vec{p}_0+\bar{\vec{q}}_1+\bar{\vec{q}}_2}
=\frac{\omega_{in}}2-\omega_{\bar{\vec{q}}_1}-\omega_{\bar{\vec{q}}_2}.
\label{resonance4=}
\end{eqnarray}\label{resonance=}\end{subequations}
The structure of the manifold defined by these conditions essentially
depends on (i)~whether we take into account the electron-hole
asymmetry, or just set
$\Re\xi_{\vec{p}}=\Re\bar\xi_{\vec{p}}=v|\vec{p}|$, and
(ii)~whether we take into account the phonon dispersion, or just
set $\omega_{\vec{q}}=\omega_0$. Indeed, subtracting
Eq.~(\ref{resonance4=}) from Eq.~(\ref{resonance3=}),
we obtain
\begin{equation}
\omega_{{\vec{q}}_1}+\omega_{{\vec{q}}_2}
+\Re\xi_{\vec{p}_0+\vec{q}_1+\vec{q}_2}
=\omega_{\bar{\vec{q}}_1}+\omega_{\bar{\vec{q}}_2}
+\Re\bar\xi_{\vec{p}_0+{\vec{q}}_1+{\vec{q}}_2},
\end{equation}
which in the phonon-dispersionless case represents either
an identity if $\xi_{\vec{p}}=\bar\xi_{\vec{p}}$, or can never
be satisfied if $\xi_{\vec{p}}\neq\bar\xi_{\vec{p}}$, thus
leading to a finite energy mismatch; only in the phonon-dispersive
case it represents a non-trivial equation.
As will be seen in Sec.~\ref{sec:dispersive}, the quantitative
condition for the phonon dispersion to matter is that the phonon
group velocity $v_{ph}>v\max\{\gamma,\Delta_{eh}\}/\omega_{in}$,
where $\gamma$~is the typical value of the electron broadening,
and $\Delta_{eh}$~is the typical value of the electron-hole asymmetry,
defined analogously to Eq.~(\ref{Deltaeh=}).

\begin{figure}
\includegraphics[width=6cm]{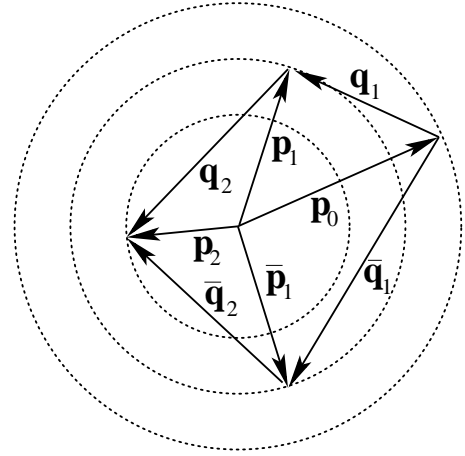}
\caption{\label{fig:resonance} Structure of the resonant manifold
of electron and phonon momenta for four-phonon Raman scattering
in the approximation
$\Re\xi_{\vec{p}}=\Re\bar\xi_{\vec{p}}=v|\vec{p}|$,
$\omega_{\vec{q}}=\omega_0$.}
\end{figure}

First, let us focus on the electron-hole symmetric and dispersionless
case, $\Re\xi_{\vec{p}}=\Re\bar\xi_{\vec{p}}=v|\vec{p}|$,
$\omega_{\vec{q}}=\omega_0$, when the structure of the resonant
manifold~(\ref{resonance=}) is the simplest.
The resonant manifold can be parametrized by
four polar angles $\varphi_0,\varphi_1,\bar\varphi_1,\varphi_2$, which
determine the positions of the four momenta $\vec{p}_0$,
$\vec{p}_1\equiv\vec{p}_0+\vec{q}_1$,
$\bar{\vec{p}}_1\equiv\vec{p}_0+\bar{\vec{q}}_1$, and
$\vec{p}_2\equiv\vec{p}_0+\vec{q}_1+\vec{q}_2=%
\vec{p}_0+\bar{\vec{q}}_1+\bar{\vec{q}}_2$ on the three circles
$v|\vec{p}_0|=\omega_{in}/2$,
$v|\vec{p}_1|=v|\bar{\vec{p}}_1|=\omega_{in}/2-\omega_0$,
$v|\vec{p}_2|=\omega_{in}/2-2\omega_0$ (Fig.~\ref{fig:resonance}).
Since the total number of independent phonon variables is~6, there
are two more variables besides the four angles, which correspond to
deviations of the phonon momenta from the resonant manifold. There
is a freedom of choice for them, and we choose them to correspond to
stretching of the vectors $\vec{p}_1,\bar{\vec{p}}_1$ (denoted by
$\deviation,\bar\deviation$, respectively):
\begin{subequations}\begin{eqnarray}
\delta\vec{q}_1&=&-[\vec{e}_z\times\vec{p}_0]\delta\varphi_0
+[\vec{e}_z\times\vec{p}_1]\delta\varphi_1+\vec{p}_1\delta\deviation,\\
\delta\vec{q}_2&=&[\vec{e}_z\times\vec{p}_2]\delta\varphi_2
-[\vec{e}_z\times\vec{p}_1]\delta\varphi_1-\vec{p}_1\delta\deviation,\\
\delta\bar{\vec{q}}_1&=&-[\vec{e}_z\times\vec{p}_0]\delta\varphi_0
+[\vec{e}_z\times\bar{\vec{p}}_1]\delta\bar\varphi_1
+\bar{\vec{p}}_1\delta\bar\deviation,\quad\\
\delta\bar{\vec{q}}_2&=&[\vec{e}_z\times\vec{p}_2]\delta\varphi_2
-[\vec{e}_z\times\bar{\vec{p}}_1]\delta\bar\varphi_1
-\bar{\vec{p}}_1\delta\bar\deviation,
\end{eqnarray}\end{subequations}
The $\vec{q}$~integration measure is transformed as
\begin{eqnarray}\nonumber
d^2\vec{q}_1\,d^2\vec{q}_2\,d^2\bar{\vec{q}}_1&=&
|\vec{p}_1|^2|\bar{\vec{p}}_1|^2
\left|[\vec{p}_0\times\vec{p}_2]_z\right|\times\\
&&{}\times{d}\varphi_0\,d\varphi_2\,d\varphi_1\,d\bar\varphi_1\,
d\deviation\,d\bar\deviation.
\end{eqnarray}
Thus, 6~independent integration variables, parametrizing the
final state, have been separated into two groups: four angles,
determining the position of the phonon momenta on the
resonant manifold, and two deviations $\deviation,\bar\deviation$
from the manifold. The dependence of the integrand on the
first group of variables is smooth, and we will call them slow
variables. The deviations from the manifold, on the contrary,
suppress the integrand dramatically, so we will call them
fast variables.

Let us now consider other 23 permutations of the phonon wave
vectors $-\vec{q}_1,-\vec{q}_2,\bar{\vec{q}}_2,\bar{\vec{q}}_1$,
which have to be added in order to obtain the correct value of the
matrix element~$\mathcal{M}$. We need to check whether
contributions to~$\mathcal{M}$ from different permutations can be
large at the same time (i.~e., for the same values of the
integration variables
$-\vec{q}_1,-\vec{q}_2,\bar{\vec{q}}_2,\bar{\vec{q}}_1$). If this
is not the case for all 23 permutations, then the summation over
all permutations in~$\mathcal{M}$ would just cancel the factor
$1/4!$ in Eq.~(\ref{I4start=}).

First, we note that if three momenta
$\vec{p}_0,\vec{p}_0+\vec{q}_1,\vec{p}_0+\vec{q}_1+\vec{q}_2$
satisfy the resonance conditions~(\ref{resonance=}) (i.~e., they
lie on the corresponding circles) for some~$\vec{p}_0$, then the
momenta
$\vec{p}_0',\vec{p}_0'+\vec{q}_2,\vec{p}_0'+\vec{q}_1+\vec{q}_2$
do not satisfy these conditions for any~$\vec{p}_0'$. Indeed, the
vector $\vec{q}_1+\vec{q}_2$ can be placed on the circles only in
two ways, which leaves only two choices for the
momentum~$\vec{p}_0'$: $\vec{p}_0'=\vec{p}_0$, or the symmetric
one. For none of them does $\vec{p}_0'+\vec{q}_2$ lie on the
circle, unless the configuration of momenta has special
symmetries. This means that the corresponding contribution to the
matrix element is small, so the permutation
$\vec{q}_1\leftrightarrow\vec{q}_2$ should be discarded. By
analogous argument we can discard all 14 permutations which leave
at least one of the $\vec{q}$'s in place (this immediately leaves
only two choices for~$\vec{p}_0'$ which can be inspected). 4
cyclic permutations are eliminated because the circles have
different radii. The permutation
$\vec{q}_1\leftrightarrow\vec{q}_2,%
\bar{\vec{q}}_1\leftrightarrow\bar{\vec{q}}_2$
is eliminated by the very first argument. Thus, we are left with four
permutations. One is
$-\vec{q}_1,-\vec{q}_2,\bar{\vec{q}}_2,\bar{\vec{q}}_1\to%
\bar{\vec{q}}_1,\bar{\vec{q}}_2,-\vec{q}_2,-\vec{q}_1$, and it
{\em does} satisfy the resonance conditions~(\ref{resonance=}) if
one chooses $\vec{p}_0'=-\vec{p}_0$. The other three are obtained
from it by swapping the first two momenta, the last two, or both,
and thus do not satisfy the resonance conditions. The permutation
$-\vec{q}_1,-\vec{q}_2,\bar{\vec{q}}_2,\bar{\vec{q}}_1\to%
\bar{\vec{q}}_1,\bar{\vec{q}}_2,-\vec{q}_2,-\vec{q}_1$ is nothing
else but the result of reversal of the electronic line direction
in the loop on Fig.~\ref{fig:Raman4ph}. As seen from
Eq.~(\ref{4phonontrace=}), this permutation gives exactly the same
contribution as the original one. Thus, their interference results
in an additional factor of~2, besides cancellation of $1/4!$ in
Eq.~(\ref{I4start=}).

\subsection{Quasiclassical representation}\label{sec:qclrep}

Now let us pass to time representation for the matrix
element~$\mathcal{M}$ by rewriting each factor in the
denominator of expression~(\ref{4phonontrace=}) analogously to
Eqs.~(\ref{timerep=}). We introduce three time variables for the
electron and three for the hole, denoted by $t_0,t_1,t_2$ and
$\bar{t}_0,\bar{t}_1,\bar{t}_2$, respectively. Next, we introduce the
deviation $\tilde{\vec{p}}=\vec{p}-\vec{p}_0$ from the
point~$\vec{p}_0$ fixed by the resonance conditions, and expand each
factor in the denominator of expression~(\ref{4phonontrace=}) as
\begin{widetext}
\begin{subequations}\begin{eqnarray}
-\frac{\omega_{in}}2+\bar\xi_{\vec{p}}&\approx&
\bar{\vec{v}}_0\tilde{\vec{p}}-i\bar\gamma_0
-\alpha_0p_0^2,\\
-\frac{\omega_{in}}2+\omega_{\bar{\vec{q}}_1}
+\bar\xi_{\vec{p}+\bar{\vec{q}}_1}
&\approx&\bar{\vec{v}}_1\tilde{\vec{p}}-i\bar\gamma_1
+\bar\deviation{v}\bar{p}_1+\delta\omega_{\bar{\vec{q}}_1}
-\alpha_0p_1^2,\\
-\frac{\omega_{in}}2+\omega_{\bar{\vec{q}}_1}+\omega_{\bar{\vec{q}}_2}
+\bar\xi_{\vec{p}+\bar{\vec{q}}_1+\bar{\vec{q}}_2}
&\approx&\bar{\vec{v}}_2\tilde{\vec{p}}-i\bar\gamma_2
+\delta\omega_{\bar{\vec{q}}_1}+\delta\omega_{\bar{\vec{q}}_2}
-\alpha_0p_2^2,\\
\frac{\omega_{in}}2-\xi_{\vec{p}}&\approx&
-\vec{v}_0\tilde{\vec{p}}+i\gamma_0-\alpha_0p_0^2,\qquad\\
\frac{\omega_{in}}2-\omega_{\vec{q}_1}-\xi_{\vec{p}+\vec{q}_1}
&\approx&-\vec{v}_1\tilde{\vec{p}}+i\gamma_1-\deviation{v}p_1
-\delta\omega_{\vec{q}_1}-\alpha_0p_1^2,\\
\frac{\omega_{in}}2-\omega_{\vec{q}_1}-\omega_{\vec{q}_2}
-\xi_{\vec{p}+\vec{q}_1+\vec{q}_2}
&\approx&-\vec{v}_2\tilde{\vec{p}}+i\gamma_2
-\delta\omega_{\vec{q}_1}-\delta\omega_{\vec{q}_2}
-\alpha_0p_2^2,
\end{eqnarray}\label{deviations=}\end{subequations}
where $\vec{v}_i,-\bar{\vec{v}}_i$ and $\gamma_i,\bar\gamma_i$ are the
velocities and the damping rates of the electron and the hole,
respectively, in the $i$th intermediate state.
In Eqs.~(\ref{deviations=}) we have also taken into
account the electron-hole asymmetry and the phonon dispersion
$\delta\omega_{\vec{q}}=\omega_{\vec{q}}-\omega_0$, both assumed to be
weak: $\alpha_0p_{0,1,2}^2\ll\omega_0$,
$|\delta\omega_{\vec{q}}|\ll\omega_0$.
Within this approximation we can take
$\vec{v}_0=\bar{\vec{v}}_0$ and $\vec{v}_2=\bar{\vec{v}_2}$.

The numerator of Eq.~(\ref{4phonontrace=}) is a
smooth function of momenta on the scale $\gamma/v$, so it can be taken
at $\tilde{\vec{p}}=0$, i.~e. on the resonant manifold. Then the
integration over $d^2\tilde{\vec{p}}/(2\pi)^2$ gives a spatial
$\delta$-function:
\begin{eqnarray}\nonumber
&&\int\frac{d^2\tilde{\vec{p}}}{(2\pi)^2}\,
e^{-i\tilde{\vec{p}}(\bar{\vec{v}}_0\bar{t}_0+\bar{\vec{v}}_1\bar{t}_1
+\bar{\vec{v}}_2\bar{t}_2+\vec{v}_0t_0+\vec{v}_1t_1+\vec{v}_2t_2)}=
\delta(\bar{\vec{v}}_0\bar{t}_0+\bar{\vec{v}}_1\bar{t}_1
+\bar{\vec{v}}_2\bar{t}_2+\vec{v}_0t_0+\vec{v}_1t_1+\vec{v}_2t_2),\quad
\end{eqnarray}
so in order to recombine, the electron and the hole should meet at the
same spatial point. As a result, the denominator of
Eq.~(\ref{4phonontrace=}), integrated over $d\ep/(2\pi)$ and
$d^2\tilde{\vec{p}}/(2\pi)^2$, is rewritten as
\begin{subequations}\begin{eqnarray}
&&\int_0^\infty
{e}^{-\eta}\,\delta(t_0+t_1+t_2-\bar{t}_0-\bar{t}_1-\bar{t}_2)\,
\delta(\bar{\vec{v}}_0\bar{t}_0+\bar{\vec{v}}_1\bar{t}_1
+\bar{\vec{v}}_2\bar{t}_2+\vec{v}_0t_0+\vec{v}_1t_1+\vec{v}_2t_2)
\,{d}t_0\,dt_1\,dt_2\,d\bar{t}_0\,d\bar{t}_1\,d\bar{t}_2,\quad\label{constrainedint=}\\
&&\eta=i\left[\deviation{v}p_1t_1+\bar\deviation{v}\bar{p}_1\bar{t}_1
+\delta\omega_{\vec{q}_1}(t_1+t_2)+\delta\omega_{\vec{q}_2}t_2
+\delta\omega_{\bar{\vec{q}}_1}(\bar{t}_1+\bar{t}_2)
+\delta\omega_{\bar{\vec{q}}_2}\bar{t}_2\right]+i\alpha_0\sum_{i=0}^2p_i^2(t_i-\bar{t}_i)
+\nonumber\\
&&\quad{}+
\sum_{i=0}^2(\gamma_it_i+\bar\gamma_i\bar{t}_i).\label{eta=}
\end{eqnarray}\end{subequations}
\end{widetext}

\subsection{Integration over deviations}\label{sec:Intdev}

The part of the summation over final phonon states in
Eq.~(\ref{I4start=}), corresponding to the integration over the
deviations~$\deviation,\bar\deviation$, can be performed explicitly.
Let us open the $\delta$-functions in Eq.~(\ref{constrainedint=})
choosing $t_0,t_1,\bar{t}_1$ as independent variables:
\begin{subequations}
\begin{eqnarray}
&&t_0+\bar{t}_0=\frac{[\vec{v}_2\times
(\vec{v}_1t_1+\bar{\vec{v}}_1\bar{t}_1)]_z}
{[\vec{v}_0\times\vec{v}_2]_z}\,,\\
&&t_2+\bar{t}_2=\frac{[\vec{v}_0\times
(\vec{v}_1t_1+\bar{\vec{v}}_1\bar{t}_1)]_z}
{[\vec{v}_2\times\vec{v}_0]_z}\,,\\
&&t_2,\bar{t}_2=\mp\left(t_0+\frac{t_1-\bar{t}_1}2\right)+\nonumber\\
&&\qquad{}+\frac{[(\vec{v}_0\mp\vec{v}_2)\times
(\vec{v}_1t_1+\bar{\vec{v}}_1\bar{t}_1)]_z}
{2[\vec{v}_2\times\vec{v}_0]_z},
\end{eqnarray}
The Jacobian of this transformation is given by
\begin{eqnarray}
&&\int{d}\bar{t}_0\,dt_2\,d\bar{t}_2\,
\delta(t_0+t_1+t_2-\bar{t}_0-\bar{t}_1-\bar{t}_2)\times\nonumber\\
&&{}\times
\delta(\vec{v}_0(t_0+\bar{t}_0)+\vec{v}_2(t_2+\bar{t}_2)
+\vec{v}_1t_1+\bar{\vec{v}}_1\bar{t}_1)=\nonumber\\
&&=\frac{1}{2|[\vec{v}_0\times\vec{v}_2]_z|}\,.
\end{eqnarray}
\end{subequations}
The detuning phase from Eq.~(\ref{eta=}) can be written as
\begin{subequations}
\begin{eqnarray}
&&\Im\eta=\deviation{v}p_1t_1+\bar\deviation{v}\bar{p}_1\bar{t}_1
+\Delta{t}_0,\\
&&\Delta\equiv
\delta\omega_{\bar{\vec{q}}_1}+\delta\omega_{\bar{\vec{q}}_2}
-\delta\omega_{\vec{q}_1}-\delta\omega_{\vec{q}_2}
+2\Delta_{eh},\label{Delta=}
\end{eqnarray}
\end{subequations}
where all detunings at $t_1,\bar{t}_1$ have been
absorbed into a shift of~$\deviation,\bar\deviation$.
The energy mismatch due to electron-hole asymmetry,
$\Delta_{eh}=\alpha_0(p_{in}^2-p_{out}^2)$, is analogous
to that defined in Eq.~(\ref{Deltaeh=}). The damping factor is
(we use the fact that $\gamma_0=\bar\gamma_0$,
$\gamma_2=\bar\gamma_2$):
\begin{eqnarray}
&&\Re\eta=\gamma_0(t_0+\bar{t}_0)+\gamma_1t_1+\bar\gamma_1\bar{t}_1
+\gamma_2(t_2+\bar{t}_2)=\nonumber\\
&&=\left(\frac{[\vec{v}_2\times\vec{v}_1]_z}
{[\vec{v}_0\times\vec{v}_2]_z}\,\gamma_0+\gamma_1
+\frac{[\vec{v}_0\times\vec{v}_1]_z}
{[\vec{v}_2\times\vec{v}_0]_z}\,\gamma_2\right)t_1+\nonumber\\
&&{}+\left(\frac{[\vec{v}_2\times\bar{\vec{v}}_1]_z}
{[\vec{v}_0\times\vec{v}_2]_z}\,\gamma_0+\bar\gamma_1
+\frac{[\vec{v}_0\times\bar{\vec{v}}_1]_z}
{[\vec{v}_2\times\vec{v}_0]_z}\,\gamma_2\right)\bar{t}_1
\equiv\nonumber\\
&&\equiv\gamma_x{t}_1+\gamma_y\bar{t}_1.
\label{gammaxy=}
\end{eqnarray}
The two times $t_1,\bar{t}_1$can be taken to vary independently
from 0~to~$\infty$.
The integration domain for~$t_0$, which we denote by~$\mathcal{O}$,
besides the condition $t_0>0$, is determined by the inequalities
\begin{subequations}\begin{eqnarray}
&&t_0<\frac{[\vec{v}_2\times
(\vec{v}_1t_1+\bar{\vec{v}}_1\bar{t}_1)]_z}
{[\vec{v}_0\times\vec{v}_2]_z},\label{t0<1}\\
&&t_0<\frac{\bar{t}_1-t_1}2
+\frac{[(\vec{v}_2-\vec{v}_0)\times
(\vec{v}_1t_1+\bar{\vec{v}}_1\bar{t}_1)]_z}
{2[\vec{v}_0\times\vec{v}_2]_z},\label{t0<2}\\
&&t_0>\frac{\bar{t}_1-t_1}2
+\frac{[(\vec{v}_2+\vec{v}_0)\times
(\vec{v}_1t_1+\bar{\vec{v}}_1\bar{t}_1)]_z}
{2[\vec{v}_0\times\vec{v}_2]_z}.\label{t0>}
\end{eqnarray}\label{t0><=}\end{subequations}

Consider integral~(\ref{constrainedint=}). Squaring
its modulus, and integrating over $\deviation,\bar\deviation$,
we obtain
\begin{eqnarray}
\mathcal{I}&\equiv&\int\left|\mbox{Eq.~(\ref{constrainedint=})}\right|^2
d\deviation\,d\bar\deviation=\nonumber\\
&=&\frac{\pi^2}{v^2p_1\bar{p}_1[\vec{v}_0\times\vec{v}_2]_z^2}
\int\limits_0^\infty{d}t_1\,d\bar{t}_1\,
e^{-2\gamma_xt_1-2\gamma_y\bar{t}_1}
\times\nonumber\\&&{}\times
\int\limits_{\mathcal{O}}dt_0\,dt_0'\,e^{i\Delta(t_0-t_0')}.
\end{eqnarray}
We pass to the polar coordinates in the $(t_1,\bar{t}_1)$ plane:
\begin{equation}
t_1=t\cos\phi,\quad\bar{t}_1=t\sin\phi,
\end{equation}
and parametrize the region~$\mathcal{O}$ as
\begin{equation}
\mathcal{O}=\left\{t_0:t\zeta_{min}(\phi)<t_0<t\zeta_{max}(\phi)\right\},
\end{equation}
where $\zeta_{min}(\phi)$ and $\zeta_{max}(\phi)$ are piecewise
functions of the form $\alpha\cos\phi+\beta\sin\phi$, corresponding
to various conditions~(\ref{t0><=}).
Performing the integration, we obtain
\begin{subequations}\begin{eqnarray}
&&\mathcal{I}=\frac{\pi^2}{v^2p_1\bar{p}_1[\vec{v}_0\times\vec{v}_2]_z^2}\,
\frac{2}{\Delta^4}\int\limits_0^{\pi/2}d\phi
\times\nonumber\\ &&{}\times
\mathcal{F}\!\left(\zeta_{max}(\phi)-\zeta_{min}(\phi),
\frac{2\gamma_x\cos\phi+2\gamma_y\sin\phi}\Delta\right),\\
&&\mathcal{F}(x,y)\equiv\int\limits_0^\infty
(1-\cos{xt}){e}^{-yt}t\,dt=
\frac{x^2(x^2+3y^2)}{y^2(x^2+y^2)^2}.
\end{eqnarray}\end{subequations}

\begin{widetext}
\subsection{Angular integration and polarization dependence}

Let us collect all the factors in the expression for~$I_{2D*}$
[we denote $\vec{e}_{0,2}=(\cos\varphi_{0,2},\sin\varphi_{0,2})$]:
\begin{eqnarray}
&&4\pi\frac{dI_{4K}}{do_{out}}=
2\pi\left(\frac{e^2}c\right)^2\frac{v^2}{c^2}
\left(\frac{\lambda_K}{2\pi}\right)^4
\frac{\omega_{out}^2(\omega_{in}+\omega_{out})^2}4
\int\frac{{d}\varphi_0\,d\varphi_2\,d\varphi_1\,d\bar\varphi_1}{|\sin(\varphi_0-\varphi_2)|}\,
\left|[\vec{e}_0\times\vec{e}_{in}]_z\right|^2
\left|[\vec{e}_2\times\vec{e}_{out}]_z\right|^2
\times\nonumber\\  
&&{}\times\sin^2\frac{\varphi_0-\varphi_1}2\sin^2\frac{\varphi_1-\varphi_2}2
\sin^2\frac{\varphi_0-\bar\varphi_1}2
\sin^2\frac{\bar\varphi_1-\varphi_2}2
\int\limits_0^{\pi/2}\frac{d\phi}{\Delta^4}\,
\mathcal{F}\!\left(\zeta_{max}(\phi)-\zeta_{min}(\phi),
\frac{2\gamma_x\cos\phi+2\gamma_y\sin\phi}\Delta\right),\label{I4Dfinal=}
\end{eqnarray}
A system of two linear equations
\begin{equation}
\vec{v}_0t_{in}+\vec{v}_1t_1+\vec{v}_2t_{out}
+\bar{\vec{v}}_1\bar{t}_1=0
\end{equation}
has solutions with $t_{in},t_{out},t_1,\bar{t}_1>0$ if and only if the
four vectors
$\vec{v}_0,\vec{v}_1,\vec{v}_2,\bar{\vec{v}}_1$
do not lie in one half-plane. Having found
$t_{in},t_{out},t_1,\bar{t}_1$, we
can always split $t_{in}=t_0+\bar{t}_0$, $t_{out}=t_2+\bar{t}_2$ in
such a way that $t_0+t_1+t_2=\bar{t}_0+\bar{t}_2+\bar{t}_2$ (since
any side of a quadrangle is shorter than the sum of other three
sides). This condition on the vectors
$\vec{v}_0,\vec{v}_1,\vec{v}_2,\bar{\vec{v}}_1$
translates into the following condition on the angles (since all
conditions are on $\varphi_2-\varphi_0$, $\varphi_1-\varphi_0$,
$\bar\varphi_1-\varphi_0$, we set $\varphi_0=0$ for brevity):
\begin{eqnarray}
&&0\leq\varphi_2<\pi:\nonumber\\
&&\begin{array}{rcccc}
\varphi_1: & [0;\varphi_2] & [\varphi_2;\pi) &
(\pi;\varphi_2+\pi) & (\varphi_2+\pi;{2}\pi] \\
\bar\varphi_1: & (\pi;\varphi_2+\pi) & (\pi;\varphi_1+\pi) &
(0;2\pi) & (\varphi_1-\pi;\varphi_2+\pi)
\end{array}\nonumber\\
&&-\pi<\varphi_2\leq{0}:\nonumber\\
&&\begin{array}{rcccc}
\varphi_1: & [0;\varphi_2+\pi) & (\varphi_2+\pi;\pi] &
(\pi;\varphi_2+2\pi] & [\varphi_2+2\pi;2\pi] \\
\bar\varphi_1: & (\varphi_2+\pi;\varphi_1+\pi) & (0;2\pi) &
(\varphi_1-\pi;\pi) & (\varphi_2+\pi;\pi)
\end{array}\nonumber\\
&& \varphi_2=\pi:\nonumber\\
&&\begin{array}{rcc}
\varphi_1: & (0;\pi) & (\pi;2\pi) \\
\bar\varphi_1: & (\pi;2\pi) & (0;\pi)
\end{array}\label{angularrange=}
\end{eqnarray}
which define the domain of the angular integration. We also show
it schematically on Fig.~\ref{fig:AngRange}.

\begin{figure}
\includegraphics[width=12cm]{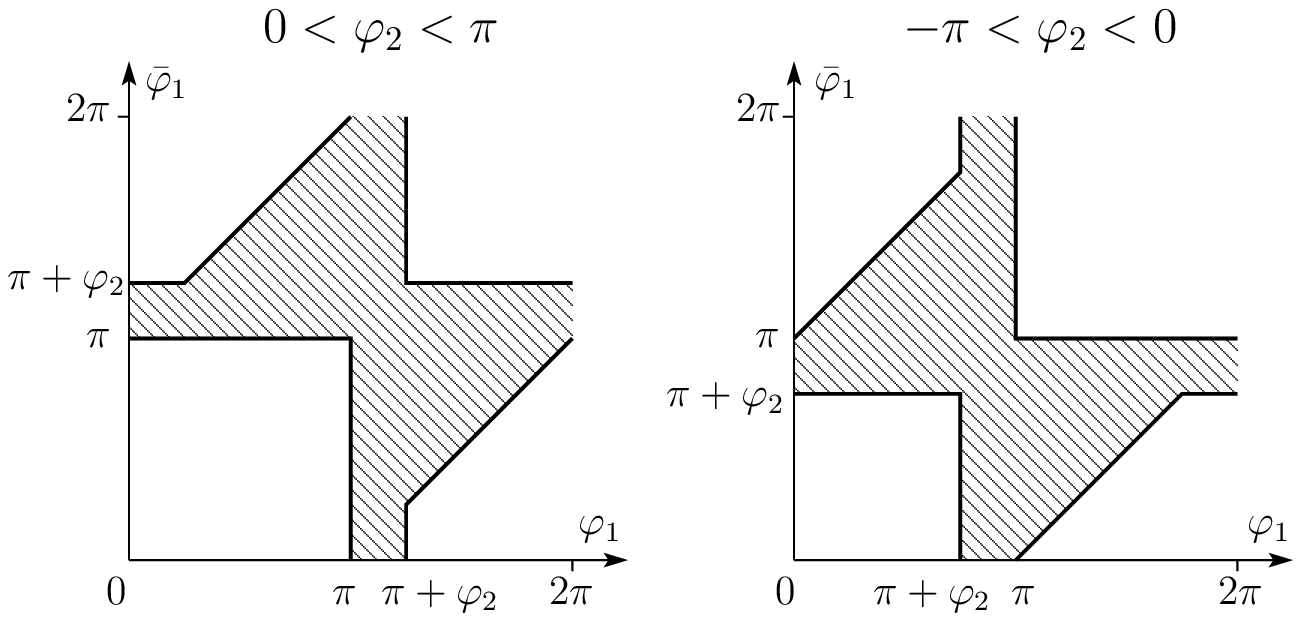}
\caption{\label{fig:AngRange} Region of integration over
$\varphi_1,\bar\varphi_1$, as determined by
 inequalities~(\ref{angularrange=}) for $0<\varphi_2<\pi$
and $\pi<\varphi_2<2\pi$.}
\end{figure}

Upon integration over $\varphi_1,\bar\varphi_1$ Eq.~(\ref{I4Dfinal=})
can be written in the form
\begin{equation}\label{Jphi=}
4\pi\frac{dI_{4K}}{do_{out}}=
\int\limits_0^{2\pi}d\varphi_0\,d\varphi_2\,
\mathcal{J}(\varphi_0-\varphi_2)\,
\left|[\vec{e}_0\times\vec{e}_{in}]_z\right|^2
\left|[\vec{e}_2\times\vec{e}_{out}]_z\right|^2.
\end{equation}
The function $\mathcal{J}(\varphi)$ is real and even,
$\mathcal{J}(\varphi)=\mathcal{J}(-\varphi)$,
as can be seen from Eq.~(\ref{4phonontrace=}),
so its Fourier series reads as
\begin{equation}
\mathcal{J}(\varphi)=\frac{\mathcal{J}_0}{2\pi}
+\sum_{n=1}^\infty\frac{\mathcal{J}_n}{\pi}\cos{n}\varphi,\quad
\mathcal{J}_n=\int\limits_0^{2\pi}d\varphi\,
\mathcal{J}(\varphi)\cos{n}\varphi.
\end{equation}
Then the angular integration gives
\begin{eqnarray}
4\pi\frac{I_{4K}}{do_{out}}&=&2\pi|\vec{e}_{in}|^2|\vec{e}_{out}|^2
\left\{\frac{\mathcal{J}_0}4+\frac{\mathcal{J}_2}8
[\cos^2(\varphi_{in}-\varphi_{out})-
\sin^2(\varphi_{in}-\varphi_{out})]\right\}=\nonumber\\
&=&2\pi\left(\frac{\mathcal{J}_0}4-\frac{\mathcal{J}_2}8\right)
|\vec{e}_{in}|^2|\vec{e}_{out}|^2
+2\pi\,\frac{\mathcal{J}_2}8
\left[|\vec{e}_{in}\cdot\vec{e}_{out}^*|^2+|\vec{e}_{in}\cdot\vec{e}_{out}|^2\right].
\label{I4Dpolarization=}
\end{eqnarray}
Comparing this with Eq.~(\ref{dIdout=}), we obtain
$I_{4K}^{\perp,\|}=2\pi(\mathcal{J}_0/4\mp\mathcal{J}_2/8)$.

\subsection{Dispersionless phonons}

For $\Delta=0$ (electron-hole symmetric case) the last integral in
Eq.~(\ref{I4Dfinal=}) can be written as
\begin{equation}
\int\limits_0^{\pi/2}{d}\phi\,\frac{3\left[\zeta_{max}(\phi)-\zeta_{min}(\phi)\right]^2}
{16\left[\gamma_x\cos\phi+\gamma_y\sin\phi\right]^4}.
\end{equation}
Numerical evaluation of the angular integral (for simplicity
we set all
$\gamma_0=\gamma_1=\bar\gamma_1=\gamma_2=\gamma$)
gives the following expression for $\mathcal{J}_0,\mathcal{J}_2$
to be substituted in Eq.~(\ref{I4Dpolarization=}):
\begin{equation}
\left.\begin{array}{c}\mathcal{J}_0 \\ \mathcal{J}_2\end{array}\right\}=
\left(\frac{e^2}c\right)^2\frac{v^2}{c^2}
\left(\frac{\lambda_K}{2\pi}\right)^4
\frac{\omega_{out}^2(\omega_{in}+\omega_{out})^2}4
\frac{3\pi}{8\gamma^4}
\left\{\begin{array}{r} 0.0440, \\ -0.0017. \end{array}\right.
\end{equation}
This results in $I_{4K}^{\perp}\approx{I}_{4K}^{\|}$,
so the polarization memory is almost completely lost. Adding
the contributions from two mutually perpendicular detection
polarizations [Eq.~(\ref{Inisotr=})], we obtain
\begin{equation}
I_{4K}=0.0440\left(\frac{e^2}c\right)^2\frac{v^2}{c^2}
\left(\frac{\lambda_K}{2\pi}\right)^4
\frac{\omega_{out}^2(\omega_{in}+\omega_{out})^2}4
\frac{\pi^2}{8\gamma^4}
\frac{4-3\cos\Theta_{det}-\cos^3\Theta_{det}}{4}\,
|\vec{e}_{in}|^2.\label{I4gammadispless=}
\end{equation}
For $|\vec{e}_{in}|=1$, $\Theta_{det}=\pi$,
$\omega_{in}\approx\omega_{out}$ this expression
corresponds\cite{mistake} to Eq.~(9) of Ref.~\onlinecite{myself}.

For $|\Delta_{eh}|\gg\gamma_x,\gamma_y$ (strong electron-hole
asymmetry)
the last integral in  Eq.~(\ref{I4Dfinal=}) can be written as
\begin{equation}
\int\limits_0^{\pi/2}\frac{d\phi}
{4\Delta^2\left[\gamma_x\cos\phi+\gamma_y\sin\phi\right]^2}.
\end{equation}
Numerical evaluation of the angular integral gives
\begin{equation}
\left.\begin{array}{c}\mathcal{J}_0 \\ \mathcal{J}_2\end{array}\right\}=
\left(\frac{e^2}c\right)^2\frac{v^2}{c^2}
\left(\frac{\lambda_K}{2\pi}\right)^4
\frac{\omega_{out}^2(\omega_{in}+\omega_{out})^2}4
\frac{\pi}{8\gamma^2\Delta_{eh}^2}
\left\{\begin{array}{r} 2.60, \\ 0.06. \end{array}\right.
\end{equation}
To obtain the expression for $I_{4K}$ one should replace
$0.0440\to{2}.60$, $\pi^2/(8\gamma^4)\to\pi^2/(24\gamma^2\Delta_{eh}^2)$
in Eq.~(\ref{I4gammadispless=}):
\begin{equation}
I_{4K}=2.60\left(\frac{e^2}c\right)^2\frac{v^2}{c^2}
\left(\frac{\lambda_K}{2\pi}\right)^4
\frac{\omega_{out}^2(\omega_{in}+\omega_{out})^2}4
\frac{\pi^2}{24\gamma^2\Delta_{eh}^2}
\frac{4-3\cos\Theta_{det}-\cos^3\Theta_{det}}{4}\,
|\vec{e}_{in}|^2.\label{I4Dehdispless=}
\end{equation}
\end{widetext}

\subsection{Dispersive phonons}\label{sec:dispersive}

In the case when $\Delta$ varies strongly compared to~$\gamma$,
but vanishes on a certain submanifold of the resonant manifold,
the appropriate way to approximate the last integral in
Eq.~(\ref{I4Dfinal=}) is
\begin{equation}
2\pi\delta(\Delta)\int\limits_0^{\pi/2}{d}\phi\,
\frac{\zeta_{max}(\phi)-\zeta_{min}(\phi)}
{8\left[\gamma_x\cos\phi+\gamma_y\sin\phi\right]^3}.
\label{intdphi=}
\end{equation}
This expression corresponds to complete suppression of interference
between trajectories of different shape by the phase mismatch
coming from the difference of the phonon frequencies. In this case
the electron-hole dynamics can be described by a kinetic equation,
analyzed in Appendix~\ref{app:kinur} (up to an overall interference
factor, see the discussion in the end of Appendix~\ref{app:kinur}).

First, let us focus on the apparent singularity at
$\sin|\varphi_2-\varphi_0|\to{0}$ in the angular integral in
Eq.~(\ref{I4Dfinal=}). For {\em generic} $\varphi_1,\bar\varphi_1$,
we see from Eq.~(\ref{gammaxy=}) that
$\gamma_{x,y}\propto{1}/\sin(\varphi_2-\varphi_0)$, and from
Eqs.~(\ref{t0><=}) -- that $\zeta_{max}(\phi)$ either stays
finite (at $\varphi_2\approx\varphi_0$), or also diverges as
${1}/\sin(\varphi_2-\varphi_0)$ (at $\varphi_2\approx\varphi_0+\pi$).
Thus, the power of $\sin(\varphi_2-\varphi_0)$ in the numerator
of expression~(\ref{intdphi=}) is sufficient to suppress the singularity.

The real danger in the angular integral comes from singularities of
the Jacobian corresponding to the resolution of $\delta(\Delta)$,
i.~e., values of $\varphi_1,\bar\varphi_1$ such that
\begin{equation}
\frac{\partial(\omega_{\vec{p}_1-\vec{p}_0}
+\omega_{\vec{p}_2-\vec{p}_1})}{\partial\varphi_1}=0,\quad
\frac{\partial(\omega_{\bar{\vec{p}}_1-\vec{p}_0}
+\omega_{\vec{p}_2-\bar{\vec{p}}_1})}{\partial\bar\varphi_1}=0.
\end{equation}
Looking at the integrals
\begin{subequations}
\begin{eqnarray}
&&\int\limits_{-1}^1{d}x\,dy\,\delta(x-y^2)=
\int\limits_{-1}^1\frac{dx}{2\sqrt{|x|}}=2,\\
&&\int\limits_{-1}^1{d}x\,dy\,\delta(x^2+y^2)
=\int\limits_{0}^{1}{2\pi}r\,dr\,\frac{\delta(r)}{2r}
=\frac\pi{2},\\
&&\int\limits_{-1}^1{d}x\,dy\,\delta(x^2-y^2)
=2\int\limits_{-1}^1\frac{dx}{2|x|}=
2\ln\frac{1}0,
\end{eqnarray}
\end{subequations}
we notice that the logarithmic divergence appears when both
$\varphi_1$~and~$\bar\varphi_1$ lie near one of these special
points. Note that both $\varphi_1$~and~$\bar\varphi_1$ should
lie near the same solution for the energy $\delta$-function itself
to be satisfied.

\begin{figure}
\includegraphics[width=8cm]{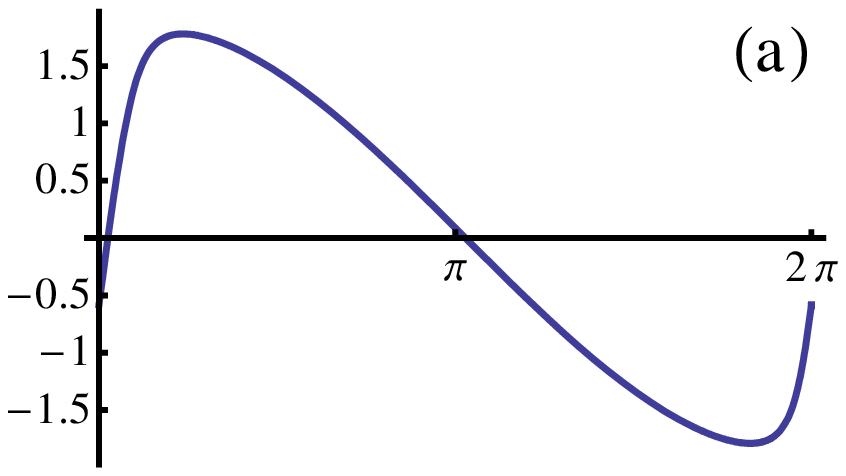}
\includegraphics[width=8cm]{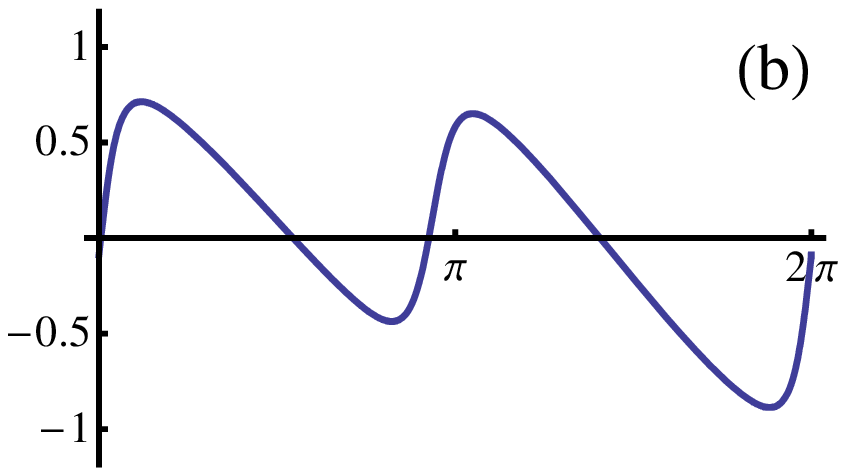}
\caption{\label{fig:singpoint} (color online).
Plot of the left-hand side of
Eq.~(\ref{zeroJacobian=}) as a function of~$\varphi_1$ for
$\varphi_0=0$, $\varphi_2=0.2$ (a) and $\varphi_2=\pi-0.2$ (b).
The incident laser frequency $\omega_{in}=2$~eV,
$\omega_0=0.17$~eV.}
\end{figure}

At this point we have to assume a particular form of the phonon
dispersion~$\omega_{\vec{q}}$. We take the conical dispersion
$\omega_{\vec{q}}=\omega_0+v_{ph}|\vec{q}|$ valid for
$\omega_0/v\ll{q}\ll{1}/a$.\footnote{At small~$q$ the phonon dispersion
is determined by the electron-phonon interaction, and can be calculated
from the polarization operator~$\Pi_{zz}(\vec{q},\omega)$. The latter
was calculated for the case $\omega=0$ in
Ref.~\onlinecite{Piscanec2004},  for abitrary $\vec{q}$ and~$\omega$
at zero doping it is given by Eq.~(\ref{Pizz=}).}
Then the singular points are determined from the equation
\begin{subequations}
\begin{eqnarray}
&&\frac{p_0}{q_1}\,\sin(\varphi_1-\varphi_0)+
\frac{p_2}{q_2}\,\sin(\varphi_1-\varphi_2)=0, \label{zeroJacobian=}\\
&& q_1=\sqrt{p_0^2+p_1^2-2p_0p_1\cos(\varphi_1-\varphi_0)},\\
&& q_2=\sqrt{p_1^2+p_2^2-2p_1p_2\cos(\varphi_2-\varphi_1)}.
\end{eqnarray}
\end{subequations}
which is the same for $\varphi_1$~and~$\bar\varphi_1$,
since $p_1=\bar{p}_1$.
Let us denote the singular points by $\varphi_1^s$, labelled by the
index~$s$. Around each singular point we can expand
\begin{subequations}
\begin{eqnarray}
&&\omega_{\vec{q}_1}+\omega_{\vec{q}_2}\approx
\omega_s+\frac{\omega_s''}{2}\,\tilde{\varphi}_1^2,\quad
\tilde{\varphi}_1=\varphi_1-\varphi_1^s,\\
&&\omega_{\bar{\vec{q}}_1}+\omega_{\bar{\vec{q}}_2}
\approx
\omega_s+\frac{\omega_s''}{2}\,\tilde{\bar\varphi}_1^2,\quad
\tilde{\bar\varphi}_1=\bar\varphi_1-\varphi_1^s.
\end{eqnarray}
\end{subequations}
The second derivative in each singular point is given by
\begin{eqnarray}
&&\frac{v_{ph}}{2q_1}\,p_1p_0\cos(\varphi_1^s-\varphi_0)
+\frac{v_{ph}}{2q_2}\,p_1p_2\cos(\varphi_1^s-\varphi_2)-\nonumber\\
&&-\frac{v_{ph}}{4q_1^3}\,p_1^2p_0^2\sin^2(\varphi_1^s-\varphi_0)
-\frac{v_{ph}}{4q_2^3}\,p_1^2p_2^2\sin^2(\varphi_1^s-\varphi_2)
\equiv\nonumber\\
&&\equiv\omega_s^{\prime\prime}(\varphi_0,\varphi_2).
\label{wspp=}
\end{eqnarray}
To get an idea of the location of singular points, we plot
the left-hand side of Eq.~(\ref{zeroJacobian=}) as a function
of~$\varphi_1$ for $\varphi_2$ close to~$\varphi_0$ and to
$\varphi_0+\pi$ (Fig.~\ref{fig:singpoint}). We see that in the
first case Eq.~(\ref{zeroJacobian=}) has two solutions, while
in the second case -- four. This means that there is a special
value of~$\varphi_2$ between $0$~and~$\pi$ (together with
the symmetric one), such that Eq.~(\ref{zeroJacobian=}) has
three solutions and at the third solution the
derivative~$\omega_s''$, defined in Eq.~(\ref{wspp=}),
vanishes. This situation will take place as long as
$\omega_0/\omega_{in}<(3-\sqrt{5})/4=0.191\ldots$,
i.~e., $\omega_{in}>0.9$~eV, as
can be established by setting $\varphi_0=0$,
$\varphi_1=\varphi_2=\pi$ [so that Eq.~(\ref{zeroJacobian=})
is satisfied], and equating $\omega_s''=0$. However, when these
conditions are fulfilled, and when $\varphi_2$ takes this special
value, the third solution for~$\varphi_1$ always lies in the
smaller sector between $\varphi_0$~and~$\varphi_2$, and does
not belong to the integration region. As a result, there is always
just one solution of Eq.~(\ref{zeroJacobian=}), which
satisfies $\pi<\varphi_1-\varphi_0<\pi+\varphi_2-\varphi_0$ for
$0<\varphi_2-\varphi_0<\pi$ and
$\varphi_2-\varphi_0-\pi<\varphi_1-\varphi_0<\pi$ for
$\pi<\varphi_2-\varphi_0<2\pi$
[i.~e., inequalities~(\ref{angularrange=})], and thus contributes
to the intensity integral. We denote this solution
by~$\varphi_1^0$.

Integration over the deviations
$\tilde\varphi_1=\varphi_1-\varphi_1^0$,
$\tilde{\bar\varphi}_1=\bar\varphi_1-\varphi_1^0$ gives
\begin{eqnarray}
&&\int
d\tilde\varphi_1\,d\tilde{\bar\varphi}_1\,
2\pi\delta(\Delta)=\nonumber\\
&&=\frac{4\pi}{|\omega^{\prime\prime}|}
\int
d\tilde\varphi_1\,d\tilde{\bar\varphi}_1\,
\delta(\tilde\varphi_1^2-\tilde{\bar\varphi}_1^2
-4\Delta_{eh}/\omega^{\prime\prime})=\nonumber\\
&&=\int
\frac{|{2\pi}/{\omega^{\prime\prime}}|\,
d\tilde\varphi_1\,d\tilde{\bar\varphi}_1}
{\sqrt{\tilde{\bar\varphi}_1^2
+4\Delta_{eh}/\omega^{\prime\prime}}}\sum_\pm
\delta\!\left(\tilde\varphi_1\pm\sqrt{\tilde{\bar\varphi}_1^2
+\frac{4\Delta_{eh}}{\omega^{\prime\prime}}}\right)\approx\nonumber\\
&&\approx\frac{4\pi}{|\omega^{\prime\prime}|}
\ln\left|\frac{\omega^{\prime\prime}}
{\max\{\gamma,\Delta_{eh}\}}\right|. \label{phi1ln=}
\end{eqnarray}
Generally, the upper and lower integration limits here are of the order
of~$\pm{1}$; more precise knowledge is not needed for for the
calculation of the leading logarithmic term. It is important that the
energy $\delta$-function has a finite width~$\sim\gamma$, which may cut
off the divergency first, if it is greater than~$\Delta_{eh}$. The
innermost integral over~$\phi$, assumed to be a non-singular function
of $\varphi_1,\bar\varphi_1$, can be taken at
$\varphi_1=\bar\varphi_1=\varphi_1^0$. To check the validity of this
assumption, we have to study in more detail the behavior of the
integral at $\varphi_2-\varphi_0\to{0},\pi$.

To make the formulas more compact, in the following we set
$\varphi_0=0$, as all angles can be counted from~$\varphi_0$.
For simplicity we also perform the calculations in the limit
$\omega_0\ll\omega_{in}$.
Then, assuming $-\pi<\varphi_2<\pi$, we simply obtain
$\varphi_1^0=\pi+\varphi_2/2$.

Let us start from the simpler case of $\varphi_2$~close to~$\pm\pi$,
denoting $\tilde\varphi_2=\varphi_2+\pi$ if $-\pi<\varphi_2<0$
and $\tilde\varphi_2=\varphi_2-\pi$ if $0<\varphi_2<\pi$. Then for
$\varphi_1=\bar\varphi_1=\varphi_0$ we have
\begin{subequations}\begin{eqnarray}
\zeta_{max}(\phi)&=&
-\frac{\cos\phi+\sin\phi}{2\sin|\tilde\varphi_2/2|}+\nonumber\\
&&{}+\frac{1}{2}\min\left\{-\cos\phi+\sin\phi\:,\,0\right\},\\
\zeta_{min}(\phi)&=&
\frac{1}{2}\max\left\{-\cos\phi+\sin\phi\:,\,0\right\}.
\end{eqnarray}\end{subequations}
For $|\tilde\varphi_2|\ll{1}$ the condition
$\zeta_{max}(\phi)>\zeta_{min}(\phi)$ severely restricts the
integration domain in~$\phi$, so $\varphi_2\approx\pm\pi$
does not introduce any extra singularities.

For $|\varphi_2|\ll{1}$ we have
\begin{subequations}
\begin{eqnarray}
&&\zeta_{max}(\phi)=\min\left\{\frac{\cos\phi+\sin\phi}2-f_1(\phi)\,,
\sin\phi\right\},\\
&&\zeta_{min}(\phi)=\max\left\{\frac{\sin\phi-\cos\phi}2-f_1(\phi)\,,0\right\},\\
&&f_1(\phi)=
\frac{\tilde\varphi_1\cos\phi+\tilde{\bar\varphi}_1\sin\phi}{\varphi_2}.
\end{eqnarray}
\end{subequations}
Again setting all $\gamma_0=\gamma_1=\bar\gamma_1=\gamma_2=\gamma$, we
obtain simply $\gamma_x=\gamma_y=2\gamma$. The requirement
$\zeta_{min}(\phi)<\zeta_{max}(\phi)$ translates into
\begin{equation}
|f_1(\phi)|<\frac{\cos\phi+\sin\phi}2.
\label{phirange=}
\end{equation}
If $|\varphi_1|,|\tilde{\bar\varphi}_1|\gg|\varphi_2|$, then only
$\tilde\varphi_1\approx-\tilde{\bar\varphi}_1$ contribute to
integral~(\ref{phi1ln=}) (for
$\tilde\varphi_1\approx\tilde{\bar\varphi}_1$ the domain of
integration is restricted by $|\tilde\varphi_1|<\varphi_2/2$). Then
constraint~(\ref{phirange=}) allows only small deviations of $\phi$
from $\pi/4$, thus we can approximate
$f_1(\phi)\approx-\sqrt{2}(\tilde\varphi_1/\varphi_2)(\phi-\pi/4)$.
In the opposite limiting case,
$|\varphi_1|,|\tilde{\bar\varphi}_1|\ll|\varphi_2|$, range of~$\phi$
is almost unrestricted, and both
$\tilde\varphi_1\approx\pm\tilde{\bar\varphi}_1$ will contribute.
The $\phi$-integral in Eq.~(\ref{intdphi=}) in these two limiting
cases is calculated to be
\begin{equation}
\int\limits_0^{\pi/2}{d}\phi\,
\frac{\zeta_{max}(\phi)-\zeta_{min}(\phi)}
{8\left[\gamma_x\cos\phi+\gamma_y\sin\phi\right]^3}
\approx\frac{1}{256\gamma^3}\min\left\{1\,,
\left|\frac{\varphi_2}{2\tilde\varphi_1}\right|\right\}.
\end{equation}
Thus, we conclude that for $|\varphi_2|\ll{1}$ the upper cutoff
in the logaritmic integral~(\ref{phi1ln=}) is~$|\varphi_2|$, and
not of the order of~1. Thus, upon integration over~$\varphi_2$
we obtain the second logarithmic divergence, which should be
cut off at $|\varphi_2|\sim{1}$ above and
$|\varphi_2|\sim\max\{\gamma,\Delta_{eh}\}/\omega''$ below.

We restrict ourselves to the calculation of the leading logarithmic
asymptotics, so the second derivative~(\ref{wspp=}) can be taken
at $\varphi_2=\varphi_0=0$ and is simply
$\omega''=-\omega_{in}v_{ph}/(4v)$. The function
$\mathcal{J}(\varphi)$, defined in Eq.~(\ref{Jphi=}), can be taken to be
\begin{eqnarray}\nonumber
\mathcal{J}(\varphi)&=&\frac{\pi^2}8\left(\frac{e^2}c\right)^2\frac{v^2}{c^2}
\left(\frac{\lambda_K}{2\pi}\right)^4
\frac{\omega_{in}^3}{\gamma^3}\frac{v}{v_{ph}}\times\nonumber\\
&&{}\times\frac{1}{|\varphi|}
\ln\frac{|\varphi|\omega_{in}v_{ph}/v}{\max\{\gamma,\Delta_{eh}\}}.
\end{eqnarray}
The final integration leads to the following expression for
$\mathcal{J}_0,\mathcal{J}_2$
to be substituted in Eq.~(\ref{I4Dpolarization=}):
\begin{eqnarray}
\mathcal{J}_0=\mathcal{J}_2&=&\frac{\pi^2}8\left(\frac{e^2}c\right)^2\frac{v^2}{c^2}
\left(\frac{\lambda_K}{2\pi}\right)^4
\frac{\omega_{in}^3}{\gamma^3}\frac{v}{v_{ph}}\times\nonumber\\
&&{}\times\ln^2\frac{\omega_{in}v_{ph}/v}{\max\{\gamma,\Delta_{eh}\}},
\end{eqnarray}
so that $I_{4K}^\|=3I_{4K}^\perp$, as for two-phonon scattering.
Thus, the polarization dependence is the same as that described in
Sec.~\ref{sec:2ramanDirac}. Using Eq.~(\ref{Inpolar=}), we obtain the
final result:
\begin{eqnarray}\nonumber
I_{4K}&=&\left(\frac{e^2}c\right)^2\frac{v^2}{c^2}
\left(\frac{\lambda_K}{2\pi}\right)^4
\frac{\omega_{in}^3}{\gamma^3}\frac{v}{v_{ph}}
\ln^2\frac{\omega_{in}v_{ph}/v}{\max\{\gamma,\Delta_{eh}\}}\times\nonumber\\
&&{}\times\frac{\pi^3}{32}
\left[\frac{|\vec{e}_{in}|^2}8
(1-\cos\Theta_{det})(3+\cos^2\Theta_{det})\right.+\nonumber\\
&&\qquad\left.{}+\frac{8-(1+\cos\Theta_{det})^3}{12}
|(\vec{e}_{in}\cdot\vec{e}_{det})|^2\right].\label{I4Kdisp=}
\end{eqnarray}
To conclude this section, we note that the leading
logarithmic term, calculated here, is not sensitive to the
assumption of the conical phonon dispersion~$\omega_{\vec{q}}$.
The same result will be obtained for dispersion of any
shape; the phonon group velocity should be taken at the
wave vector $q=\omega_{in}/v$, corresponding to electron
backscattering.

\section{Renormalization of the coupling constants}\label{sec:RG}

\subsection{Coulomb renormalization}\label{sec:CoulombRG}

As discussed in the beginning of Sec.~\ref{sec:warping}, the measured
ratio of the integrated intensities
of $2K$ and $2\Gamma$ peaks at 2700~cm$^{-1}$ and 3250~cm$^{-1}$,
$I_{2K}/I_{2\Gamma}\approx{20}$ (Ref.~\onlinecite{Ferrari2006}) is
in noticeable disagreement with the calculated values of
electron-phonon coupling constants.\cite{Piscanec2004}
In this section we investigate how the electron-phonon coupling
constants are renormalized by the Coulomb interaction between
electrons. A brief account of this part has been reported in the
short publication (Ref.~\onlinecite{us}).

We are going to consider only the long-range part of the
Coulomb interaction (i.~e., smooth on the length scale of
the lattice constant). Such interaction does not mix the
states in different valleys, so the interaction hamiltonian
can be written as
\begin{equation}
\hat{H}_{ee}=\frac{e^2}2\int{d}^2\vec{r}\,{d}^2\vec{r}'\,
\frac{\hat\rho(\vec{r})\,\hat\rho(\vec{r}')}{|\vec{r}-\vec{r}'|},
\quad
\hat\rho(\vec{r})=\hat\psi^\dagger(\vec{r})\hat\psi(\vec{r}).
\end{equation}
In this equation we have not included explicitly the screening
by the background dielectric constant of the substrate~$\vep_\infty$
(the high-frequency value), which can be taken
into account by incorporating it into~$e^2$.

\begin{figure}
\includegraphics[width=8cm]{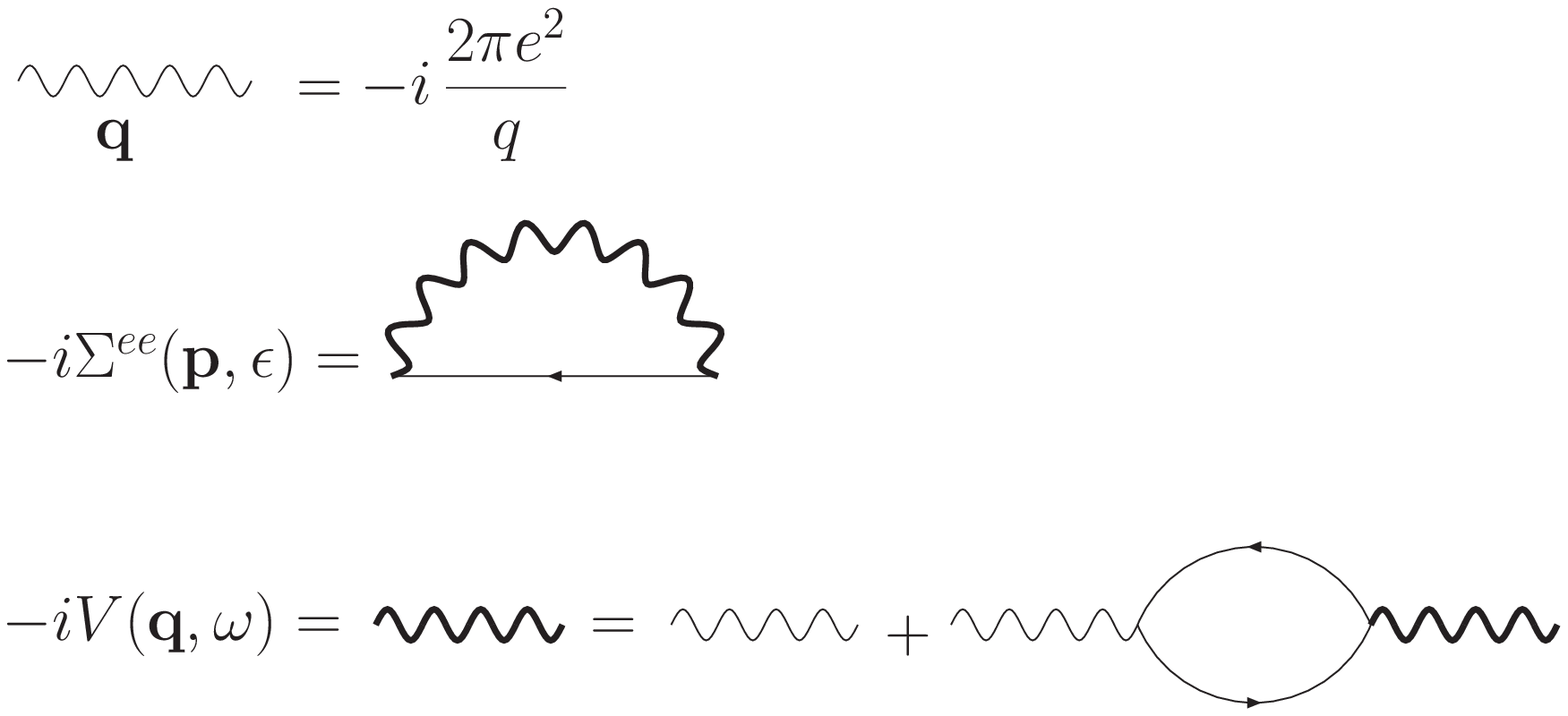}
\caption{\label{fig:RPA} Electronic self-energy from the
RPA-screened Coulomb interaction.}
\end{figure}

The electronic self-energy due to the Coulomb interaction (the Fock
term), $\Sigma^{ee}(\vec{p},\ep)$  is shown in Fig.~\ref{fig:RPA}.
Its leading logarithmic asymptotics is given by:\cite{Guinea99}
\begin{eqnarray}
&&\Sigma^{ee}(\vec{p},\ep)=
i\int\frac{d\omega}{2\pi}\frac{d^2\vec{q}}{(2\pi)^2}\,
V(\vec{q},\omega)\,G(\vec{p}-\vec{q},\ep-\omega)\approx\nonumber\\
&&\approx\frac{8}{\pi^2\mathcal{N}}\left[f(g)(2\ep-v\vec{p}\vec\Sigma)
-\tilde{f}(g)(\ep-v\vec{p}\vec\Sigma)\right]
\ln\frac{\xi_{max}}{\xi_{min}},\nonumber\\ &&\label{SigmaCoul=}\\
&&f(g)=1-\frac{\pi}{2g}+\frac{\arccos{g}}{g\sqrt{1-g^2}},\;\;\;
\label{fg=}
\tilde{f}(g)=\frac{g\arccos{g}}{\sqrt{1-g^2}},\\
&&V(\vec{q},\omega)= \frac{16g}{\mathcal{N}}\frac{v}q\,
\frac{\sqrt{(vq)^2-\omega^2}}{gvq+\sqrt{(vq)^2-\omega^2}}. \label{VRPA=}
\end{eqnarray}
Here we have introduced the total number of the Dirac species,
$\mathcal{N}=4$, which takes into account the valley and the
spin degeneracy (the latter enters as factor of~$2$ multiplying
the electron polarization operator).
The lower cutoff $\xi_{min}\sim\max\{vp,\ep\}$, the upper cutoff
$\xi_{max}\sim{v}/a$ is of the order of the electronic bandwidth,
and the dimensionless Coulomb coupling constant is defined as
\begin{equation}
g=\frac{\pi\mathcal{N}e^2}{8v}\,.
\end{equation}
The derivation of Eq.~(\ref{SigmaCoul=}) is given in
Appendix~\ref{app:logarithms}. The logarithmic divergence in the
Fock self-energy~$\Sigma^{ee}$ is due to the long-distance
nature of the Coulomb interaction, and thus is not picked
up by local approximations such as LDA or GGA.

The random phase approximation (RPA) for $V(\vec{q},\omega)$,
shown in Fig.~\ref{fig:RPA}, corresponds to expansion of the coefficient
in front of the logarithm to the leading order in the parameter
$1/\mathcal{N}=0.25$, assumed to be small. This assumption is better
justified than the expansion in~$g$, which would be obtained if
we used the bare coupling $2\pi{e}^2/q$ instead of the RPA-dressed
one $V(\vec{q},\omega)$. Indeed, for $\mathcal{N}=4$ we
have $g=(\pi/2)(e^2/v)\approx{3}.4$; taking into account the
background dielectric screening reduces it to $g\sim{1}$.

The presence of the large logarithm invalidates the simple
first-order expansion in $1/\mathcal{N}$, and makes it necessary
to sum all leading logaritmic terms
$\sim(1/\mathcal{N})^n\ln^n(\xi_{max}/\xi_{min})$ of the
perturbation theory. This summation is performed using the
standard renormalization group (RG) procedure. Let us
introduce the running cutoff $\xi_{max}e^{-\ell}$. One RG step
consists of reducing the cutoff,
$\ell\to\ell+\delta\ell$, so that $e^{\delta\ell}\gg{1}$, while
$(1/\mathcal{N})\delta{\ell}\ll{1}$.
The inverse Green's function transforms as
\begin{equation}
\ep-v\vec{p}\cdot\vec\Sigma-\Sigma(\vec{p},\ep)=
\frac{\ep-(v+\delta{v})\vec{p}\cdot\vec\Sigma}{1+\delta{Z}}\,,
\end{equation}
where $Z$ is chosen to preserve the coefficient at $\ep$ upon
rescaling of the electronic fields,
$\psi\to(1+\delta{Z}/2)\psi$:
\begin{equation}
\frac{1}{1+\delta{Z}}=1-\frac{\partial\Sigma}{\partial\ep}.
\end{equation}
The renormalization of the velocity is then given by
\begin{equation}
\frac{\delta{v}}{v}=\frac{\partial\Sigma}{\partial\ep}
+\frac{\partial\Sigma}{\partial(v\vec{p}\cdot\vec\Sigma)}.
\end{equation}


\begin{figure}
\includegraphics[width=2cm]{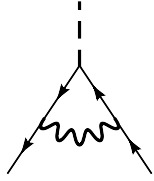}
\caption{\label{fig:PhononVertexCoul} Diagram describing the
logarithmic correction to the electron-phonon vertex due to
the Coulomb interaction in the order $O(1/\mathcal{N})$.}
\end{figure}

Next, we need to determine renormalization of the coupling
constants. The electron charge is not renormalized, as guaranteed
by the gauge invariance, so the renormalization of the Coulomb coupling
constant~$g$ is determined by renormalization of the velocity~$v$.
The correction to the electron-phonon coupling constants is determined
by the diagram in Fig.~\ref{fig:PhononVertexCoul}, and is evaluated in
Appendix~\ref{app:logarithms}. The result is
\begin{subequations}
\begin{eqnarray}
&&\frac{\delta{F}_{\Gamma}}{F_{\Gamma}}=\delta{Z}+
\frac{8}{\pi^2\mathcal{N}}
\left[\tilde{f}(g)-f(g)\right]\ln\frac{\xi_{max}}{\xi_{min}},\qquad\\
&&\frac{\delta{F}_K}{F_K}=\delta{Z}+
\frac{8}{\pi^2\mathcal{N}}\,\tilde{f}(g)\ln\frac{\xi_{max}}{\xi_{min}}.
\end{eqnarray}
\end{subequations}
Let us pass to dimensionless electron-phonon coupling constants
$\lambda_{\Gamma}$, $\lambda_K$, introduced in
Eq.~(\ref{lambdaphonon=}). Then the equations for the RG flow are
the following:
\begin{subequations}
\begin{eqnarray}
\frac{d\ln{g}}{d\ell}&=&-\frac{8}{\pi^2\mathcal{N}}\,f(g),\label{RGflowg=}\\
\frac{d\ln\lambda_{\Gamma}}{d\ell}&=&0,\label{RGflowE2=}\\
\frac{d\ln\lambda_K}{d\ell}&=&\frac{16}{\pi^2\mathcal{N}}\,f(g).
\end{eqnarray}\label{RGflow=}
\end{subequations}

\begin{figure}
\includegraphics[width=8cm]{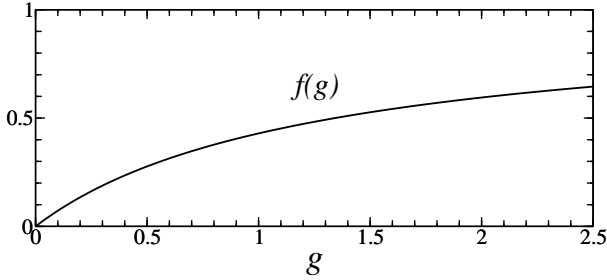}
\caption{\label{fig:plotf} Plot of the function $f(g)$, defined in
  Eq.~(\ref{fg=}).}
\end{figure}

As $f(g)$ is positive and monotonous (see Fig.~\ref{fig:plotf}),
$g$~flows to
weak coupling:\cite{Guinea99} if the initial value of~$g$ is
large,
\begin{equation}
f(g)=1-\frac\pi{2g}+O(g^{-2})\;\;\;\Rightarrow\;\;\;
g(\ell)=g(0)\,{e}^{-8\ell/(\pi^2\mathcal{N})},
\end{equation}
while at small~$g$ we have
\begin{equation}
f(g)=\frac{\pi{g}}4+O(g^2)\;\;\;\Rightarrow\;\;\;
g(\ell)=\frac{g(0)}{1+2\ell\,g(0)/(\pi\mathcal{N})}.
\end{equation}
Integration of Eqs.~(\ref{RGflow=}) yields the following relation:
\begin{equation}
\frac{\lambda_K(\ell)}{\lambda_K(0)}
=\left[\frac{g(0)}{g(\ell)}\right]^2
=\left[\frac{v(\ell)}{v(0)}\right]^2,
\end{equation}
which, in principle, can be checked experimentally.
Thus, $\lambda_K$~is enhanced, which is in qualitative
agreement with the Raman data:  according to the results of
Sec.~\ref{sec:2ramanDirac}, the ratio of the intensities of the
two-phonon peaks is
$I_{2K}/I_{2\Gamma}=2(\lambda_{A_1}/\lambda_{\Gamma})^2$.

\begin{figure}
\includegraphics[width=8cm]{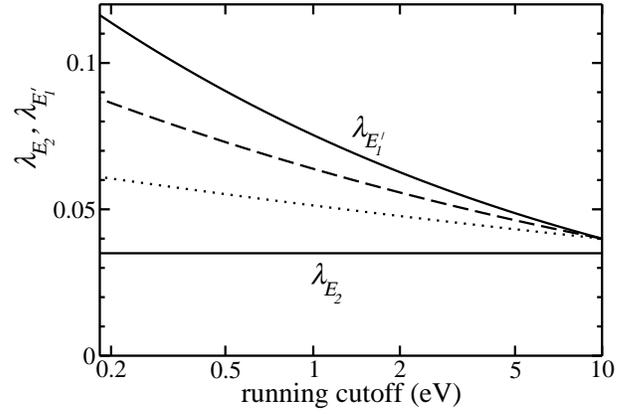}
\caption{\label{fig:RGflow} Dependence of the dimensionless
coupling constant $\lambda_K$ on the running cutoff
$\xi_{max}e^{-\ell}$ (to be identified with the electronic
energy), represented  by three upper curves starting from the
bare value 0.04 at 10~eV, for three values of the
bare Coulomb coupling $g(0)=3.4,\:1.5,\:0.5$ (solid, dashed, and
dotted curves, corresponding to the substrate dielectric constant
$\vep_\infty=1,\,2.3,\,6.8$, respectively)
as determined by Eqs.~(\ref{RGflow=}). The
constant $\lambda_{\Gamma}=0.035$ is unchanged.}
\end{figure}

To study the behavior of the coupling constants quantitatively, we
solve Eqs.~(\ref{RGflow=}) numerically. The largest value of~$\ell$
is determined by the lower cutoff
$\xi_{min}\sim\omega_\phonon\sim{0}.2\:\mbox{eV}$.  In
Fig.~\ref{fig:RGflow}, we show the flow of~$\lambda_K$ for
three values of the bare Coulomb coupling constant: $g(0)=3.4$
(corresponding to no dielectric screening at all), $g(0)=1.5$, and
$g(0)=0.5$. The bare values of the the electron-phonon coupling
constants $\lambda_{\Gamma}(0)=0.035$, $\lambda_K(0)=0.040$
were chosen (i)~to satisfy the relation
$\lambda_{\Gamma}(0)/\lambda_K(0)=\omega_K/\omega_{\Gamma}$,
valid in the tight-binding approximation, (ii)~to reproduce the
experimental value $\lambda_{\Gamma}\approx{0}.035$. Note that the RPA
calculation without the RG collection of all leading logarithmic terms,
would give all dependencies on Fig.~\ref{fig:RGflow} to be straight
lines with slopes fixed at 10~eV. A comparable error would be
produced by the GW approximation, which neglects vertex corrections,
and thus picks up correctly only the first term of the logarithmic series.

To estimate the EPC strength relevant for Raman scattering, we identify
the running cutoff with the typical electronic energy, involved in the
process, thus stopping
the RG flow at electronic energies $\ep\sim{1}\:\mbox{eV}$
(half of the incident laser frequency).
In the unscreened case, $g(0)=3.4$, it gives
$\lambda_{A_1}/\lambda_{\Gamma}\approx{3.2}$, in agreement with the
observed ratio $I_{2K}/I_{2\Gamma}\approx{20}$.

Finally, we wish to note that the cancellation of the self-energy
and vertex corrections, leading to $d\lambda_{\Gamma}/d\ell=0$ in
Eq.~(\ref{RGflowE2=}), is
not occasional. Indeed, comparing Eqs.~(\ref{Heph=})
and~(\ref{Heem=}), we can see that coupling to the
$E_2$~phonon displacement~$\vec{u}_{E_2}$
has the same form  as the coupling to the vector potential~$\vec{A}$
with the correspondence $u_{E_2x}\leftrightarrow{A}_y$,
$-u_{E_2y}\leftrightarrow{A}_x$ (up to the sign, different for the
$\Kpnt,\Kpnt'$ valleys). This means that a uniform phonon
displacement~$\vec{u}_{E_2}$ can be gauged out of the
electronic hamiltonian, which should hold for both intial and
renormalized hamiltonians.
Thus, gauge invariance requires that
$F_{\Gamma}$~is renormalized in the same way as the velocity~$v$.
Since $\lambda_{\Gamma}\propto{F}_{\Gamma}^2/v^2$, it must remain
constant.

\subsection{Renormalization due to the electron-phonon coupling}\label{sec:phononRG}

It turns out that the Coulomb interaction is not the only source
of renormalizations. Since the electron-phonon self-energy
$\Sigma^{ph}(\vec{p},\ep)$, calculated in Sec.~\ref{sec:inelastic},
also has a logarithmic divergence, renormalizations due to
electron-phonon interaction should be taken into account as well.
However, in practice, the electron-phonon coupling is so weak
($\lambda_\mu\ll{1}$, see the previous subsection), that its
effect is negligible, so that the previous subsection contains
all the practical information. Still, for the sake of completeness,
in this subsection we describe the theory of renormalizations due
to electron-phonon coupling.

Let us return to the electron-phonon self-energy
$\Sigma^{ph}(\vec{p},\ep)=
\Sigma^{\Gamma}(\vec{p},\ep)+\Sigma^K(\vec{p},\ep)$,
calculated in Sec.~\ref{sec:inelastic}.
The leading logarithmic asymptotics of $\Sigma^{ph}$ is given
by (see also Appendix~\ref{app:logarithms}):
\begin{eqnarray}
\Sigma^{ph}(\vec{p},\ep)&=&
i\int\frac{d\omega}{2\pi}\frac{d^2\vec{q}}{(2\pi)^2}\sum_\phonon
\frac{F^2_\phonon}{2M\omega_\phonon}\frac{\sqrt{27}a^2}4\,
D_\phonon(\omega)\times\nonumber\\ &&\quad\qquad{}\times
(\Lambda\Sigma)_\phonon{G}(\vec{p}-\vec{q},\ep-\omega)
(\Lambda\Sigma)_\phonon \approx\nonumber\\
&\approx&\frac{\lambda_{\Gamma}+\lambda_K}{2\pi}\:\ep
\ln\frac{\xi_{max}}{\xi_{min}}.\label{SigmaEPC=}
\end{eqnarray}
Here the phonon mode index $\phonon$ runs over the two modes
belonging to the $E_2$~representation and the two modes belonging
to the $E_1'$ representation, the corresponding matrices
$(\Lambda\Sigma)_\phonon$ being
$-\Lambda_z\Sigma_y$, $\Lambda_z\Sigma_x$, $\Lambda_x\Sigma_z$,
$\Lambda_y\Sigma_z$. The upper and lower cutoff is given by
$\xi_{max}\sim{v}/a$, $\xi_{min}\sim\max\{\ep,\omega_\phonon\}$.
The dimensionless constants $\lambda_{\Gamma},\lambda_K$,
defined in Eq.~(\ref{lambdaphonon=}), will be
treated as small parameters.

The latter statement deserves some discussion. In principle, one
could proceed analogously to the Coulomb case: instead of doing
the perturbative expansion in $\lambda_\phonon$, one could dress
the bare phonon propagators by the appropriate polarization
operators $\Pi(\vec{q},\omega)$, corresponding to $1/\mathcal{N}$
expansion (the polarization operators for different matrix vertices
are calculated in Appendix~\ref{app:polarization}).
Since $\Pi(\vec{q},\omega)\propto{q}$ at large~$q$,
the dressed phonon frequency would grow as~$\sqrt{q}$, and
$\Sigma^{ph}$ would no longer diverge logarithmically. However,
the inelastic X-ray scattering data for the phonon
dispersion~\cite{Maultzschexp} show that the phonon dispersion is
smaller than the phonon frequency itself. Thus, the
renormalization of the phonon frequency remains small even at
$q\sim{1}/a$, so the perturbative expansion in~$\lambda_\phonon$
is more justified, and we neglect the phonon dispersion.

The logarithmically divergent integrals in Eqs.~(\ref{SigmaCoul=})
and~(\ref{SigmaEPC=}) have different structure due to different
form of the screened interaction $V(\vec{q},\omega)$ and the
phonon propagator $D_\phonon(\omega)$. In Eq.~(\ref{SigmaCoul=})
the integral is dominated by the frequencies $|\omega|\sim{v}q$,
while in Eq.~(\ref{SigmaEPC=}) it is $|\omega|\sim\omega_\phonon$,
since $D_\phonon(\omega)\propto{1}/\omega^2$ at
$|\omega|\gg\omega_\phonon$.
Thus, in the calculation of the leading logarithmic asymptotics it
is sufficient to approximate
$D_\phonon(\omega)\approx -2\pi{i}\,\delta(\omega)$
(see Appendix~\ref{app:logarithms}).
This substitution makes the phonon propagator (combined with
electron-phonon vertices) formally analogous to the correlator of
a static disorder potential (in other words, from the point of view
of electrons with $\ep\gg\omega_\phonon$ the lattice is effectively
frozen).
Thus, renormalizations due to electron-phonon interaction at
$\ep\gg\omega_\phonon$ are equivalent to those due to static
disorder~\cite{Ye,Guinea2005,AleinerEfetov,AleinerFoster}. This
equivalence holds only in the leading order in electron-phonon
coupling, since in higher orders the phonon propagator is dressed
by polarization loops, and the static disorder correlator is not.

\begin{figure}
\includegraphics[width=8cm]{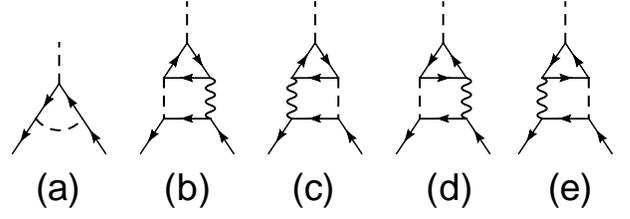}
\caption{\label{fig:PhononVertexRG} Diagrams describing the
logarithmic correction to the electron-phonon vertex due to
the electron-phonon interaction in the order
$O(1/\mathcal{N},\lambda_\phonon^2)$. Diagrams (b)--(e)
vanish.}
\end{figure}

\begin{figure}
\includegraphics[width=5cm]{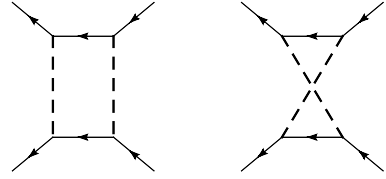}
\caption{\label{fig:Phonon2LineRG} Logarithmic diagrams of the
order $O(\lambda^2)$ not reduced to a renormalization of the
electron-phonon vertex.}
\end{figure}

Logarithmic corrections to the electron-phonon vertex in the
order $O(1/\mathcal{N},\lambda_\phonon^2)$ are shown in
Fig.~\ref{fig:PhononVertexRG}. All diagrams vanish, except the
first one, which gives a nonzero correction to~$\lambda_K$:
\begin{subequations}
\begin{eqnarray}
&&\frac{\delta{F}_{\Gamma}}{F_{\Gamma}}=\delta{Z}+0\,,\\
&&\frac{\delta{F}_K}{F_K}=\delta{Z}
-\frac{\lambda_{\Gamma}}{2\pi}\ln\frac{\xi_{max}}{\xi_{min}}.
\end{eqnarray}
\end{subequations}
The diagrams of Fig.~\ref{fig:PhononVertexRG}, however, do not
exhaust all logarithmic vertex corrections. In addition, one
has to consider two diagrams, shown in
Fig.~\ref{fig:Phonon2LineRG}, as they are of the same order
$O(\lambda_\phonon^2)$, and also logarithmically divergent.
They may be viewed as a correction to the two-electron vertex
$\Gamma^{(2)}(\vec{p},\ep,\vec{p}',\ep';\vec{q},\omega)$.
The bare value of~$\Gamma^{(2)}$ is given just by the phonon
single phonon propagator, combined with the electron-phonon
vertices. Diagrams of Fig.~\ref{fig:Phonon2LineRG} are
evaluated in Appendix~\ref{app:logarithms}, and give:
\begin{widetext}
\begin{subequations}
\begin{eqnarray}
&&\Gamma^{(2)}(\vec{p},\ep,\vec{p}',\ep';\vec{q},\omega)
=\frac{v^2}2
\left[\lambda_{\Gamma}\,D_{\Gamma}(\omega)
\left(\Lambda_z\Sigma_y\otimes\Lambda_z\Sigma_y
+\Lambda_z\Sigma_x\otimes\Lambda_z\Sigma_x\right)
\right.+\quad\nonumber\\ &&\hspace*{3.7cm}
+\left.\lambda_{E_1}\,D_K(\omega)
\left(\Lambda_y\Sigma_z\otimes\Lambda_y\Sigma_z
+\Lambda_x\Sigma_z\otimes\Lambda_x\Sigma_z\right)
\right],\label{Gamma2=}\\
&&\delta\Gamma^{(2)}(\vec{p},\ep,\vec{p}',\ep';\vec{q},\omega)
=\frac{v^2}2\ln\frac{\xi_{max}}{\xi_{min}}
\left[\frac{\lambda_K^2}{2\pi}\,
D_{E_1'+E_1'}(\omega)
\left(\Lambda_z\Sigma_y\otimes\Lambda_z\Sigma_y
+\Lambda_z\Sigma_x\otimes\Lambda_z\Sigma_x\right)
\right.+\quad\nonumber\\ && \hspace*{5.3cm}
+\left.\frac{\lambda_{\Gamma}\lambda_{E_1}}{\pi}\,
D_{E_2+E_1'}(\omega)
\left(\Lambda_y\Sigma_z\otimes\Lambda_y\Sigma_z
+\Lambda_x\Sigma_z\otimes\Lambda_x\Sigma_z\right)
\right],\label{deltaGamma2=}\\
&&{D}_{\phonon+\phonon'}(\omega)\equiv
\frac{2(\omega_\phonon+\omega_{\phonon'})}
{\omega^2-(\omega_\phonon+\omega_{\phonon'}-io)^2}.
\end{eqnarray}
\end{subequations}
\end{widetext}
The following features of the expression for
$\delta\Gamma^{(2)}$ are worth noting: (i)~the matrix
structure of~$\delta\Gamma^{(2)}$ is identical to that
of~$\Gamma^{(2)}$; (ii)~$\delta\Gamma^{(2)}$ depends on
electronic energies and momenta only through~$\xi_{min}$,
i.~e., logarithmically; (iii)~the $\omega$~dependence
of~$\delta\Gamma^{(2)}$ is analogous to that
of~$\Gamma^{(2)}$, but the pole is at the sum of two
phonon frequencies.

The latter fact has a simple physical meaning.
Before the reduction of the ultraviolet cutoff~$\xi_{max}$
the excitations of the lattice were phonons with momenta
$|\vec{q}|<\xi_{max}/v$, as well as their combinations.
In the theory with the reduced cutoff $\xi_{max}e^{-\ell}$,
besides phonons with momenta
$|\vec{q}|<(\xi_{max}/v)e^{-\ell}$,
one has to consider also pairs of phonons with large and
almost opposite momenta $\vec{q}'$~and $\vec{q}-\vec{q}'$,
so that the total momentum of the pair is
$|\vec{q}|<\xi_{max}e^{-\ell}$. Each of the phonons
constituting the pair has
$(\xi_{max}/v)e^{-\ell}<|\vec{q}'|,|\vec{q}-\vec{q}'|<\xi_{max}/v$
and thus has been integrated out. The pair, however, having
small total momentum, has to be included into the low-energy
theory as a single excitation. Thus, $D_{\mu+\mu'}(\omega)$
has the meaning of the propagator of this excitation, and
comparing expressions (\ref{Gamma2=})~and~(\ref{deltaGamma2=}),
one can define electron-two-phonon vertex by analogy with the
electron-phonon one.

Iterations of the RG procedure will generate electron
coupling to excitations with larger number of phonons,
hence all these excitations with frequencies
$n\omega_{\Gamma}+n'\omega_{E_1}'$ have to be included in
the low-energy theory separately. Obviously, electron
coupling to excitations of the type $n{\Gamma}+2k{E_1}'$
will have the same matrix structure as coupling
to the $E_2$~phonons, while coupling to excitations
$n{\Gamma}+(2k+1){E_1}'$ -- the same as to the
$E_1'$~phonons. Denoting the dimensionless coupling
constant for the excitation $n\omega_{\Gamma}+n'\omega_{E_1}'$
by $\lambda_{n,n'}$, we can generalize the RG
equations~(\ref{RGflow=}) as follows:
\begin{subequations}
\begin{eqnarray}
\frac{1}{g}\frac{dg}{d\ell}&=&-\frac{8f(g)}{\pi^2\mathcal{N}}
+\sum_{n,k=0}^\infty\frac{\lambda_{n,2k}+\lambda_{n,2k+1}}{2\pi},\\
\frac{d\lambda_{n,2k}}{d\ell}&=&
\frac{1}{2\pi}\sum_{n'=0}^n\sum_{k'=1}^k\lambda_{n',\,2k'-1}\,
\lambda_{n-n',\,2k-2k'+1},\nonumber\\ &&\label{RGflowPh1=}\\
\frac{d\lambda_{n,2k+1}}{d\ell}&=&
\left[\frac{16f(g)}{\pi^2\mathcal{N}}
-\sum_{n',\,k'=0}^\infty\frac{\lambda_{n',2k'}}{\pi}\right]
\lambda_{n,2k+1}
+\nonumber\\&&{}+
\frac{1}{\pi}\sum_{n'=0}^n\sum_{k'=0}^k\lambda_{n',\,2k'}\,
\lambda_{n-n',\,2k-2k'+1}.\nonumber\\ &&\label{RGflowPh2=}
\end{eqnarray}\label{RGflowPh=}
\end{subequations}
These equations can be simply related to Eqs.~(12) of
Ref.~\onlinecite{us},
where $\lambda_{\Gamma}$ and $\lambda_{A_1}$ denoted the total
oscillator strength of all $E_2$-like and all $E_1'$-like excitations,
respectively:
\[
\lambda_{E_2}^{(\mbox{\scriptsize Ref. \onlinecite{us}})}=
\sum_{n,k=0}^\infty\lambda_{n,2k},\quad
\lambda_{A_1}^{(\mbox{\scriptsize Ref. \onlinecite{us}})}=
\sum_{n,k=0}^\infty\lambda_{n,2k+1}.
\]
Namely, Eqs.~(12b), (12c) of Ref.~\onlinecite{us} are obtained
by summing Eqs.~(\ref{RGflowPh1=}), (\ref{RGflowPh2=}) over
$n$~and~$k$.

We will not solve Eqs.~(\ref{RGflowPh=}), since in graphene
the effect of the
electron-phonon interaction turns out to be negligibly small
due to the smallness of $\lambda_{\Gamma},\lambda_K$, as
we mentioned in the beginning of this subsection. In particular,
the modification of the plot in Fig.~\ref{fig:RGflow} would
not be noticeable by naked eye.

\section{Summary}

In this paper we have calculated the frequency-integrated intensities
of two- and four-phonon Raman peaks in disorder-free graphene.
We started by writing down the low-energy hamiltonian
of the interaction of electrons with the crystal vibrations and the
electromagnetic field from pure symmetry considerations; as a result,
we obtained a description of the system in terms of just a few
independent coupling constants, considered to be parameters of the
theory. Another parameter of the theory, introduced phenomenologically,
is the electron scattering rate~$2\gamma$.

First, we analyzed the one-phonon peak at $1580\:\mathrm{cm}^{-1}$
and have shown that the scattering is completely off-resonant: the
intermediate electron and hole states in the whole first Brillouin zone
contribute to the Raman scattering amplitude.
As a result, the intensity of the peak is expected to be insensitive
to most external parameters: polarization, electron concentration,
degree of disorder, etc. However, according to our results, it is
proportional to the fourth power of the excitation frequency.

Then we calculated the intensities $I_{2K}$~and~$I_{2\Gamma}$
of the two-phonon peaks at $2700\:\mathrm{cm}^{-1}$ and
$3250\:\mathrm{cm}^{-1}$, respectively. We have shown that
two-phonon scattering is fully resonant, so that the intermediate
electron and hole states correspond to real particles, propagating
along the quasiclassical trajectories and subject to scattering
processes. As a result, the intensities are determined by the
electron scattering rate. Besides, the quasiclassical character
of the process imposes a severe restriction on the electron
and hole trajectories which can contribute to the two-phonon
Raman scattering: upon the phonon emission the electron and
the hole must be scattered backwards. This restriction results
in a significant polarization memory: it is almost three times more
probable for the scattered photon to have the same polarization
as the incidend photon than to have the orthogonal polarization.

We have also calculated the intensity $I_{4K}$ of the most intense
four-phonon peak at $5400\:\mathrm{cm}^{-1}$. The four-phonon
Raman scattering is also fully resonant. As a consequence, we have
shown that measurement of the ratio $I_{4K}/I_{2K}$ enables one
to extract information about the relative contributions of different
processes to the electron scttering rate.

Having compared the experimental two-phonon peak intensities, we
extracted the ratio of the corresponding electron-phonon coupling
constants. This ratio turned out to be significantly different from
that obtained earlier from the density-functional theory calculations.
We have shown that the reason for this discrepancy is the
renormalization of the coupling constants due to the Coulomb
interaction between electrons, missed by DFT calculations based
on local or semi-local approximations for the exchange-correlation
functional.
In particular, we found that the
constant, responsible for the peak at $3250\:\mathrm{cm}^{-1}$,
is enhanced, and this enhancement is in quantitative agreement
with the experimental Raman data, provided that the screening
of the Coulomb interaction by the substrate is weak.

\section{Acknowledgements}

The author is grateful to I.~L.~Aleiner for numerous discussions
and for collaboration in writing Secs.~\ref{sec:hamiltonian}
and~\ref{sec:1raman} of this paper.
The author also acknowledges stimulating discussions with
J.~Yan, M.~S.~Foster, F.~Guinea, S.~Piscanec, A.~C.~Ferrari,
F.~Mauri, M.~Calandra, and M.~Lazzeri.

\appendix

\section{Matrix algebra}\label{app:algebra}

The natural basis in the 16-dimensional space of all $4\times{4}$
hermitian matrices is represented by the
$\tau_i^{\Kpnt\Kpnt'}\reptimes\tau_j^{AB}$, $i,j=x,y,z,0$, where
$\tau_i^{\Kpnt\Kpnt'},\tau_j^{AB}$, $i,j=x,y,z$, are the Pauli
matrices acting in the corresponding subspaces, respectively, and
$\tau_0^{\Kpnt\Kpnt'},\tau_0^{AB}$ are the $2\times{2}$ unit
matrices. The square of each matrix is equal to~$\unitmatrix$ -- the
$4\times{4}$ unit matrix.

These 16 matrices can be split into two sets: those diagonal and
those off-diagonal in the $\Kpnt\Kpnt'$ subspace (i.~e.,
containing $\tau_0^{\Kpnt\Kpnt'},\tau_z^{\Kpnt\Kpnt'}$ and
$\tau_x^{\Kpnt\Kpnt'},\tau_y^{\Kpnt\Kpnt'}$, respectively). These
two sets are invariant with respect to~$C_{6v}$ because any
transformation from $C_{6v}$ either (i)~leaves
$\Kpnt$~and~$\Kpnt'$ in place, thus belonging to~$C_{3v}$, or
(ii)~swaps between $\Kpnt$~and~$\Kpnt'$ (belonging to
$C_2C_{3v}$), but never mixes them. Thus, these two sets form two
8-dimensional representations. Both are reduced as
$A_1\repplus{A}_2\repplus{B}_1\repplus{B}_2\repplus{E}_1\repplus{E}_2$,
since each set contains a matrix transforming according to~$A_1$:
for the valley-diagonal set it is $\unitmatrix$, while for the
off-diagonal one it is the matrix~$\parity$ of the $C_2$~rotation
which commutes with all $C_{6v}$, and is hermitian since
$C_2^{-1}=C_2$.

Let us focus on the valley-diagonal set. The two matrices which
transform according to $E_1$~(vector) representation will be
denoted by $\Sigma_x,\Sigma_y$. To establish their multiplication
rules, we form the direct product
$E_1\reptimes{E}_1=A_1\repplus{A}_2\repplus{E}_2$. The corresponding linear
combinations are
\begin{eqnarray}
&&\Sigma_x\Sigma_x+\Sigma_y\Sigma_y\sim{A}_1,\\
&&\Sigma_x\Sigma_y-\Sigma_y\Sigma_x\sim{A}_2,\\
&&\{\Sigma_x\Sigma_x-\Sigma_y\Sigma_y,\Sigma_x\Sigma_y+\Sigma_y\Sigma_x\}
\sim{E}_2.
\end{eqnarray}
Since $\Sigma_x^2=\Sigma_y^2=\unitmatrix$, the proportionality
coefficient in the last line must be zero, so
$\Sigma_x\Sigma_y=-\Sigma_y\Sigma_x=i\Sigma_z$. The matrix $\Sigma_z$
defined in this way (i)~is hermitian, (ii)~transforms according
to~$A_2$, and (iii)~$\Sigma_z^2=\unitmatrix$. Thus, it must coincide
with the corresponding matrix from the basis (the sign may need to be
changed). In other words, the set $\{\Sigma_x,\Sigma_y,\Sigma_z\}$
satisfies the usual Pauli matrix algebra.

As the subgroup $C_{2v}=\{E,\sigma_v,\sigma_v',C_2\}$ is a direct
product, the matrices of $B_1,B_2,A_2$ representations of the
valley-diagonal set must commute, and the product of any two will give
the third one. Let us denote by~$\Lambda_z$ the matrix of the
$B_1$~representation, then that of~$B_2$ is $\Lambda_z\Sigma_z$.

The two matrices transforming according to the $E_2$ (pseudovector
or tensor) representation are denoted by
$\{T_x,T_y\}\equiv\vec{T}$. Just like for~$\vec\Sigma$, using
$E_2\reptimes{E}_2=A_1\repplus{A}_2\repplus{E}_2$, we establish
the Pauli matrix algebra for the set $\{T_x,T_y,\Sigma_z\}$. It is
important that matrices of $\sigma_a',\sigma_b',\sigma_c'$
reflections are expressed in terms of $T_x,T_y$. Indeed,
(i)~$\sigma_a'=(\sigma_a')^{-1}$, so its matrix is hermitian and
its square is equal to~$\unitmatrix$; (ii)~this matrix is diagonal
in the $\Kpnt\Kpnt'$~indices;
(iii)~$C_3^{-1}\sigma_a'C_3=\sigma_b'$, so the matrix
of~$\sigma_a'$ cannot belong to a one-dimensional representation;
(iv)~$C_2\sigma_a'C_2=\sigma_a'$, so it must be the
$E_2$~representation. Since we have a rotational arbitrariness in
the choice of $T_x,T_y$, we simply fix $T_y$~to be the matrix of
the $\sigma_a'$~reflection.

The products of $\vec\Sigma$ and $\vec{T}$ can be analyzed by
using $B_1\reptimes{E}_1=B_2\reptimes{E}_1=E_2$. The rotational
arbitrariness in the choice of $\Sigma_x,\Sigma_y$ can be removed
by fixing their behavior under the $\sigma_a'$~reflection:
$T_y\Sigma_xT_y=\Sigma_x$, $T_y\Sigma_yT_y=-\Sigma_y$. The
algebraic relations established earlier leave us with
\begin{equation}\label{STalgebra=}
T_x=-\zeta\Lambda_z\Sigma_y,\quad
T_y=\zeta\Lambda_z\Sigma_x,\quad\zeta^4=1\,.
\end{equation}
The sign of~$\zeta$ is not important as one can always redefine
$\Lambda_z\to-\Lambda_z$. The difference between real and
imaginary~$\zeta$ is essential, as it determines the symmetry
under the time reversal, and will be discussed below.

The off-diagonal set can be obtained from the diagonal one by
simply multiplying it by~$\parity$, the matrix of~$C_2$, which
transforms according to the identical representation of~$C_{6v}$, but
swaps $\Kpnt$~and~$\Kpnt'$. According to the representation algebra,
$\parity$~commutes with $\Sigma_z,T_x,T_y$, and anticommutes with
$\Lambda_z,\Lambda_z\Sigma_z,\Sigma_x,\Sigma_y$. Generally, if two
hermitian operators commute, their product is hermitian, while if they
anticommute, their product is antihermitian, so the proportionality
coefficient in the corresponding algebraic relation must be
imaginary. Hence, the matrices of the off-diagonal set can be written
as $\parity$, $i\parity\Lambda_z$, $\parity\Sigma_z$,
$i\parity\Lambda_z\Sigma_z$, $i\parity\Sigma_x$, $i\parity\Sigma_y$,
$-\parity{T}_x$, $-\parity{T}_y$. Denoting
$\parity\Sigma_z=\Lambda_x$, we recover the bottom row of
Table~\ref{tab:matrices}, provided that $\zeta=1$.

To establish the form of the matrix of the $C_3$ rotation, we note
that (i)~it must be diagonal in the valley subspace, (ii)~commute
with~$\parity$, (ii)~its third power should be equal
to~$\unitmatrix$. This fixes $U_{C_3}=e^{\pm({2}\pi{i}/3)\Sigma_z}$.

Now let us establish the symmetry properties of the matrices with
respect to the time reversal. The explicit form of the corresponding
matrix $U_{\mathcal{T}}$ need not be specified. We only notice that
time reversal must commute with any spatial transformation. Applying
this condition to the matrices $\parity$, $\vec{T}$,
$e^{{2}\pi{i}\Sigma_z/3}$, we obtain
$\parity\mapsto\parity$, $\vec{T}\mapsto\vec{T}$,
$\Sigma_z\mapsto-\Sigma_z$.
Next, the time reversal swaps $\Kpnt$~and~$\Kpnt'$, so
$\Lambda_z\mapsto-\Lambda_z$. Applying the algebraic relations
established above, we obtain
\begin{equation}
\Lambda_z\mathop{\mapsto}_{\mathcal{T}}-\Lambda_z,\quad
\vec\Lambda\mathop{\mapsto}_{\mathcal{T}}-\vec\Lambda,\quad
\Sigma_z\mathop{\mapsto}_{\mathcal{T}}-\Sigma_z,\quad
\vec\Sigma\mathop{\mapsto}_{\mathcal{T}}-\zeta^2\vec\Sigma.
\end{equation}
Next, since the matrices of $\sigma_a,\sigma_b,\sigma_c$
reflections are expressed in terms of $\parity\vec{T}$, we must have
$\Lambda_y\vec\Sigma\mapsto\Lambda_y\vec\Sigma$, which fixes
$\zeta=\pm{1}$.

To help those readers who prefer to work with a particular
representation, rather than basis-independent algebraic relations,
we give specific expressions for the matrices defined above.
In Ref.~\onlinecite{Falko} the representation is introduced by
defining the column state vector as
\begin{equation}
\psi=\left[\begin{array}{c} \psi_{A\Kpnt} \\ \psi_{B\Kpnt} \\
\psi_{B\Kpnt'} \\ \psi_{A\Kpnt'} \end{array}\right],
\end{equation}
and the $\Sigma_i,\Sigma_j'$ matrices have the form:
\begin{eqnarray}
\Sigma_x=\tau_z^{\Kpnt\Kpnt'}\tau^{AB}_x,&&
\Lambda_x=\tau_x^{\Kpnt\Kpnt'}\tau^{AB}_z, \nonumber\\
\Sigma_y=\tau_z^{\Kpnt\Kpnt'}\tau^{AB}_y,&&
\Lambda_y=\tau_y^{\Kpnt\Kpnt'}\tau^{AB}_z, \nonumber\\
\Sigma_z=\tau^{\Kpnt\Kpnt'}_0\tau_z^{AB},&&
\Lambda_z=\tau_z^{\Kpnt\Kpnt'}\tau^{AB}_0,
\end{eqnarray}
where $\tau_i^{\Kpnt\Kpnt'},\tau_j^{AB}$, $i,j=x,y,z$, are the Pauli
matrices acting in the corresponding subspaces, respectively, and
$\tau_0^{\Kpnt\Kpnt'},\tau_0^{AB}$ are the $2\times{2}$ unit
matrices.

In Ref.~\onlinecite{AleinerEfetov} the column state vector is defined
as
\begin{equation}
\psi=\left[\begin{array}{c} \psi_{A\Kpnt} \\ \psi_{B\Kpnt} \\
\psi_{B\Kpnt'} \\ -\psi_{A\Kpnt'} \end{array}\right],
\end{equation}
and it is assumed that
$\bloch_{\Kvec',A(B)}(\vec{r})=\bloch_{\Kvec,A(B)}^*(\vec{r})$.
In this representation the electronic matrices acquire an especially
simple form:
$\Sigma_i=\tau_0^{\Kpnt\Kpnt'}\tau^{AB}_i$,
$\Lambda_i=\tau_i^{\Kpnt\Kpnt'}\tau^{AB}_0$, and the time
reversal matrix
$U_{\mathcal{T}}=\tau_y^{\Kpnt\Kpnt'}\tau^{AB}_y$.

\section{Effective hamiltonian in an external  field}
\label{app:extfield}

The theory of electrons in a crystal lattice subject to a magnetic
field was developed long ago.\cite{Blount} Here we consider the
specific case of the two-dimensional graphene crystal.

Let us start from the simpler case of the scalar
potential~$\varphi(\vec{r},z)$, assumed to be smooth on the
scale of the lattice constant.
The definition~(\ref{Heff=}) of the effective hamiltonian
can be written as
\begin{eqnarray}
&&\int\psi^*_\alpha(\vec{r})\,{H}_{\alpha\beta}^\varphi\,\psi_\beta(\vec{r})\,
d^2\vec{r}=\int\psi^*_\alpha(\vec{r})\,e\varphi(\vec{r},z)\,\psi_\beta(\vec{r})
\times\nonumber\\
&&\qquad{}\times
\left[{e}^{i(\Kvec_\beta-\Kvec_\alpha)\vec{r}}\,
\bloch_\alpha^*(\vec{r},z)\,
\bloch_\beta(\vec{r},z)\right]d^2\vec{r}\,dz.\label{Heffphi=}
\end{eqnarray}
Here $\alpha,\beta=1,2,3,4$ label the four states with zero
energy, $\Kvec_{\alpha,\beta}$ are $\Kvec$ or $\Kvec'$, correspondingly,
and summation over repeating
indices is assumed hereafter, unless stated explicitly otherwise.

The integration over~$z$ in Eq.~(\ref{Heffphi=}) is straightforward
due to confinement provided by the Bloch functions. As for the integration
over $x,y$, we note that the combination of functions in the square
brackets can be represented as
$\delta_{\alpha\beta}\chi_\alpha(z)+\tilde\chi_{\alpha\beta}(\vec{r},z)$,
where the second function (i)~has zero spatial average, and (ii)~is
periodic in $x,y$ with the period corresponding to the tripled unit
cell. The  first term is written using the normalization~(\ref{Blochnorm=})
of the Bloch functions (no summation over~$\alpha$ is assumed).
The rest of the integrand,
$\psi^*_\alpha(\vec{r})\,\varphi(\vec{r},z)\,\psi_\beta(\vec{r})$, is
a smooth function of~$x,y$, i.~e. its spatial harmonics have wave vectors
much smaller than the inverse lattice constant.

To study the integral of an arbitrary smooth function~$f(\vec{r})$ with the
periodic function $\tilde\chi_{\alpha\beta}(\vec{r},z)$, we expand the
latter in the Fourier sum over the reciprocal lattice vectors~$\vec{G}$:
\begin{equation}
\tilde\chi_{\alpha\beta}(\vec{r},z)=
\sum_{\vec{G}}\sum_{j=0,\pm{1}}C_{\alpha\beta}^j(\vec{G},z)
\,e^{i(\vec{G}+j\Kvec)\vec{r}}.
\end{equation}
Then the integral can be rewritten in the Fourier space:
\begin{equation}
\int{f}(\vec{r})\,\tilde\chi_{\alpha\beta}(\vec{r},z)\,{d}^2\vec{r}=
\sum_{\vec{G}}\sum_{j=0,\pm{1}}\tilde{f}^*(\vec{G}+j\Kvec)\,
C_{\alpha\beta}^j(\vec{G},z).
\label{smoothbloch=}
\end{equation}
The sum does not contain the term with $\vec{G}=0$, $j=0$, which
has been excluded from $\tilde\chi_{\alpha\beta}(\vec{r},z)$
by construction.
The Fourier transform of $f(\vec{r})$,
\begin{equation}
\tilde{f}(\vec{p})=\int{e}^{-i\vec{p}\vec{r}}f(\vec{r})\,d^2\vec{r},
\end{equation}
is rapidly decaying away from~$\vec{p}=0$. If all spatial
derivatives of $f(\vec{r})$ are continuous, this decay is exponential.
This leads to the following expression for the effective hamiltonian:
\begin{subequations}\begin{eqnarray}
&&H^\varphi_{\alpha\beta}=
e\varphi(\vec{r},z=0)\delta_{\alpha\beta}
-d_z\mathcal{E}_z(\vec{r},z=0)\delta_{\alpha\beta},\\
&&d_z=\int{ez}\,|\bloch_\alpha(\vec{r},z)|^2\,\frac{d^2\vec{r}\,dz}{L_xL_y},
\end{eqnarray}\end{subequations}
where $\mathcal{E}_z=-\partial_z\varphi$. The fact that $d_z$~does not
depend on~$\alpha$ follows from the transformation properties of the
Bloch functions under the $\sigma_v,\sigma_v'$ reflections.
The matrix element between
states with $\Kvec_\alpha\neq\Kvec_\beta$ is small as $e^{-1/(pa)}$,
and thus cannot be included in the regular expansion of the effective
hamiltonian in the parameter $pa\ll{1}$.

\begin{widetext}

To describe the effect of an external magnetic field, we introduce
the vector potential $\vec{A}(\vec{r},z)$ in the gauge $A_z=0$.
The microscopic hamiltonian is given by
\begin{eqnarray}
&&\mathcal{H}^A=-\frac{e}{2c}
\left[A_i(\vec{r},z)\,\hat{v}_i+\hat{v}_i\,A_i(\vec{r},z)\right]
+\frac{e^2}{2mc^2}\,\vec{A}^2(\vec{r},z)
,\quad
\hat{\vec{v}}\equiv\frac{-i\vec\nabla}{m}.
\end{eqnarray}
In the first order of the $\vec{k}\cdot\vec{p}$ perturbation theory
we obtain
\begin{eqnarray}
&&-\frac{e}{2c}\int\psi^*_\alpha(\vec{r})\,\bloch_\alpha^*(\vec{r},z)\,
{e}^{-i\Kvec_\alpha\vec{r}}
\left[A_i(\vec{r},z)\,\hat{v}_i+\hat{v}_iA_i(\vec{r},z)\,\right]
e^{i\Kvec_\beta\vec{r}}\,\bloch_\beta(\vec{r},z)\,\psi_\beta(\vec{r})\,
d^2\vec{r}\,dz=\nonumber\\
&&=-\frac{e}{c}\,\langle\alpha|\hat{v}_i|\beta\rangle
\int{d}^2\vec{r}\,\psi^*_\alpha(\vec{r})\,A_i(\vec{r},0)\,\psi_\beta(\vec{r})
-\frac{e}{2c}\,\delta_{\alpha\beta}
\int{d}^2\vec{r}\,\psi^*_\alpha(\vec{r})
\left[A_i(\vec{r},0)\,\hat{v}_i+\hat{v}_iA_i(\vec{r},0)\,\right]
\psi_\beta(\vec{r})-\nonumber\\
&&\quad{}
-\frac{e}{c}\,\langle\alpha|z\hat{v}_i|\beta\rangle
\int{d}^2\vec{r}\,\psi^*_\alpha(\vec{r})\,
\frac{\partial{A}_i(\vec{r},0)}{\partial{z}}\,\psi_\beta(\vec{r})
,\label{Akp1xy=}
\end{eqnarray}
where we introduced the following notation for the matrix elements of
an arbitrary operator~$\mathcal{O}$ between the Bloch functions:
\begin{equation}
\langle\alpha|\mathcal{O}|\beta\rangle=
\int{e}^{-i\Kvec_\alpha\vec{r}}\,
\bloch_\alpha^*(\vec{r},z)\,\mathcal{O}\,e^{i\Kvec_\beta\vec{r}}
\,\bloch_\beta(\vec{r},z)\,\frac{d^2\vec{r}\,dz}{L_xL_y}.
\end{equation}
We again encounter integrals of smooth functions with periodic ones,
such as $\bloch_\alpha^*{e}^{-i\Kvec_\alpha\vec{r}}%
\hat{v}_ie^{i\Kvec_\beta\vec{r}}\,\bloch_\beta$, so the matrix
elements are different from zero only if $\Kvec_\alpha=\Kvec_\beta$.
Recalling that in the first order of the $\vec{k}\cdot\vec{p}$ perturbation
theory the Dirac hamiltonian is obtained as
\begin{equation}
[H_1(\vec{p})]_{\alpha\beta}=
\langle\alpha|\hat{\vec{v}}|\beta\rangle\cdot\vec{p}=
v\vec\Sigma_{\alpha\beta}\cdot\vec{p},
\end{equation}
the first term on the right-hand side of Eq.~(\ref{Akp1xy=}) can be
identified with the gauge elongation $\vec{p}\to-i\vec\nabla-(e/c)\vec{A}$
in the Dirac hamiltonian, Eq.~(\ref{Heem=}).

In the second order of the
$\vec{k}\cdot\vec{p}$ perturbation theory one has to include
the contribution of remote bans $b\neq\alpha\;\forall\alpha=1,2,3,4$
with energies~$E_b$, taken at the points $\Kvec,\Kvec'$.
The correction to the Dirac hamiltonian is then given by
\begin{eqnarray}
&&[H_2(\vec{p})]_{\alpha\beta}
=\frac{p^2}{2m}\,\delta_{\alpha\beta}
+p_ip_j{\sum_b}^\prime
\frac{\langle\alpha|\hat{v}_i|b\rangle\langle{b}|\hat{v}_j|\beta\rangle
+(i\leftrightarrow{j})}{2E_b}.
\label{Hkp2=}
\end{eqnarray}
The gauge elongation $\vec{p}\to-i\vec\nabla-(e/c)\vec{A}$
of the first term on the right-hand side corresponds to the second
term on the right-hand side of Eq.~(\ref{Akp1xy=}) and the $\vec{A}^2$
term of~$\mathcal{H}^A$.
The effective mass tensor originating from remote bands in
Eq.~(\ref{Hkp2=}) is symmetrized with respect to~$i,j$ in
order to ensure the hermiticity of the effective hamiltonian upon
the gauge elongation.
Taking $\mathcal{H}^A$ in the first order
(the contribution of the $\Div\vec{A}$ vanishes, as it is a smooth function),
and $\vec{p}_i\hat{v}_i$
in the first order, we obtain the following contribution of the remote
bands to the effective hamiltonian:
\begin{eqnarray}
&&-\frac{e}{c}\,{\sum_b}^\prime
\frac{\langle\alpha|\hat{v}_i|b\rangle\langle{b}|\hat{v}_j|\beta\rangle}{E_b}\,
\left[A_i(\vec{r},0)\,m\hat{v}_j+m\hat{v}_i\,A_j(\vec{r},0)\right].
\end{eqnarray}
The $ij$-symmetric part 
of this expression corresponds to the
gauge elongation of the last term in Eq.~(\ref{Hkp2=}), while
the antisymmetric part can be rewritten as
\begin{eqnarray}
&&-\frac{em}{2c}\,{\sum_b}^\prime
\frac{\langle\alpha|\hat{v}_i|b\rangle\langle{b}|\hat{v}_j|\beta\rangle
-(i\leftrightarrow{j})}{E_b}\,
\left[A_i(\vec{r},0)\,\hat{v}_j+\hat{v}_i\,A_j(\vec{r},0)\right]=\nonumber\\
&&=\frac{ie}{4c}\,{\sum_b}^\prime
\frac{\langle\alpha|\hat{v}_i|b\rangle\langle{b}|\hat{v}_j|\beta\rangle
-(i\leftrightarrow{j})}{E_b}
\left[\frac{\partial{A}_i(\vec{r},0)}{\partial{x}_j}
-\frac{\partial{A}_j(\vec{r},0)}{\partial{x}_i}\right].
\end{eqnarray}
The expression in the square brackets is an antisymmetric tensor
whose $xy$ and $yx$ components are equal to $\mathcal{B}_z$
and $-\mathcal{B}_z$, respectively. Thus, this term corresponds to
$-\mu_z\mathcal{B}_z\Sigma_z$ term in Eq.~(\ref{Heemp=}).

Since $\mathcal{B}_x=-\partial{A_y}/\partial{z}$,
$\mathcal{B}_y=\partial{A_x}/\partial{z}$, the last term in
Eq.~(\ref{Akp1xy=}) corresponds to the
$\mu_{xy}(\mathcal{B}_x\Sigma_y-\mathcal{B}_y\Sigma_x)$
term in Eq.~(\ref{Heemp=}).
Using the facts that (i)~$\langle\alpha|z|\beta\rangle=(d_z/e)\delta_{\alpha\beta}$,
(ii)~$z\hat{v}_i=\hat{v}_iz$,
(iii)~$\langle\alpha|z|b\rangle=(i/2E_b)\langle\alpha|\hat{v}_z|b\rangle$
(the latter follows from the commutation relation $[\mathcal{H},z]=-(i/2)\hat{v}_z$,
valid in the absence of the field), we can write
\begin{equation}
\langle\alpha|z\hat{v}_i|\beta\rangle=
\frac{d_z}e\,\langle\alpha|\hat{v}_i|\beta\rangle+\frac{i}{2}\,{\sum_b}^\prime
\frac{\langle\alpha|\hat{v}_z|b\rangle\langle{b}|\hat{v}_i|\beta\rangle}{E_b}=
\frac{d_z}e\,\langle\alpha|\hat{v}_i|\beta\rangle-\frac{i}{2}\,{\sum_b}^\prime
\frac{\langle\alpha|\hat{v}_i|b\rangle\langle{b}|\hat{v}_z|\beta\rangle}{E_b}.
\end{equation}
The contribution of the
$(d_z/e)\,\langle\alpha|\hat{v}_i|\beta\rangle$ term to the effective
hamiltonian has the form
$-d_z(v\Sigma_x\mathcal{B}_y-v\Sigma_y\mathcal{B}_x)$ and may be viewed as
the Lorentz force contribution to the $-d_z\mathcal{E}_z$ term.



\section{Kinetic equation}\label{app:kinur}

Kinetic equation is the most natural way to describe the dynamics
when real quasiparticles (electrons and holes) move along quasiclassical
trajectories, and interference between different trajectories is
suppressed. Since electrons and holes are
created in pairs, the kinetic equation should be written for the joint
distribution function
$\df_{\vec{p}_e,\vec{p}_h}(\vec{r}_e,\vec{r}_h;t)$ --
the joint probability for the electron to be in the elementary
volume of phase space $d^2\vec{r}_e\,d^2\vec{p}_e/(2\pi)^2$
around $\vec{r}_e,\vec{p}_e$ and for the hole in the elementary
volume $d^2\vec{r}_h\,d^2\vec{p}_h/(2\pi)^2$ around
$\vec{r}_h,\vec{p}_h$. We write the kinetic equation as
\begin{subequations}\begin{eqnarray}
&&\frac{\partial\df}{\partial{t}}
+\frac{\partial\Re\xi_{\vec{p}_e}}{\partial\vec{p}_e}
\frac{\partial\df}{\partial\vec{r}_e}
+\frac{\partial\Re\bar\xi_{-\vec{p}_h}}{\partial\vec{p}_h}
\frac{\partial\df}{\partial\vec{r}_h}
=\Stoss^{e,out}_{\vec{p}_e,\vec{p}_h}f
+\Stoss^{e,in}_{\vec{p}_e,\vec{p}_h}f
+\Stoss^{h,out}_{\vec{p}_e,\vec{p}_h}f
+\Stoss^{h,in}_{\vec{p}_e,\vec{p}_h}f,\\
&&\Stoss^{e,out}_{\vec{p}_e,\vec{p}_h}f=
-2\pi\int\frac{d^2\vec{q}}{(2\pi)^2}\,
\frac{F^2_K}{M\omega_\vec{q}}\frac{\sqrt{27a^2}}{4}\,
\sin^2\frac{\varphi_{\vec{p}_e}-\varphi_{\vec{p}_e-\vec{q}}}2\,
\delta(\Re\xi_{\vec{p}_e}-\Re\xi_{\vec{p}_e-\vec{q}}-\omega_{\vec{q}})
\,\df_{\vec{p}_e,\vec{p}_h},\\
&&\Stoss^{e,in}_{\vec{p}_e,\vec{p}_h}f=
2\pi\int\frac{d^2\vec{q}}{(2\pi)^2}\,
\frac{F^2_K}{M\omega_\vec{q}}\frac{\sqrt{27a^2}}{4}\,
\sin^2\frac{\varphi_{\vec{p}_e+\vec{q}}-\varphi_{\vec{p}_e}}2\,
\delta(\Re\xi_{\vec{p}_e+\vec{q}}-\Re\xi_{\vec{p}_e}-\omega_{\vec{q}})\,
\df_{\vec{p}_e+\vec{q},\vec{p}_h},\\
&&\Stoss^{h,out}_{\vec{p}_e,\vec{p}_h}f=
-2\pi\int\frac{d^2\vec{q}}{(2\pi)^2}\,
\frac{F^2_K}{M\omega_\vec{q}}\frac{\sqrt{27a^2}}{4}\,
\sin^2\frac{\varphi_{-\vec{p}_h}-\varphi_{-\vec{p}_h+\vec{q}}}2\,
\delta(\Re\bar\xi_{-\vec{p}_h}
-\Re\bar\xi_{-\vec{p}_h+\vec{q}}-\omega_{\vec{q}})
\,\df_{\vec{p}_e,\vec{p}_h},\\
&&\Stoss^{h,in}_{\vec{p}_e,\vec{p}_h}f=
2\pi\int\frac{d^2\vec{q}}{(2\pi)^2}\,
\frac{F^2_K}{M\omega_\vec{q}}\frac{\sqrt{27a^2}}{4}\,
\sin^2\frac{\varphi_{-\vec{p}_h-\vec{q}}-\varphi_{-\vec{p}_h}}2\,
\delta(\Re\bar\xi_{-\vec{p}_h-\vec{q}}
-\Re\bar\xi_{-\vec{p}_h}-\omega_{\vec{q}})\,
\df_{\vec{p}_e,\vec{p}_h+\vec{q}}.
\end{eqnarray}\label{kinetic=}\end{subequations}
The left-hand side of the kinetic equation (Liouville operator
acting on the distribution function) represents the free propagation
of the electron and the hole with the corresponding (group) velocities
$\vec{v}_e=\partial\Re\xi_{\vec{p}_e}/\partial\vec{p}_e$ and
$\vec{v}_h=\partial\Re\bar\xi_{-\vec{p}_h}/\partial\vec{p}_h$.
The right-hand side (collision integral) describes emission of
phonons; assuming to be in the linear regime, we have
neglected the Fermi statistics of electrons and holes.
In what follows, we will write the out-scattering part of
the collision integral as
$-2(\gamma_{\vec{p}_e}+\bar\gamma_{-\vec{p}_h})
\df_{\vec{p}_e,\vec{p}_h}$, and include it in the
Liouville operator. In the scattering rates
$2\gamma_{\vec{p}_e},2\bar\gamma_{-\vec{p}_h}$
we also include those for emission of electron-hole
pairs, or other excitations with broad spectrum; the
contribution of such processes to the in-scattering part of
the collision integral is neglected (see discussion in
Sec.~\ref{sec:fullresonance}).

The Raman signal is treated as a weak probe not affecting
the electron and hole population, and thus is not included
into the kinetic equation. It is determined by the radiative
recombination rate, and can be calculated from the Fermi
Golden Rule. The probability of emission of a photon with
a given polarization~$\vec{e}_{out}$ per unit solid angle, per
unit frequency interval, and {\em per unit time} is expressed
in terms of the joint distribution function as
\begin{equation}
\frac{4\pi\:dI}{do_{out}\,d\omega_{out}\,dt}=2\pi
\left(\frac{ev}c\right)^2\frac{2\pi{c}^2}{\omega_{out}}
\frac{\omega_{out}^2}{2\pi^2c^3}
\int{d}^2\vec{r}\,\frac{d^2\vec{p}}{(2\pi)^2}\,
\delta(\omega_{out}-\Re\xi_{\vec{p}}-\Re\bar\xi_{\vec{p}})\,
|[\vec{e}_{out}\times\vec{e}_{\vec{p}}]_z|^2
\df_{\vec{p},-\vec{p}}(\vec{r},\vec{r}).
\end{equation}
The easiest way to arrive at this expression is to calculate the
electronic radiative self-energy:
\begin{equation}
\Sigma^{\mathrm{rad}}(\vec{p},\ep)=
i\left(\frac{ev}c\right)^2\int\frac{d^3\vec{Q}}{(2\pi)^3}
\frac{d\Omega}{2\pi}\frac{2\pi{c}^2}{cQ}
\frac{2cQ}{\Omega^2-(cQ-io)^2}
\,(\vec{e}_{out}\cdot\vec\Sigma)\,
\frac{\ep-\Omega+v\vec{p}\cdot\vec\Sigma}
{(\ep-\Omega+io)^2-(vp)^2}
\,(\vec{e}_{out}^*\cdot\vec\Sigma).
\label{Sigmarad=}
\end{equation}
Note that the imaginary shift of the poles in the electron
Green's function is different from the prescription~(\ref{Gpe=}).
Indeed, the latter corresponds to the full valence band and empty
conduction band, while radiative recombination requires the
valence band to be empty, hence the shift of both poles to the
lower half-plane of~$\ep$ in Eq.~(\ref{Sigmarad=}).

Kinetic equation~(\ref{kinetic=}) contains no term corresponding
to the generation of electron-hole pairs by incoming photons.
Instead, we prefer to choose the initial condition at $t=0$,
which would correspond to the electron-hole population upon
arrival of a short pulse of the electromagnetic field of the
total energy~$\omega_{in}$, i.~e., containing a single photon:
\begin{equation}
\df_{\vec{p}_e,\vec{p}_h}(\vec{r}_e,\vec{r}_h;t=0)
=\frac{\pi{e}^2}c\,
2|[\vec{e}_{in}\times\vec{e}_{\vec{p}_e}]_z|^2
\frac{\delta(\vec{r}_e-\vec{r}_h)}{L_xL_y}\,
\frac{8\pi{v}^2}{\omega_{in}}\,
\delta(\Re\xi_{\vec{p}_e}+\Re\bar\xi_{-\vec{p}_h}-\omega_{in})\,
(2\pi)^2\delta(\vec{p}_e+\vec{p}_h).\label{ft=0}
\end{equation}
The last $\delta$-function takes care of momentum conservation
during photon absorption: since the photon momentum is very small,
the electron and the hole must have opposite momenta.
The energy $\delta$-function ensures that the total energy of the
electron-hole pair is equal to the energy of the absorbed photon;
the coefficient in front of it is just the inverse density of states
of electron-hole pairs with zero total momentum, necessary to
preserve the normalization of
$\df_{\vec{p}_e,\vec{p}_h}(\vec{r}_e,\vec{r}_h)$.
The factor $\delta(\vec{r}_e-\vec{r}_h)/(L_xL_y)$ reflects the
fact that the electron and the hole are born at the same spatial
point, which can be located anywhere in the sample. The
angular factor $2|[\vec{e}_{in}\times\vec{e}_{\vec{p}}]_z|^2$
shows that the transition dipole is perpendicular to the electron
momentum, and the factor of~2 fixes the average to unity.
As a result, the integral of
$\df_{\vec{p}_e,\vec{p}_h}(\vec{r}_e,\vec{r}_h)$ over the whole
phase space equals $\pi{e}^2/c$, which is
nothing else but the total probability for the incident photon to
be absorbed; all subsequent factors, discussed above, represent
the partitioning of this probability over different states of the
electron-hole pairs.
%
The total probability can be found directly from the Fermi Golden Rule,
and is given by
\[
\frac{L_z}c\,4\sum_\vec{p}
\left|\frac{ev}c\sqrt{\frac{2\pi{c}^2}{V\omega_{in}}}\right|^2
2\pi\delta(\Re\xi_\vec{p}+\Re\bar\xi_\vec{p}-\omega_{in}),
\]
where the factor of~$4$ keeps track of the valley and spin degeneracy,
and $L_z/c$ is the attempt period.

Solution of the kinetic equation~(\ref{kinetic=}), corresponding
to emission of~$n$ phonons, is obtained by $n$~iterations of the
collision integral. First of all, we find the inverse of the
Liouville operator, acting on a source
$J_{\vec{p}_e,\vec{p}_h}(\vec{r}_e,\vec{r}_h;t)$:
\begin{equation}
\left[\frac\partial{\partial{t}}+2(\gamma_e+\gamma_h)
+\vec{v}_e\frac\partial{\partial\vec{r}_e}
+\vec{v}_h\frac\partial{\partial\vec{r}_h}\right]^{-1}
J_{\vec{p}_e,\vec{p}_h}(\vec{r}_e,\vec{r}_h;t)
=\int\limits_{-\infty}^tdt'\,
J_{\vec{p}_e,\vec{p}_h}(\vec{r}_e-\vec{v}_e(t-t'),%
\vec{r}_h-\vec{v}_h(t-t');t')\,
e^{-2(\gamma_e+\gamma_h)(t-t')}.
\end{equation}

Let us follow evolution of a single electron-hole pair, created
at $t=0$ at the point $\vec{r}=0$ (evolution of the initial
condition~(\ref{ft=0}) can be obtained by a simple convolution).
The zero-approximation distribution function is given by:
\begin{equation}
\df^{(0)}_{\vec{p}_e,\vec{p}_h}(\vec{r}_e,\vec{r}_h;t)
=(2\pi)^2\delta(\vec{p}_e-\vec{p}_0)\,
(2\pi)^2\delta(\vec{p}_h+\vec{p}_0)\,
\delta(\vec{r}_e-\vec{v}_et)\,\delta(\vec{r}_h-\vec{v}_ht)\,
e^{-2\gamma_{\vec{p}_0}t-2\bar\gamma_{\vec{p}_0}t}.
\end{equation}
After one iteration of the in-scattering part of the collision
integral we obtain the first-approximation correction -- the
contribution from electrons and holes which have emitted one phonon:
\begin{subequations}\begin{eqnarray}
\df^{(1)}_{\vec{p}_e,\vec{p}_h}(\vec{r}_e,\vec{r}_h;t)
&=&\df^{(e)}_{\vec{p}_e,\vec{p}_h}(\vec{r}_e,\vec{r}_h;t)
+\df^{(h)}_{\vec{p}_e,\vec{p}_h}(\vec{r}_e,\vec{r}_h;t)\,,\\
\df^{(e)}_{\vec{p}_e,\vec{p}_h}(\vec{r}_e,\vec{r}_h;t)
&=&(2\pi)^d\delta(\vec{p}_h+\vec{p}_0)\,
\delta(\vec{r}_h+\bar{\vec{v}}_0t)\,e^{-2\bar\gamma_{\vec{p}_0}t}
\times\nonumber\\
&&{}\times\frac{2\pi{F}^2_K}{M\omega_{\vec{p}_0-\vec{p}_e}}
\frac{\sqrt{27a^2}}{4}\,
\sin^2\frac{\varphi_{\vec{p}_0}-\varphi_{\vec{p}_e}}2\,
\delta(\Re\xi_{\vec{p}_0}-\Re\xi_{\vec{p}_e}
-\omega_{\vec{p}_0-\vec{p}_e})
\times\nonumber\\ &&{}\times
\int\limits_0^tdt_0\,\delta(\vec{r}_e-\vec{v}_e(t-t_0)-\vec{v}_0t_0)\,
e^{-2\gamma_{\vec{p}_{e}}(t-t_0)-2\gamma_{\vec{p}_0}t_0},\\
\df^{(h)}_{\vec{p}_e,\vec{p}_h}(\vec{r}_e,\vec{r}_h;t)&=&
(2\pi)^d\delta(\vec{p}_e-\vec{p}_0)\,\delta(\vec{r}_e-\vec{v}_0t)\,
e^{-2\gamma_{\vec{p}_0}t}
\times\nonumber\\ &&{}\times
\frac{2\pi{F}^2_K}{M\omega_{\vec{p}_0-\vec{p}_e}}
\frac{\sqrt{27a^2}}{4}\,
\sin^2\frac{\varphi_{\vec{p}_0}-\varphi_{-\vec{p}_h}}2\,
\delta(\Re\bar\xi_{\vec{p}_0}-\Re\bar\xi_{-\vec{p}_h}
-\omega_{-\vec{p}_0-\vec{p}_h})
\times\nonumber\\ &&{}\times
\int\limits_0^td\bar{t}_0\,
\delta(\vec{r}_h-\vec{v}_h(t-\bar{t}_0)+\bar{\vec{v}}_0\bar{t}_0)\,
e^{-2\bar\gamma_{-\vec{p}_h}(t-\bar{t}_0)
-2\bar\gamma_{\vec{p}_0}\bar{t}_0}.
\end{eqnarray}\end{subequations}
After the second iteration we have\footnote{
Upon applying the operator $\Stoss^{e,in}+\Stoss^{h,in}$
twice one would expect a binomial type expression:
$\df^{(2)}=\df^{(ee)}+\df^{(eh)}+\df^{(he)}+\df^{(hh)}$.
However, in this form $\df^{(eh)}$ and $\df^{(he)}$
correspond to different ordering of $t_0$ and $\bar{t}_0$.
We prefer to join the terms without assuming any relation
between emission events for the electron and the hole.}
\begin{subequations}\begin{eqnarray}
\df^{(2)}_{\vec{p}_e,\vec{p}_h}(\vec{r}_e,\vec{r}_h;t)
&=&\df^{(ee)}_{\vec{p}_e,\vec{p}_h}(\vec{r}_e,\vec{r}_h;t)
+\df^{(eh)}_{\vec{p}_e,\vec{p}_h}(\vec{r}_e,\vec{r}_h;t)
+\df^{(hh)}_{\vec{p}_e,\vec{p}_h}(\vec{r}_e,\vec{r}_h;t)\,,\\
\df^{(ee)}_{\vec{p}_e,\vec{p}_h}(\vec{r}_e,\vec{r}_h;t)
&=&(2\pi)^d\delta(\vec{p}_h+\vec{p}_0)\,
\delta(\vec{r}_h+\bar{\vec{v}}_0t)\,e^{-2\bar\gamma_{\vec{p}_0}t}
\times\nonumber\\&&{}\times
\int\frac{d^2\vec{p}_1}{(2\pi)^2}\,
\frac{2\pi{F}^2_K}{M\omega_{\vec{p}_0-\vec{p}_1}}
\frac{\sqrt{27a^2}}{4}\,
\sin^2\frac{\varphi_{\vec{p}_0}-\varphi_{\vec{p}_1}}2\,
\delta(\Re\xi_{\vec{p}_0}-\Re\xi_{\vec{p}_1}
-\omega_{\vec{p}_0-\vec{p}_1})
\times\nonumber\\&&{}\times
\frac{2\pi{F}^2_K}{M\omega_{\vec{p}_1-\vec{p}_e}}
\frac{\sqrt{27a^2}}{4}\,
\sin^2\frac{\varphi_{\vec{p}_1}-\varphi_{\vec{p}_e}}2\,
\delta(\Re\xi_{\vec{p}_1}-\Re\xi_{\vec{p}_e}
-\omega_{\vec{p}_1-\vec{p}_e})
\times\nonumber\\ &&{}\times
\int\limits_0^tdt_1\int\limits_0^{t_1}dt_0\,
\delta(\vec{r}_e-\vec{v}_e(t-t_1)-\vec{v}_1(t_1-t_0)-\vec{v}_0t_0)
e^{-2\gamma_{\vec{p}_e}(t-t_1)-2\gamma_{\vec{p}_1}(t_1-t_0)
-2\gamma_{\vec{p}_0}t_0},\\
\df^{(eh)}_{\vec{p}_e,\vec{p}_h}(\vec{r}_e,\vec{r}_h;t)&=&
\frac{2\pi{F}^2_K}{M\omega_{\vec{p}_0-\vec{p}_e}}
\frac{\sqrt{27a^2}}{4}\,
\sin^2\frac{\varphi_{\vec{p}_0}-\varphi_{\vec{p}_e}}2\,
\delta(\Re\xi_{\vec{p}_0}-\Re\xi_{\vec{p}_e}
-\omega_{\vec{p}_0-\vec{p}_e})
\times\nonumber\\ &&{}\times
\int\limits_0^tdt_0\,\delta(\vec{r}_e-\vec{v}_e(t-t_0)-\vec{v}_0t_0)\,
e^{-2\gamma_{\vec{p}_{e}}(t-t_0)-2\gamma_{\vec{p}_0}t_0}
\times\nonumber\\ &&{}\times
\frac{2\pi{F}^2_K}{M\omega_{\vec{p}_0-\vec{p}_e}}
\frac{\sqrt{27a^2}}{4}\,
\sin^2\frac{\varphi_{\vec{p}_0}-\varphi_{-\vec{p}_h}}2\,
\delta(\Re\bar\xi_{\vec{p}_0}-\Re\bar\xi_{-\vec{p}_h}
-\omega_{-\vec{p}_0-\vec{p}_h})
\times\nonumber\\ &&{}\times
\int\limits_0^td\bar{t}_0\,
\delta(\vec{r}_h-\vec{v}_h(t-\bar{t}_0)+\bar{\vec{v}}_0\bar{t}_0)\,
e^{-2\bar\gamma_{-\vec{p}_h}(t-\bar{t}_0)
-2\bar\gamma_{\vec{p}_0}\bar{t}_0},\\
\df^{(hh)}_{\vec{p}_e,\vec{p}_h}(\vec{r}_e,\vec{r}_h;t)
&=&(2\pi)^d\delta(\vec{p}_e-\vec{p}_0)\,
\delta(\vec{r}_e-{\vec{v}}_0t)\,e^{-2\gamma_{\vec{p}_0}t}
\times\nonumber\\&&{}\times
\int\frac{d^2\bar{\vec{p}}_1}{(2\pi)^2}\,
\frac{2\pi{F}^2_K}{M\omega_{\vec{p}_0-\vec{p}_1}}
\frac{\sqrt{27a^2}}{4}\,
\sin^2\frac{\varphi_{\vec{p}_0}-\varphi_{\bar{\vec{p}}_1}}2\,
\delta(\Re\bar\xi_{\vec{p}_0}
-\Re\bar\xi_{\bar{\vec{p}}_1}
-\omega_{-\vec{p}_0+\bar{\vec{p}}_1})
\times\nonumber\\&&{}\times
\frac{2\pi{F}^2_K}{M\omega_{-\bar{\vec{p}}_1-\vec{p}_h}}
\frac{\sqrt{27a^2}}{4}\,
\sin^2\frac{\varphi_{\bar{\vec{p}}_1}-\varphi_{-\vec{p}_h}}2\,
\delta(\Re\bar\xi_{\bar{\vec{p}}_1}-\Re\bar\xi_{-\vec{p}_h}
-\omega_{-\bar{\vec{p}}_1-\vec{p}_h})
\times\nonumber\\ &&{}\times
\int\limits_0^td\bar{t}_1\int\limits_0^{\bar{t}_1}d\bar{t}_0\,
\delta(\vec{r}_h-\vec{v}_h(t-\bar{t}_1)
+\bar{\vec{v}}_1(\bar{t}_1-\bar{t}_0)+\bar{\vec{v}}_0\bar{t}_0)
e^{-2\bar\gamma_{-\vec{p}_h}(t-\bar{t}_1)
-2\bar\gamma_{\bar{\vec{p}}_1}(\bar{t}_1-\bar{t}_0)
-2\bar\gamma_{\vec{p}_0}\bar{t}_0}.
\end{eqnarray}\end{subequations}
After four iterations we have $\df^{(4)}=\df^{(eeee)}
+\df^{(eeeh)}+\df^{(eehh)}+\df^{(ehhh)}+\df^{(hhhh)}$. The
term contributing to the four-phonon Raman signal is given by
\begin{eqnarray}
\df^{(eehh)}_{\vec{p}_e,\vec{p}_h}(\vec{r}_e,\vec{r}_h;t)
&=&\int\frac{d^2\vec{p}_1}{(2\pi)^2}\,
\frac{2\pi{F}^2_K}{M\omega_{\vec{p}_0-\vec{p}_1}}
\frac{\sqrt{27a^2}}{4}\,
\sin^2\frac{\varphi_{\vec{p}_0}-\varphi_{\vec{p}_1}}2\,
\delta(\Re\xi_{\vec{p}_0}-\Re\xi_{\vec{p}_1}
-\omega_{\vec{p}_0-\vec{p}_1})
\times\nonumber\\&&{}\times
\frac{2\pi{F}^2_K}{M\omega_{\vec{p}_1-\vec{p}_e}}
\frac{\sqrt{27a^2}}{4}\,
\sin^2\frac{\varphi_{\vec{p}_1}-\varphi_{\vec{p}_e}}2\,
\delta(\Re\xi_{\vec{p}_1}-\Re\xi_{\vec{p}_e}
-\omega_{\vec{p}_1-\vec{p}_e})
\times\nonumber\\ &&{}\times
\int\limits_0^tdt_1\int\limits_0^{t_1}dt_0\,
\delta(\vec{r}_e-\vec{v}_e(t-t_1)-\vec{v}_1(t_1-t_0)-\vec{v}_0t_0)
e^{-2\gamma_{\vec{p}_e}(t-t_1)-2\gamma_{\vec{p}_1}(t_1-t_0)
-2\gamma_{\vec{p}_0}t_0}
\times\nonumber\\&&{}\times
\int\frac{d^2\bar{\vec{p}}_1}{(2\pi)^2}\,
\frac{2\pi{F}^2_K}{M\omega_{\vec{p}_0-\vec{p}_1}}
\frac{\sqrt{27a^2}}{4}\,
\sin^2\frac{\varphi_{\vec{p}_0}-\varphi_{\bar{\vec{p}}_1}}2\,
\delta(\Re\bar\xi_{\vec{p}_0}
-\Re\bar\xi_{\bar{\vec{p}}_1}
-\omega_{-\vec{p}_0+\bar{\vec{p}}_1})
\times\nonumber\\&&{}\times
\frac{2\pi{F}^2_K}{M\omega_{-\bar{\vec{p}}_1-\vec{p}_h}}
\frac{\sqrt{27a^2}}{4}\,
\sin^2\frac{\varphi_{\bar{\vec{p}}_1}-\varphi_{-\vec{p}_h}}2\,
\delta(\Re\bar\xi_{\bar{\vec{p}}_1}-\Re\bar\xi_{-\vec{p}_h}
-\omega_{-\bar{\vec{p}}_1-\vec{p}_h})
\times\nonumber\\ &&{}\times
\int\limits_0^td\bar{t}_1\int\limits_0^{\bar{t}_1}d\bar{t}_0\,
\delta(\vec{r}_h-\vec{v}_h(t-\bar{t}_1)
+\bar{\vec{v}}_1(\bar{t}_1-\bar{t}_0)+\bar{\vec{v}}_0\bar{t}_0)
e^{-2\bar\gamma_{-\vec{p}_h}(t-\bar{t}_1)
-2\bar\gamma_{\bar{\vec{p}}_1}(\bar{t}_1-\bar{t}_0)
-2\bar\gamma_{\vec{p}_0}\bar{t}_0}.
\end{eqnarray}
In terms of this correction, the Raman scattering probability can be
expressed as
\begin{eqnarray}
4\pi\,\frac{dI_{4K}}{do_{out}}&=&2\pi
\left(\frac{ev}c\right)^2\frac{2\pi{c}^2}{\omega_{out}}
\frac{\omega_{out}^2}{2\pi^2c^3}\,
\frac{2\pi{e}^2}c\,\frac{8\pi{v}^2}{\omega_{in}}
\times\nonumber\\ &&{}\times
\int{d}^2\vec{r}\,\frac{d^2\vec{p}_0}{(2\pi)^2}\,
\frac{d^2\vec{p}_2}{(2\pi)^2}\,
|[\vec{e}_{in}\times\vec{e}_{\vec{p}_0}]_z|^2
|[\vec{e}_{out}\times\vec{e}_{\vec{p}_2}]_z|^2
\delta(\Re\xi_{\vec{p}_0}+\Re\bar\xi_{\vec{p}_0}-\omega_{in})\,
\df^{(eehh)}_{\vec{p}_2,-\vec{p}_2}(\vec{r},\vec{r}).
\end{eqnarray}
Let us change time integration variables according to
\begin{eqnarray}\nonumber
&&\int\limits_0^\infty{dt}
\left[\int\limits_0^tdt_1\int\limits_0^{t_1}dt_0\,\mathcal{F}(t-t_1,t_1-t_0,t_0)\right]
\left[\int\limits_0^td\bar{t}_1\int\limits_0^{\bar{t}_1}d\bar{t}_0\,
\bar{\mathcal{F}}(t-\bar{t}_1,\bar{t}_1-\bar{t}_0,\bar{t}_0)\right]=\\
&&=\int\limits_{0}^\infty{d}t_0\,dt_1\,dt_2\,d\bar{t}_0\,d\bar{t}_1\,d\bar{t}_2\,
\delta(t_0+t_1+t_2-\bar{t}_0-\bar{t}_1-\bar{t}_2)\,\mathcal{F}(t_2,t_1,t_0)\,
\bar{\mathcal{F}}(\bar{t}_2,\bar{t}_1,\bar{t}_0),
\end{eqnarray}
and evaluate the time integral, following Sec.~\ref{sec:Intdev}. We obtain
\begin{equation}
\frac{1}{2|[\vec{v}_0\times\vec{v}_2]_z|}
\int\limits_0^\infty{d}t_1\,d\bar{t}_1\,e^{-2\gamma_x{t}_1-2\gamma_y\bar{t}_1}
\int\limits_\mathcal{O}dt_0=
\frac{1}{|[\vec{v}_0\times\vec{v}_2]_z|}\int\limits_0^{\pi/2}d\phi\,
\frac{\zeta_{max}(\phi)-\zeta_{min}(\phi)}
{8\left[\gamma_x\cos\phi+\gamma_y\sin\phi\right]^3}.
\end{equation}
Rearranging the energy $\delta$-functions and we arrive at
\begin{eqnarray}
4\pi\,\frac{dI_{4K}}{do_{out}}&=&16\pi^2\left(\frac{e}c\right)^4
\left(\frac{\lambda_K}{2\pi}\right)^4v^8\int
d^2\vec{p}_0\,d^2\vec{p}_1\,d^2\bar{\vec{p}}_1\,d^2\vec{p}_2\,
\left|[\vec{v}_0\times\vec{e}_{in}]_z\right|^2
\left|[\vec{v}_2\times\vec{e}_{out}]_z\right|^2
\times\nonumber\\ &&  {}\times
\sin^2\frac{\varphi_0-\varphi_1}2\sin^2\frac{\varphi_1-\varphi_2}2
\sin^2\frac{\varphi_0-\bar\varphi_1}2
\sin^2\frac{\bar\varphi_1-\varphi_2}2
\times\nonumber\\ &&{}\times
2\delta(\Re\xi_{\vec{p}_0}+\Re\bar\xi_{\vec{p}_0}-\omega_{in})\,
\delta(\Re\xi_{\vec{p}_1}-\Re\xi_{\vec{p}_0}+\omega_{\vec{p}_0-\vec{p}_1})\,
\delta(\Re\bar\xi_{\bar{\vec{p}}_1}-\Re\bar\xi_{\vec{p}_0}
+\omega_{-\vec{p}_0+\bar{\vec{p}}_1})
\times\nonumber\\ &&{}\times{2}
\delta(\Re\xi_{\vec{p}_2}+\Re\bar\xi_{\vec{p}_2}-\omega_{in}
+\omega_{\vec{p}_0-\vec{p}_1}+\omega_{\vec{p}_1-\vec{p}_2}
+\omega_{-\vec{p}_0+\bar{\vec{p}}_1}+\omega_{-\bar{\vec{p}}_1+\vec{p}_2})
\times\nonumber\\ &&{}\times
\delta(\Re\xi_{\vec{p}_2}-\Re\bar\xi_{\vec{p}_2}
-\Re\xi_{\vec{p}_0}+\Re\bar\xi_{\vec{p}_0}
+\omega_{\vec{p}_0-\vec{p}_1}+\omega_{\vec{p}_1-\vec{p}_2}
-\omega_{-\vec{p}_0+\bar{\vec{p}}_1}-\omega_{-\bar{\vec{p}}_1+\vec{p}_2})
\times\nonumber\\ &&{}\times
\frac{1}{|[\vec{v}_0\times\vec{v}_2]_z|}\int\limits_0^{\pi/2}d\phi\,
\frac{\zeta_{max}(\phi)-\zeta_{min}(\phi)}
{8\left[\gamma_x\cos\phi+\gamma_y\sin\phi\right]^3}.\label{I4Dkinur=}
\end{eqnarray}
The first four $\delta$-functions constrain the electronic momenta
to lie on the resonant manifold (Fig.~\ref{fig:resonance}). The
argument of the fifth $\delta$-function is nothing but~$\Delta$,
defined in Eq.~(\ref{Delta=}). For $\omega_{\vec{q}}=\omega_0$
this $\delta$-function equals either~$\infty$ in electron-hole symmetric
case, or~0 in the electron-hole asymmetric case. Thus,
Eq.~(\ref{I4Dkinur=}) makes sense only for dispersive phonons.
In this case it gives the probability 4~times smaller than that given
by Eqs.~(\ref{I4Dfinal=}),~(\ref{intdphi=}). One factor of~2 is due
to the constructive interference of the amplitudes for the two spin
projections: the spin degeneracy multiplies the {\em matrix element},
obtained by tracing in the electron loop. The other factor of two is
due to the constructive interference of two time-reversed processes,
described in the end of Sec.~\ref{sec:resonant}, which may also be
viewed as the interference of the processes in the two electronic valleys.



\section{Logarithmic terms in self-energy and vertex corrections}%
\label{app:logarithms}

The RPA-dressed Coulomb propagator~(\ref{VRPA=}) determines the
electron self-energy:
\begin{eqnarray}
&&\Sigma^{ee}(\vec{p},\ep)=
i\int\frac{d\omega}{2\pi}\frac{d^2\vec{q}}{(2\pi)^2}\,
V(\vec{q},\omega)\,G(\vec{p}-\vec{q},\ep-\omega)=
-\int\frac{d\varpi}{2\pi}\frac{d^2\vec{q}}{(2\pi)^2}\,
V(\vec{q},i\varpi)\,G(\vec{p}-\vec{q},\ep-i\varpi)\approx\nonumber\\
&&\approx\frac{16g}{\mathcal{N}}
\int\frac{d\varpi}{2\pi}\frac{d^2(v\vec{q})}{(2\pi)^2}\,
\frac{1}{vq}\,\frac{\sqrt{(vq)^2+\varpi^2}}{gvq+\sqrt{(vq)^2+\varpi^2}}
\left[-\frac{i\varpi+v\vec{q}\vec\Sigma}{\varpi^2+(vq)^2}
+\frac{\ep+v\vec{p}\vec\Sigma}{\varpi^2+(vq)^2}
+\frac{2(i\ep\varpi-v\vec{p}\vec{q})(i\varpi+v\vec{q}\vec\Sigma)}
{(\varpi^2+(vq)^2)^2}\right]=\nonumber\\
&&=\frac{4g}{\pi^2\mathcal{N}}
\int\limits_{\xi_{min}}^{\xi_{max}}d\xi
\int\limits_{-\infty}^\infty{d}\varpi\,
\frac{\sqrt{\xi^2+\varpi^2}}{g\xi+\sqrt{\xi^2+\varpi^2}}
\left[\frac{\xi^2-\varpi^2}{(\varpi^2+\xi^2)^2}\,\ep
+\frac{\varpi^2}{(\varpi^2+\xi^2)^2}\,v\vec{p}\vec\Sigma\right]
\approx\nonumber\\
&&\approx\frac{8}{\pi^2\mathcal{N}}
\left(\ln\frac{\xi_{max}}{\xi_{min}}\right)
\int\limits_0^\pi{d}\varphi\left[
(-\ep+v\vec{p}\vec\Sigma)\,\frac{g}{2(1+g\sin\varphi)}
+(2\ep-v\vec{p}\vec\Sigma)\,\frac{g\sin^2\varphi}{2(1+g\sin\varphi)}\right]
=\nonumber\\
&&=\frac{8}{\pi^2\mathcal{N}}
\left(\ln\frac{\xi_{max}}{\xi_{min}}\right)
\left[(-\ep+v\vec{p}\vec\Sigma)\,\frac{g\arccos{g}}{\sqrt{1-g^2}}
+(2\ep-v\vec{p}\vec\Sigma)
\left(1-\frac{\pi}{2g}+\frac{\arccos{g}}{g\sqrt{1-g^2}}\right)\right].
\end{eqnarray}
In the first line we have performed the Wick rotation
$\omega=i\varpi$ (corresponding to the Matsubara representation
for zero temperature). In the second line we have expanded
$G(\vec{p}-\vec{q},\ep-i\varpi)$ to the first order in $\ep$ and
$v\vec{p}$, since the integral is dominated by
$\varpi\sim{v}q\gg\ep,vp$. In the third line we have integrated over
the directions of~$\vec{q}$. In the fourth line we have replaced the
integration region with the momentum cutoff $\xi_{min}<vq<\xi_{max}$,
$-\infty<\varpi<\infty$ by the region
$\xi_{min}<\sqrt{(vq)^2+\varpi^2}<\xi_{max}$, which does not change
the leading logarithmic asymptotics. The lower cutoff
$\xi_{min}\sim\max\{vp,\ep\}$, the upper cutoff $\xi_{max}\sim{v}/a$
is of the order of the electronic bandwidth.

\begin{figure}
\includegraphics[width=8cm]{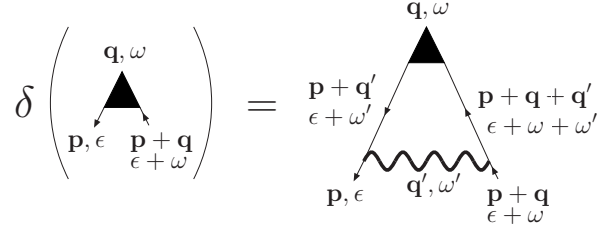}
\caption{\label{fig:CoulGamma} Coulomb correction to a generic vertex
$-i\Gamma(\vec{p},\ep;\vec{q},\omega)$, shown by the triangle.}
\end{figure}

Let us now calculate the Coulomb correction to an arbitrary matrix
vertex $-i\Gamma(\vec{p},\ep;\vec{q},\omega)$, corresponding to the
process, shown in Fig.~\ref{fig:CoulGamma}:
\begin{eqnarray}
&&\delta\Gamma(\vec{p},\ep;\vec{q},\omega)=\nonumber\\
&&=i\int\frac{d\omega'}{2\pi}\frac{d^2\vec{q}'}{(2\pi)^2}\,
V(\vec{q}',\omega')\,G(\vec{p}+\vec{q}',\ep-\omega')\,
\Gamma(\vec{p}+\vec{q}',\ep+\omega';\vec{q},\omega)\,
G(\vec{p}+\vec{q}+\vec{q}',\ep+\omega+\omega')=\nonumber\\
&&=-\int\frac{d\varpi}{2\pi}
\frac{d^2\vec{q}'}{(2\pi)^2}\frac{2\pi{e}^2}{q'}\,
\frac{\sqrt{(vq')^2+\varpi^2}}{gvq'+\sqrt{(vq')^2+\varpi^2}}
\times\nonumber\\&&\qquad{}\times
\frac{\ep-i\varpi+v(\vec{p}+\vec{q}')\cdot\vec\Sigma}
{(-i\ep+\varpi)^2+v^2|\vec{p}+\vec{q}'|^2}\,
\Gamma(\vec{p}+\vec{q}',\ep+\omega';\vec{q},\omega)\,
\frac{\ep-\omega-i\varpi+v(\vec{p}+\vec{q}+\vec{q}')\cdot\vec\Sigma}
{(-i\ep+i\omega+\varpi)^2+v^2|\vec{p}+\vec{q}+\vec{q}'|^2}
\approx\nonumber\\ &&\approx\frac{e^2}{2\pi{v}}
\int\limits_{\xi_{min}}^{\xi_{max}}d\xi\int\limits_{-\infty}^\infty{d}\varpi\,
\frac{\sqrt{\xi^2+\varpi^2}}{g\xi+\sqrt{\xi^2+\varpi^2}}
\frac{\varpi^2\Gamma-\xi^2\vec\Sigma\Gamma\vec\Sigma/2}
{(\varpi^2+\xi^2)^2}
=\nonumber\\ &&=\frac{4g}{\pi^2\mathcal{N}}\ln\frac{\xi_{max}}{\xi_{min}}
\int\limits_0^\pi{d}\varphi\,
\frac{(1-\sin^2\varphi)\Gamma-\sin^2\varphi\,\vec\Sigma\Gamma\vec\Sigma/2}
{1+g\sin\varphi}=\nonumber\\
&&=\frac{8}{\pi^2\mathcal{N}}\ln\frac{\xi_{max}}{\xi_{min}}\left[
\tilde{f}(g)\,\Gamma-f(g)\,(\Gamma+\vec\Sigma\Gamma\vec\Sigma/2)\right].
\end{eqnarray}
Here we have assumed that the dependence of
$\Gamma(\vec{p},\ep;\vec{q},\omega)$ on $\vec{p}$~and~$\ep$
is weak. If this dependence comes entirely from the renormalization,
which is true in our case, its weakness is due to the smallness of the
parameter $1/\mathcal{N}$.


Let us repeat the calculation of the self-energy~(\ref{SigmaE2=}) due
to electron-phonon interaction:
\begin{eqnarray}
\Sigma^{ph}(\vec{p},\ep)&=&
-\int\frac{d\varpi}{2\pi}\frac{d^2\vec{q}}{(2\pi)^2}\sum_\phonon
\frac{F_\phonon^2}{2M\omega_\phonon}\frac{\sqrt{27}a^2}{4}\,
D_\phonon(\vec{q},i\varpi)
(\Lambda\Sigma)_\phonon{G}(\vec{p}-\vec{q},\ep-i\varpi)\,
(\Lambda\Sigma)_\phonon
\approx\nonumber\\ &\approx&
(\lambda_{\Gamma}+\lambda_K)
\int\frac{d^2(v\vec{q})}{(2\pi)^2}\,\frac{\ep}{(vq)^2}=
(\lambda_{\Gamma}+\lambda_K)
\left(\ln\frac{\xi_{max}}{\xi_{min}}\right)
\frac\ep{2\pi},\label{SigmaE2log=}
\end{eqnarray}
where the phonon mode index $\phonon$ runs over the two modes
belonging to the $E_2$~representation and the two modes belonging
to the $E_1'$ representation, the corresponding matrices
$(\Lambda\Sigma)_{\phonon}$ being $-\Lambda_z\Sigma_y$,
$\Lambda_z\Sigma_x$, $\Lambda_x\Sigma_z$, $\Lambda_y\Sigma_z$.
Note that the here the integral is dominated by
$\varpi\sim\max\{\ep,\omega_\phonon\}$, in contrast with the Coulomb
self-energy where we had $\varpi\sim{v}q$. Thus, to calculate the
integral in Eq.~(\ref{SigmaE2log=}) we simply approximated
\begin{equation}\label{D=delta}
D_\phonon(\vec{q},i\varpi)\to
\frac{2\omega_\phonon}{-\varpi^2-\omega_\phonon^2}\approx
-2\pi\,\delta(\varpi),\quad |\varpi|\gg\omega_\phonon.
\end{equation}

\newcommand{\up}{\underline{p}}
\newcommand{\uq}{\underline{q}}
The sum of the two diagrams in Fig.~\ref{fig:Phonon2LineRG}
gives the following expression for the effective two-electron vertex
[we denoted $\up\equiv(\vec{p},\ep)$, $\uq\equiv(\vec{q},\omega)$
for compactness]:
\begin{eqnarray}
\delta\Gamma^{(2)}(\up,\up';\uq)&=&i\int\frac{d^3\uq'}{(2\pi)^3}
\sum_{\phonon,\phonon'}
\frac{F_\phonon^2}{2M\omega_\phonon}\frac{\sqrt{27}a^2}{4}\,
\frac{F_{\phonon'}^2}{2M\omega_{\phonon'}}\frac{\sqrt{27}a^2}{4}\,
D_{\phonon}(\uq-\uq')\,D_{\phonon'}(\uq')\times\nonumber\\
&&{}\times(\Lambda\Sigma)_{\phonon}G(\up+\uq')\,
(\Lambda\Sigma)_{\phonon'}\otimes
\left[(\Lambda\Sigma)_{\phonon}G(\up'-\uq')\,
(\Lambda\Sigma)_{\phonon'}
+(\Lambda\Sigma)_{\phonon'}G(\up'-\uq+\uq')\,
(\Lambda\Sigma)_{\phonon}\right],
\end{eqnarray}
Due to the condition~(\ref{D=delta}), we can approximate
\begin{equation}
G(\up+\uq')\approx -G(\up'-\uq')\approx{G}(\up'-\uq+\uq')\approx
-\frac{v\vec{q}\cdot\vec\Sigma}{(vq)^2}\,.
\end{equation}
Then the frequency integral is simply
\begin{equation}
i\int\frac{d\omega'}{2\pi}\,
D_{\phonon}(\omega-\omega')\,D_{\phonon'}(\omega')=
\frac{2(\omega_\phonon+\omega_{\phonon'})}
{\omega^2-(\omega_\phonon+\omega_{\phonon'}-io)^2}\equiv
{D}_{\phonon+\phonon'}(\omega)\,.
\end{equation}
Taking the $\vec{q}'$-integral, we obtain
\begin{equation}\label{Gamma2mat=}
\delta\Gamma^{(2)}(\up,\up';\uq)=
\frac{v^2}{4}\ln\frac{\xi_{max}}{\xi_{min}}\sum_{\phonon,\phonon'}
\frac{\lambda_\phonon\lambda_{\phonon'}}{2\pi}
\left\{
\frac{1}{2}\,(\Lambda\Sigma)_{\phonon}
\Sigma_i(\Lambda\Sigma)_{\phonon'}\otimes
\left[-(\Lambda\Sigma)_{\phonon}\Sigma_i(\Lambda\Sigma)_{\phonon'}
+(\Lambda\Sigma)_{\phonon'}\Sigma_i(\Lambda\Sigma)_{\phonon}\right]
\right\}.
\end{equation}
The matrix expression appearing in the braces is evaluated for
each of the 16 combinations of the indices $\phonon,\phonon'$
in Table~\ref{tab:vertexRG}. Adding up the contributions, we
obtain Eq.~(\ref{deltaGamma2=}).

\begin{table*}
\begin{tabular}[t]{|c||c|c|c|c|} \hline
$\phonon,\phonon'$ & $-\Lambda_z\Sigma_y$ & $\Lambda_z\Sigma_x$ &
$\Lambda_x\Sigma_z$ & $\Lambda_y\Sigma_z$ \\ \hline\hline
$-\Lambda_z\Sigma_y$ & 0 & 0 &
$\Lambda_y\Sigma_z\otimes\Lambda_y\Sigma_z$ &
$\Lambda_x\Sigma_z\otimes\Lambda_x\Sigma_z$ \\ \hline
$\Lambda_z\Sigma_x$ & 0 & 0 &
$\Lambda_y\Sigma_z\otimes\Lambda_y\Sigma_z$ &
$\Lambda_x\Sigma_z\otimes\Lambda_x\Sigma_z$ \\ \hline
$\Lambda_x\Sigma_z$ &
$\Lambda_y\Sigma_z\otimes\Lambda_y\Sigma_z$ &
$\Lambda_y\Sigma_z\otimes\Lambda_y\Sigma_z$ & 0 &
$\Lambda_z\Sigma_y\otimes\Lambda_z\Sigma_y
+\Lambda_z\Sigma_x\otimes\Lambda_z\Sigma_x$\\ \hline
$\Lambda_y\Sigma_z$ &
$\Lambda_x\Sigma_z\otimes\Lambda_x\Sigma_z$ &
$\Lambda_x\Sigma_z\otimes\Lambda_x\Sigma_z$ &
$\Lambda_z\Sigma_y\otimes\Lambda_z\Sigma_y
+\Lambda_z\Sigma_x\otimes\Lambda_z\Sigma_x$ & 0\\ \hline
\end{tabular}
\caption{\label{tab:vertexRG} The matrix expression
in the braces in Eq.~(\ref{Gamma2mat=}), evaluated for
all 16 combinations of the phonon indices
$\phonon,\phonon'$.}
\end{table*}

\section{Polarization operator}\label{app:polarization}

Polarization operator with arbitrary matrix vertices
$\Sigma_i$, $\Sigma_j$, ${i,j}=x,y,z,0$ (we denoted
$\Sigma_0\equiv\unitmatrix$), can be conveniently
calculated in the coordinate representation:\cite{Foster}
\begin{eqnarray}
&&-i\Pi_{ij}(\vec{r}-\vec{r}',t-t')=
\frac{\mathcal{N}}{2}\Tr_{4\times{4}}
\{\Sigma_iG(\vec{r}-\vec{r}',t-t')\,
\Sigma_jG(\vec{r}'-\vec{r},t'-t)\},\\
&&(i\partial_t+iv\vec\Sigma\cdot\vec\nabla)\,
G(\vec{r},t)
=\unitmatrix\,\delta(\vec{r})\,\delta(t).
\end{eqnarray}
If $\Lambda$~matrices are also present in the vertices, tracing
of them is trivial: the matrices in the two vertices must coincide
for the trace not to vanish, then their product is equal to the unit
matrix.

It is convenient to switch to the imaginary time $\tau=it$,
$i\partial_t\to-\partial_\tau$. Then the Green's function in the
coordinate representation can be found by using the analogy
with the 3D Coulomb problem. Namely, we introduce the third
dimension $z=v\tau$, and notice that
\begin{equation}
(-\partial_\tau+iv\vec\Sigma\cdot\vec\nabla)
(-\partial_\tau-iv\vec\Sigma\cdot\vec\nabla)=v^2\nabla_{3D}^2.
\end{equation}
Since the inverse of the 3D Laplacian is the Coulomb potential,
we obtain
\begin{eqnarray}
G(\vec{r},\tau)=(\partial_\tau+iv\vec\Sigma\cdot\vec\nabla)\,
\frac{1}{4\pi{v}\sqrt{v^2\tau^2+r^2}}=
-\frac{v\tau+i\vec\Sigma\cdot\vec{r}}{4\pi(v^2\tau^2+r^2)^{3/2}}\,.
\end{eqnarray}
Using auxiliary relations (here $i,j=x,y,z$),
\begin{subequations}\begin{eqnarray}
&&\Tr_{4\times{4}}\left\{\Sigma_i\Sigma_k\Sigma_j\Sigma_l\right\}
=4\delta_{ik}\delta_{jl}+4\delta_{il}\delta_{jk}-4\delta_{ij}\delta_{kl}\,,\\
&&\partial_i\partial_j\frac{1}{R^2}
=\frac{8x_ix_j-2\delta_{ij}R^2}{R^6},\;\;\;
R^2=x^2+y^2+z^2,\qquad
\end{eqnarray}\end{subequations}
we calculate the polarization operator (here $i,j=x,y$):
\begin{subequations}\begin{eqnarray}
&&\Pi_{00}(\vec{r},\tau)=
-\mathcal{N}\,\frac{v^2\tau^2-r^2}{8\pi^2(v^2\tau^2+r^2)^3}=
\frac{\mathcal{N}}{32\pi^2}\,(\partial_x^2+\partial_y^2)\,\frac{1}{v^2\tau^2+r^2},\\
&&\Pi_{zz}(\vec{r},\tau)=
-\mathcal{N}\,\frac{v^2\tau^2+r^2}{8\pi^2(v^2\tau^2+r^2)^3}=
-\frac{2\mathcal{N}}{32\pi^2}\,\nabla_{3D}^2
\frac{1}{v^2\tau^2+r^2},\\
&&\Pi_{ij}(\vec{r},\tau)=
\mathcal{N}\,\frac{4x_ix_j-2\delta_{ij}(v^2\tau^2+r^2)}{16\pi^2(v^2\tau^2+r^2)^3}
=
\frac{\mathcal{N}}{32\pi^2}\,\partial_i\partial_j\,
\frac{1}{v^2\tau^2+r^2}+\frac{\delta_{ij}}2\,\Pi_{zz},\\
&&\Pi_{0i}(\vec{r},\tau)=
-\mathcal{N}\,\frac{izx_i}{8\pi^2(v^2\tau^2+r^2)^3}=
-\frac{\mathcal{N}}{32\pi^2}\,\frac{i}v\,\partial_\tau\partial_i\,
\frac{1}{v^2\tau^2+r^2}.
\end{eqnarray}\end{subequations}
Using the 3D Fourier transform
\begin{equation}
\int{d}^3\vec{R}\,\frac{e^{i\vec{Q}\vec{R}}}{32\pi^2{R}^2}=
\int\frac{d^3\vec{R}}{(2\pi)^3}\frac{\pi}{4R^2}\,e^{i\vec{Q}\vec{R}}
=\frac{1}{16}\,\frac{1}Q,
\end{equation}
we obtain $\Pi(\vec{q},\omega)$ (up to a $\vec{q},\omega$-independent
constant coming from $r,\tau\to{0}$):
\begin{subequations}\begin{eqnarray}
\Pi_{00}(\vec{q},\omega)&=&
-\frac{\mathcal{N}}{16v^2}\frac{v^2q^2}{\sqrt{v^2q^2-\omega^2}},\\
\Pi_{0i}(\vec{q},\omega)&=&
-\frac{\mathcal{N}}{16v^2}\frac{\omega{v}q_i}{\sqrt{v^2q^2-\omega^2}},\\
\Pi_{ij}(\vec{q},\omega)&=&
-\frac{q_iq_j}{q^2}\frac{\mathcal{N}}{16v^2}\frac{\omega^2}{\sqrt{v^2q^2-\omega^2}}
+\left(\delta_{ij}-\frac{q_iq_j}{q^2}\right)\frac{\mathcal{N}}{16v^2}\,
\sqrt{v^2q^2-\omega^2},
\label{Pi_ij=}\\
\Pi_{zz}(\vec{q},\omega)&=&
\frac{2\mathcal{N}}{16v^2}\,{\sqrt{v^2q^2-\omega^2}},\label{Pizz=}\\
\Pi_{0z}(\vec{q},\omega)&=&0,\\
\Pi_{iz}(\vec{q},\omega)&=&0.
\end{eqnarray}\end{subequations}
Given the polarization operator, one can find corrections to the optical
phonon frequencies due the electron-phonon interaction:
\begin{subequations}\begin{eqnarray}
D_K^{-1}(\vec{q},\omega)-\frac{\lambda_K}2\,v^2\Pi_{zz}(\vec{q},\omega)=0
&\Rightarrow&
\delta\omega_K(\vec{q})\approx\frac{\lambda_K}4\,\sqrt{v^2q^2-\omega_K^2},\\
D_{\Gamma,L}^{-1}(\vec{q},\omega)-\frac{\lambda_\Gamma}2\,v^2\Pi_T(\vec{q},\omega)=0
&\Rightarrow&
\delta\omega_{\Gamma,L}(\vec{q})\approx
\frac{\lambda_\Gamma}8\,\sqrt{v^2q^2-\omega_\Gamma^2},\\
D_{\Gamma,T}^{-1}(\vec{q},\omega)-\frac{\lambda_\Gamma}2\,v^2\Pi_L(\vec{q},\omega)=0
&\Rightarrow&
\delta\omega_{\Gamma,T}(\vec{q})\approx
-\frac{\lambda_\Gamma}8\,\frac{\omega_\Gamma^2}{\sqrt{v^2q^2-\omega_\Gamma^2}}.
\end{eqnarray}\end{subequations}
At $vq<\omega_\phonon$ the square roots are imaginary which
corresponds to the decay of phonons into the continuum of electron-hole
pairs.

\end{widetext}


\begin{thebibliography}{99}
\bibitem{mistake}
{There is an overall factor of $2\pi$ missing in Eq.~(6) of
Ref.~\onlinecite{myself}, which affects all $I_{2K}$, $I_{2\Gamma}$,
and $I_{4K}$, but not their ratios. Besides, there is a factor
of~$\pi/4$ missing in Eq.~(9) of Ref.~\onlinecite{myself}, which
affects the central result [Eq.~(1) of Ref.~\onlinecite{myself}].
Fortunately, $\pi/4\approx{0}.79$ is not very much different from
unity.}
\end{thebibliography}
\end{document}